\documentclass[a4paper,11pt]{article}
\usepackage{jcappub} 
\usepackage{enumitem}
\usepackage{hyperref}
\newcommand\Tstrut{\rule{0pt}{3.1ex}} 

\RequirePackage[top=10mm,bottom=10mm,left=22mm,right=22mm,head=12mm,includeheadfoot]{geometry}

\hypersetup{
    colorlinks = true,
    linkcolor = blue,
    citecolor = blue,
    urlcolor = Blue
}
\usepackage{fontawesome5}

\makeatletter
\newcommand{\github}[1]{%
   \href{#1}{\faGithubSquare}%
}
\makeatother

\usepackage{cancel}
\usepackage{etoolbox}

\usepackage{fancyhdr}

\usepackage{amssymb}
\usepackage{amsmath,bm}
\usepackage{mathtools}
\hypersetup{colorlinks=True, citecolor=blue} 
\usepackage{amsfonts}
\usepackage{amssymb}
\usepackage{caption}
\usepackage{float}
\usepackage{bm}
\usepackage{graphicx}
\usepackage{color}
\usepackage{verbatim}
\usepackage{blindtext, multicol}
\usepackage{multirow}
\usepackage{epigraph}
\usepackage[dvipsnames]{xcolor}
\usepackage{subcaption}
\def\beq{\begin{equation}}
\def\eeq{\end{equation}}
\def\ber{\begin{eqnarray}}
\def\eer{\end{eqnarray}}
\def\benu{\begin{enumerate}}
\def\eenu{\end{enumerate}}

\def\l{\left}
\def\r{\right}
\def\f{\frac}
\def\d{{\rm d}}

\def\mpc{\rm Mpc^{-1}}

\def\cosmolattice{${\mathcal C}$osmo${\mathcal L}$attice}
\def \grad{\bm{\nabla}}

\def \red {\color{red}}

\def \fgreen {\color{ForestGreen}}

\def \violet {\color{violet}}
\def \burntor {\color{BurntOrange}}

\def \Sepia {\color{Sepia}}

\def \Blue {\color{Blue}}

\usepackage[sorting=none,style=numeric-comp]{biblatex}
\usepackage[english]{babel}
\usepackage{csquotes}
\DeclareUnicodeCharacter{0301}{\'{i}}
\usepackage[nottoc]{tocbibind}

\usepackage{academicons}
\usepackage{xcolor}
\usepackage{fontawesome5}
\definecolor{orcidlogocol}{rgb}{0.65, 0.807, 0.223}

\newcommand{\orcid}[1]{$\,$\href{https://orcid.org/#1}{\textcolor{orcidlogocol}{\faOrcid}}}

\title{ Formation and decay of oscillons after inflation in the presence of an external coupling, Part-I: Lattice simulations}

\author[a]{Mohammed Shafi\orcid{0000-0003-0438-4155},}
\author[b]{Edmund J. Copeland\orcid{0000-0003-3959-6051},}
\author[c]{Rafid Mahbub\orcid{0000-0003-2665-2798},}
\author[b]{Swagat S. Mishra\orcid{0000-0003-4057-145X},}
\author[a]{Soumen Basak}
\affiliation[a]{School of Physics, Indian Institute of Science Education and Research, Thiruvananthapuram, 695551, India.}
\affiliation[b]{School of Physics and Astronomy,  University of Nottingham, Nottingham, NG7 2RD, UK.}
\affiliation[c]{Department of Physics, Gustavus Adolphus College, Saint Peter, MN 56082, USA.}

\emailAdd{mohammedshafi18@alumni.iisertvm.ac.in}
\emailAdd{edmund.copeland@nottingham.ac.uk}
\emailAdd{mahbub@gustavus.edu}
\emailAdd{swagat.mishra@nottingham.ac.uk}
\emailAdd{sbasak@iisertvm.ac.in}

\abstract{We investigate the formation and decay of oscillons during the post-inflationary reheating epoch from inflaton oscillations around asymptotically flat potentials $V(\varphi)$ in the presence of an external coupling of the form $\frac{1}{2}\, g^2 \, \varphi^2 \, \chi^2$. It is well-known that in the absence of such an external coupling, the attractive self-interaction term in the potential leads to the formation of copious amounts of long-lived oscillons both for symmetric and asymmetric plateau potentials.  We perform a detailed numerical analysis to study the formation of oscillons in the $\alpha$-attractor E- and T-model potentials using the publicly available lattice simulation code ${\cal C}$osmo${\cal L}$attice. We observe the formation of nonlinear oscillon-like structures with the average equation of state $\langle w_\varphi\rangle \simeq 0$ for a range of values of the inflaton self-coupling $\lambda$ and the external coupling $g^2$. Our results demonstrate that oscillons form even in the presence of an external coupling and we determine the upper bound on  $g^2$ which facilitates oscillon formation. We also find that eventually, these oscillons decay into the scalar inflaton radiation as well as into the quanta of the offspring field  $\chi$.  Thus, we establish the possibility that reheating could have proceeded through the channel of oscillon decay, along with the usual decay of the  oscillating  inflaton condensate into $\chi$ particles.  For a given value of the self-coupling $\lambda$, we notice  that  the lifetime of a population of oscillons  decreases with an increase in the strength of the external coupling, following an (approximately) {\em inverse power-law} dependence on $g^2$.}


\begin{document}
\maketitle

\section*{Units, Notations and Convention}
\begin{itemize}
\item Einstein's gravity ({\bf GR}) with sign $(-,+,+,+)$
\item  Natural units $\hbar,c =1$, reduced Planck mass $m_p = \f{1}{\sqrt{8\pi G}}$
\item Conformal time $ \int \, {\rm d} \tau = \int \, \f{{\rm d}t}{a(t)}$
\item Derivative \textit{w.r.t} coordinate time $\frac{\d \Upsilon}{\d t} \equiv \dot{\Upsilon}$ and \textit{w.r.t} conformal time $\frac{\d \Upsilon}{\d \tau} \equiv \Upsilon^\prime$
\item $3d$ vectors are denoted in \textbf{boldface}, \textit{e.g.}  the comoving spatial coordinate vector is denoted as $\bm{x}$, while the comoving momentum is denoted as $\bm{k}$
\item Inflaton field $\varphi$, homogeneous inflaton condensate $\phi$ and inflaton fluctuations $\delta\varphi$
\end{itemize}

\newpage

\section{Introduction}
\label{sec:intro}

Cosmic inflation is a rapid accelerated expansion of space in the very early universe which provides natural initial conditions for the hot Big Bang phase~\cite{Starobinsky:1980te,Guth:1980zm,Linde:1981mu,Albrecht:1982wi,Linde:1983gd,Baumann_TASI}. In the simplest inflationary scenario, this acceleration is facilitated by a single canonical scalar field $\varphi$, called the inflaton field, which slowly rolls down a potential $V(\varphi)$. Scalar quantum fluctuations~\cite{Mukhanov:1981xt,Hawking:1982cz,Starobinsky:1982ee,Guth:1982ec,Baumann_TASI} generated during  inflation provide the primordial seeds for the formation of structure in the universe, while tensor fluctuations~\cite{Starobinsky:1979ty,Sahni:1990tx,Baumann_TASI,Baumann:2022mni} constitute a background of stochastic gravitational waves (GWs) in the post-inflationary universe.

Over the past two decades, strong constraints have been imposed on the functional form of the inflaton potential by precision measurements of the anisotropies of the Cosmic Microwave Background (CMB) radiation~\cite{Planck:2018vyg,Planck:2018jri}. The latest CMB observations by the BICEP/Keck collaboration~\cite{BICEP:2021xfz}, combined with the Planck measurements~\cite{Planck:2018jri}, constrain the ratio of the tensor to scalar power spectra, called the tensor-to-scalar ratio, to be $r \leq 0.036$ on large cosmological scales with $95\%$ confidence. The aforementioned bound on $r$ translates into an upper bound on the Hubble parameter during inflation, thus directly related to its energy scale. Equivalently, the latest CMB observations favour relatively shallow potentials, while ruling out a large number of popular models of inflation~\cite{Martin:2013tda}. For example, simple monomial potentials of the form $V(\varphi) \propto \varphi^n$ are disfavoured due to their tendency to overproduce primordial gravitational waves~\cite{Mishra:2022ijb}, whereas those with asymptotically flat plateaus (symmetric or asymmetric) are currently the most favoured models due to their suppressed levels of such primordial gravitational waves~\cite{Kallosh:2016gqp,Mishra:2022ijb}. 


Towards the end of inflation, the slow-roll conditions are violated as the inflaton field begins to descend  more rapidly down its potential. After the end of inflation, the inflaton begins to oscillate coherently around the minimum of its potential. The universe transitions to the reheating epoch characterised by the decay of the rapidly oscillating inflaton condensate, and the inflaton eventually transfers most of its energy into the Standard Model particles, marking the beginning of radiation domination. Historically, particle production during reheating was modelled by the slow perturbative decay of massive inflaton particles, eventually leading to the thermalisation of the universe to the radiation-dominated epoch~\cite{Albrecht:1982mp,Kolb:1990vq}. However, it was later realised that non-perturbative effects usually dominate during the initial regime of reheating due to the coherent nature of the inflaton  oscillations~\cite{Turner:1983he}. In particular, the phenomenon of \textit{parametric resonance}  results in the amplification of perturbations of the inflaton field and other fields coupled to it, resulting in copious particle production~\cite{Kofman:1994rk,Shtanov:1994ce,Kofman:1996mv,Kofman:1997yn,Greene:1997fu,Amin:2014eta,Lozanov:2019jxc}. In general, reheating is often divided into three distinct regimes -- \textit{(i)} a {\em preheating} phase where the non-perturbative effects lead to an exponential growth of the field fluctuations in distinct resonance bands,  \textit{(ii)} a {\em backreaction} phase where the produced field quanta strongly backreact on the inflaton condensate, thereby quenching the parametric resonance,  and \textit{(iii)} the {\em thermalisation} phase where a cascade of perturbative processes lead to the eventual thermalisation of the universe, thus facilitating the commencement of the hot Big Bang phase. 


During preheating in non-linear potentials, the oscillating inflaton condensate drives the amplification of perturbations of the inflaton field itself (aptly called \textit{self-resonance}) and other coupled bosonic fields (called \textit{external resonance}). This leads to the exponential growth of field fluctuations as $e^{\mu_k t}$, where $\mu_k$, known as the Floquet exponent, gives the associated timescale for the rapid growth. Extensive analyses on the subject have been performed in Refs.~\cite{Kofman:1994rk,Shtanov:1994ce,Kofman:1996mv,Kofman:1997yn,Greene:1997fu}. Of particular importance is the formation of dense and localised field configurations that oscillate in time, called \textit{oscillons}~\cite{Amin:2011hj,Zhou:2013tsa,Lozanov:2014zfa,Kim:2017duj,Lozanov:2017hjm,Amin:2018xfe,Lozanov:2019ylm,Zhang:2020bec,Kim:2021ipz,Mahbub:2023faw}. Remarkably, formation of oscillons is a prediction of generic real scalar field theories  with  
self-interaction potentials that are shallower than  $\varphi^2$ around the minimum because such potentials allow the scalar field to feel the effect of the attractive self-interaction as it oscillates about its minimum. Oscillons do not strictly possess a conserved charge, hence they are technically not stable. Nevertheless,  due to the fact that they are non-linear configurations,  they can be exceptionally long-lived (hence \textit{quasi-stable}), decaying slowly via the emission of radiation~\cite{Fodor:2008du,Hertzberg:2010yz,Salmi:2012ta,Zhang:2020bec}.  By relaxing the reality condition on $\varphi$, thereby allowing it to be a complex field, it is possible to form solitonic objects known as \textit{Q-balls}\footnote{Q-balls have a conserved charge in the particle number associated with a global U(1) symmetry. An approximate conserved charge is often defined for oscillons in analogy with Q-balls.}~\cite{Kusenko:1997si,Enqvist:1999mv,Kasuya:1999wu} which have been studied in the context of supersymmetric dark matter, and  more recently as possible sources to explain boson stars and superradiance phenomena therein~\cite{Almumin:2023wwi,Gao:2023gof}.


 Oscillons were extensively studied in condensed matter and nonlinear systems in Refs.~\cite{PhysRevE.53.2972,1996Natur.382..793U}.  In the context of early universe physics, they were initially considered in scenarios of cosmological phase transitions~\cite{Copeland:1995fq,Adib:2002ff,Farhi:2005rz}, while the study of oscillons has more recently been extended to other cosmological phenomenon, ranging from providing the seeds for primordial black holes (PBHs)~\cite{Cotner:2018vug,Cotner:2019ykd,Widdicombe:2019woy,Nazari:2020fmk}, of primordial gravitational waves~\cite{Easther:2006gt,Dufaux:2007pt,Antusch:2016con,Liu:2017hua,Amin:2018xfe,Helfer:2018vtq,Hiramatsu:2020obh} and for the formation of non-linear structure in ultra-light dark matter (ULDM)~\cite{Hu:2000ke,Olle:2019kbo,Kawasaki:2019czd,Ferreira:2020fam}. From the preheating perspective, since the formation of these structures require  potentials close to the minimum that are shallower than quadratic in $\varphi$, initial analytical and numerical investigations were carried out in theories with an  even  polynomial potential containing terms up to $\mathcal{O}(\varphi^6)$ of the form $V(\varphi) = m^2\varphi^2/2 - \lambda \, \varphi^4/4 + g \, \varphi^6/6$~\cite{Amin:2010dc,Amin:2010jq,Amin:2010xe}, where \textit{flat-top} oscillons were shown to exist. They have subsequently been studied in axion monodromy potentials~\cite{Amin:2011hj,Zhou:2013tsa,Lozanov:2014zfa,Sang:2019ndv}, hilltop potentials~\cite{Antusch:2017vga,Antusch:2019qrr} and in $\alpha$-attractor models~\cite{Lozanov:2017hjm,Lozanov:2019ylm,Mahbub:2023faw}. In Ref.~\cite{Lozanov:2017hjm}, for potentials of the form $V(\varphi) \propto \varphi^{2n}$ close to its minimum, the authors used the T-model to quantify the differences between $n=1$ and $n>1$  (where the minima are quadratic and non-quadratic respectively at leading order), leading to quite different field configurations as end products. They demonstrated that oscillons can form when $n=1$ such that the time-averaged equation of state parameter $\langle w_\varphi \rangle \rightarrow 0$. In cases where $n>1$,  dense transients form  initially which quickly fragment into radiation within a few $e$-folds with $\langle w_\varphi \rangle \rightarrow \frac{1}{3}$. More recently, Ref.~\cite{Mahbub:2023faw} demonstrated that oscillons can also form in asymmetric potentials by using lattice simulations of the E-model $\alpha$-attractor potential, obtaining results that are qualitatively similar to those of the T-model. 


Most of the oscillon literature has studied their formation by explicitly assuming the absence of alternate decay channels, such as neglecting the coupling of the inflaton to an external field. However, it is important to include such interactions and couplings since the universe needs to be reheated to a thermal plasma state before the commencement of Big Bang Nucleosynthesis (BBN). In principle, the inflaton can have couplings to both fermionic and bosonic degrees of freedom. However, the phenomenon of parametric resonance is only applicable to the production of bosons, due to Bose enhancement effects. Furthermore, most  UV complete theories of inflation, such as those arising from string compactifications, predict the existence of multiple scalar fields (\textit{moduli fields})~\cite{Baumann:2014nda,Cicoli:2023opf,Apers:2024ffe} along with the inflaton. Hence, it is natural to expect the inflaton to be coupled to one or more scalar degrees of freedom.


In this paper, we consider such a case where the inflaton field $\varphi$ is coupled to a light scalar field $\chi$, which we call the \textit{offspring field}. We carry out detailed (3+1)-dimensional lattice simulations of the inflaton dynamics in the E- and T-model $\alpha$-attractor potentials with a quadratic-quadratic external coupling to $\chi$ of the simplest form $\mathcal{L}_{\rm int}\supset\f{1}{2} \, g^2 \, \varphi^2 \chi^2$. By  considering a range of possible values of the coupling parameter $g^2$, we carry out a thorough numerical investigation of inflaton decay with the primary objective of quantifying the conditions for the formation and decay of oscillons. The presence of the two coupling parameters, namely the self-coupling $\lambda$ appearing in the inflaton potential (which determines the tensor-to-scalar ratio during inflation), and the external coupling $g^2$, allows us to determine the subspace of the parameter space, $\lbrace \lambda, \, g^2 \rbrace$, leading to oscillon formation as well as to study their subsequent decay due to the presence of such simple external interactions; all the while ensuring that the  inflationary predictions are consistent with the latest CMB constraints~\cite{Planck:2018jri,BICEP:2021xfz,Mishra:2022ijb}.  Consequently,  we establish an upper bound on the external coupling $g^2$ for which oscillons form for a given value of the self-coupling $\lambda$, both for the E- and the T-model potentials. Furthermore, in reference to the decay of oscillons, we establish the novel possibility of a substantial production of the offspring  particles during preheating to have arisen via  oscillon decay, in addition to the  the usual channel of  inflaton condensate decay. We observe that the lifetime of a population of  oscillons decreases with the increase in the strength of the external coupling, following an (approximately) inverse power-law dependence on $g^2$.  Our upcoming paper~\cite{Mishra:2024Part2} will be dedicated to an analytical treatment of oscillon formation and decay in asymptotically flat potentials in the presence of such an external coupling.

\medskip


This paper  is organized  as follows:   
in Sec.~\ref{sec:system},  after providing a brief introduction to the scalar field dynamics during inflation, we set up the complete set of coupled  field equations describing the expansion of space and the decay of the inflaton in the post-inflationary epoch.  We introduce both E- and T-model $\alpha$-attractors  as representatives for  asymptotically flat potentials of  symmetric and asymmetric types, respectively.   In Sec.~\ref{sec:analyt}, a Floquet analysis is carried out to characterize the  inflaton decay via  self and external  parametric resonances  in the linear regime  where we generate the instability chart for the quanta of the decay modes  during the preheating phase in order  to demonstrate the existence of a broad-band resonance region followed by multiple narrow ones. In Sec.~\ref{sec:simulations} we carry out  full $(3+1)d$ lattice simulations for  the E- and T-model potentials and confirm the formation of  oscillons for a subspace of the parameter space $\lbrace \lambda, \, g^2 \rbrace$. We also  study the lifetime of oscillons  due to their decay via scalar inflaton radiation as well as  into the quanta of the offspring field.  We conclude in Sec.~\ref{sec:discussion}  with a discussion of the primary inferences of this work, and its possible future extension, especially in the context of analytical treatment of the  oscillon lifetime.  Various appendices provide  supplementary information on the   inflationary and post-inflationary scalar field dynamics of asymptotically flat potentials.

\medskip

\section{Dynamics of the post-inflationary epoch}
\label{sec:system}
\subsection{Scalar field dynamics during inflation}
\label{sec:inf_dyn_during_inflation}
In the simplest inflationary scenario which we consider in this work, the quasi-exponential expansion of space is sourced by a single canonical scalar field called the \textit{inflaton} $\varphi(t,\bm{x})$ with a potential $V(\varphi)$ which is minimally coupled to gravity. In this standard paradigm,   inflaton couplings to the external fields are assumed to be very weak,  so that they can be neglected during inflation, ensuring that the dynamics is completely dictated by the inflaton field.  Hence, during inflation,  the system is described by the action 
\beq
S[g_{\mu\nu},\varphi] = \int \,  {\rm d}^4x \,  \sqrt{-g} \,  \l( \, \f{m_p^2}{2} \, R - \f{1}{2}  \, \partial_{\mu}\varphi \,  \partial_{\nu}\varphi  \, g^{\mu\nu}-V(\varphi)\r) \, .
\label{eq:Action_phi_g_munu}
\eeq
In linear perturbation theory during inflation, we split the metric and the inflaton field into their corresponding background and fluctuations, namely
$$g_{\mu\nu}(t,\bm{x}) = \bar{g}_{\mu\nu}(t) + \delta g_{\mu\nu}(t,\bm{x}) \, ; \quad \varphi(t,\bm{x}) = \phi(t) + \delta\varphi(t,\bm{x}) \, .$$
The  perturbed  line element in the  ADM formalism~\cite{Arnowitt:1962hi,Maldacena:2002vr} can be written as   
\beq
 {\rm d}s^2 = -  \alpha^2 \, {\rm d}t^2 + \gamma_{ij} \l( {\rm d}x^i + \beta^i \, {\rm d}t \r) \l( {\rm d}x^j + \beta^j \, {\rm d}t \r) \, ,
\label{eq:metric_ADM}
\eeq
where the non-dynamical  Lagrange multipliers, $\alpha = 1 + \delta\alpha$ and $\beta^i$, are the lapse and shift functions respectively, which are completely fixed by the  Hamiltonian and momentum constraints, while $\gamma_{ij}$ are the true dynamical metric perturbations.  During inflation, it is convenient to work in the {\em comoving gauge}\footnote{ Note that we do not work in the comoving gauge while describing the post-inflationary dynamics in Sec.~\ref{sec:dyn_inf_post}. In fact, the metric perturbations can be safely ignored while studying  the  preheating dynamics  which occurs  over shorter time scales  and on small scales $k > aH$.} where the inflaton and metric fluctuations are given by~\cite{Maldacena:2002vr,Baumann_TASI,Baumann:2018muz}
\beq 
\delta\varphi(t,\bm{x}) = 0 \, ; \quad \gamma_{ij}(t,\bm{x}) = a^2 \bigl[ \bigl(  1 + 2 \, \zeta(t,\bm{x}) \bigr) \, \delta_{ij} \, + \,  h_{ij}(t,\bm{x}) \bigr]  \, ,
\label{eq:metric_momov_gauge}
\eeq
where  $\zeta(t,\bm{x})$ and $ h_{ij}(t,\bm{x})$ are the gauge-invariant comoving curvature perturbation (scalar-type), and transverse and traceless perturbations (tensor-type), respectively.
The scalar perturbations eventually induce density and temperature fluctuations in the hot Big Bang phase, which subsequently facilitates the formation of the large-scale structure in the universe. Meanwhile, the tensor fluctuations propagate as gravitational waves (GWs) which, at late times, constitute a stochastic background of GWs.

The perturbed action in the comoving gauge during inflation can be expressed as~\cite{Baumann:2018muz} 
\beq
S[g_{\mu\nu},\varphi] = S_{B}[\bar{g}_{\mu\nu},\phi] + S^{(2)}[\zeta] + S^{(2)}[h_{ij}] + S_{\rm int}[\zeta, h_{ij}] \, ,
\label{eq:Action_perturbed_scalar-tensor}
\eeq
where $S_{B}[\bar{g}_{\mu\nu},\phi]$ is the background action, while $S^{(2)}[\zeta], \,  S^{(2)}[h_{ij}]$ are the quadratic actions for the scalar and tensor fluctuations (which results in linear field equations for the fluctuations).  The term $S_{\rm int}[\zeta, h_{ij}]$ is the action for fluctuations beyond the linear order which leads to primordial non-Gaussianity (detailed discussions on this topic can be found in   Refs.~\cite{Baumann_TASI,Maldacena:2002vr,Baumann:2018muz}).  

For the  background evolution of $\lbrace \bar{g}_{\mu\nu},\phi \rbrace$ during inflation, specializing to a spatially flat FLRW metric
and the homogeneous part $\phi$ of the  scalar field $\varphi$, one gets
\beq 
{\rm d}s^2  =  -{\rm d}t^2 + a^{2}(t) \;  \left[{\rm d}x^2 + {\rm d}y^2 + {\rm d}z^2\right] \, ,
\label{eq:FRW}
\eeq
\beq
T^{\mu}_{\;\:\;\nu} \equiv \mathrm{diag}\left(-\rho_{_{\phi}}, \,  p_{_{\phi}}, \,    \, p_{_{\phi}}, \,  p_{_{\phi}}\right) \, ,
\eeq
where the energy density $\rho_{_{\phi}}$, and pressure $p_{_{\phi}}$, of the homogeneous inflaton  condensate  are given by
\beq
\rho_{_{\phi}} = \frac{1}{2}{\dot\phi}^2 +\;  V(\phi) \, ; \quad 
p_{_{\phi}} = \frac{1}{2}{\dot\phi}^2 -\; V(\phi) \, .
\label{eq:rho_p_phi}
\eeq
Evolution of the scale factor $a(t)$ is governed by the Friedmann equations
\begin{align}
H^2 = \frac{1}{3m_p^2} \, \rho_{\phi} &= \frac{1}{3m_p^2} \left[\frac{1}{2}{\dot\phi}^2 +V(\phi)\right] \, ,
\label{eq:friedmann1}\\
\dot{H} \equiv \frac{\ddot{a}}{a}-H^2 &= -\frac{1}{2m_p^2}\, \dot{\phi}^2 \, ,
\label{eq:friedmann2}
\end{align} 
while  $\phi$ satisfies the equation of motion
\beq
{\ddot \phi}+ 3\, H {\dot \phi} + V_{,\phi} = 0 \, ,
\label{eq:phi_EOM}
\eeq
where $V_{,\phi}\equiv \d V/\d \phi$. Evolution of various physical quantities during inflation is usually described with respect to the number of $e$-folds of expansion which is defined by $N = {\rm ln}(a/a_i)$, where $a_i$  is the scale factor at some arbitrary epoch at very early times during inflation. 
A better physical  quantity that quantifies the extent of inflation is the  {\em  number of $e$-folds  before the end of inflation}, which is defined as 
\beq
N_e(a)  = \ln{ \l( \frac{a_e}{a} \r) }=\int_{t}^{t_e} H(t) \, {\rm d}t,
\label{eq:efolds}
\eeq
where $H(t)$ is the Hubble parameter during inflation. Note that  $a_e$ denotes the scale factor at the  
end of inflation, hence
$N_e  = 0$ corresponds to the end of inflation. Typically a period of quasi-de Sitter (exponential) inflation lasting for at least 60-70 $e$-folds is required in order to address the problems of the standard hot Big Bang model. We denote $N_\star$ as the number of $e$-folds (before the end of inflation) when the CMB pivot scale 
\beq
k_\star=(aH)_\star=0.05~\mpc
\label{eq:k_pivot_CMB}
\eeq
left the comoving Hubble radius during inflation. Typically  $N_\star \in [50,\,60]$ depending on the details of reheating after inflation (see Ref.~\cite{Liddle:2003as,Mishra:2021wkm}). In this work, we fix $N_\star = 60$.

Background kinematics during inflation is usually characterised by the first  two Hubble slow-roll parameters $\epsilon_H$, $\eta_H$, defined as ~\cite{Liddle:1994dx,Baumann_TASI} 
\begin{align}
\epsilon_H &= -\frac{\dot{H}}{H^2}= -\frac{{\rm d}\ln{H}}{{\rm d}N} = \frac{1}{2m_p^2} \, \frac{\dot{\phi}^2}{H^2} \, ,
\label{eq:epsilon_H}\\
\eta_H &= -\frac{\ddot{\phi}}{H\dot{\phi}}=\epsilon_H  - \frac{1}{2\epsilon_H} \, \frac{{\rm d}\epsilon_H}{{\rm d}N} \, .
\label{eq:eta_H}
\end{align}
For asymptotically flat  potentials,  there exists a  slow-roll regime of inflation (ensured by the presence of the Hubble friction term~\cite{Brandenberger:2016uzh, Mishra:2018dtg}  in Eq.~(\ref{eq:phi_EOM})), defined by
\beq
\epsilon_H,~|\eta_H| \ll 1 \, .
\label{eq:slow-roll_condition}
\eeq
Using the definition of the Hubble parameter, $H=\dot{a}/a$, we have $\ddot{a}/a=\dot{H}+H^2=H^2(1 + \dot{H}/H^2)$. From the expression for $\epsilon_{H}$ in Eq.~(\ref{eq:epsilon_H}), it is easy to see that 
\beq
\f{\ddot{a}}{a} =  H^2  \, \big( 1 - \epsilon_H \big) \, ,
\label{eq:inf_acc_cond}
\eeq
which implies that the universe accelerates, ${\ddot a} > 0$, when $\epsilon_{H} < 1$.
Using Eq.~(\ref{eq:friedmann1}), the expression for  $\epsilon_{H}$ in Eq.~(\ref{eq:epsilon_H}) reduces to $\epsilon_{H} \simeq \f{3}{2}\f{\dot{\phi}^2}{V}$  when ${\dot\phi}^2 \ll V$.  In fact, under the slow-roll conditions in Eq.~(\ref{eq:slow-roll_condition}), the Friedmann Eqs.~(\ref{eq:friedmann1}) and (\ref{eq:phi_EOM}) take the form
\beq
H^2 \simeq \f{1}{3m_p^2} V(\phi) \, ; \quad \dot{\phi} \simeq  - \f{V_{,\phi}}{3H} \, .
\label{eq:friedmann_SR1_SR2}
\eeq

For  asymptotically flat potentials the power spectra of scalar and tensor perturbations generated {\em via} vacuum quantum fluctuations  during slow-roll inflation  agree well with the latest CMB observations~\cite{Planck_inflation,BICEP:2021xfz} as discussed in App.~\ref{app:CMB_T_E_models}. The end of inflation is marked by the violation of slow roll, with  $\epsilon_H =1$,  which corresponds to $\dot{\phi}_{\rm end}^2 = V(\phi_{\rm end})$ as can be inferred from Eqs.~(\ref{eq:epsilon_H}) and (\ref{eq:inf_acc_cond}). After the end of inflation, the homogeneous inflaton condensate begins to oscillate around the minimum of its potential.

\subsection{Dynamics of the post-inflationary universe}
\label{sec:dyn_inf_post}
The inflationary paradigm is formulated to naturally facilitate the transition into the radiation-dominated hot Big Bang phase during the post-inflationary epoch. The key element of this transient reheating phase, is the ability of the inflaton field $\varphi$ to decay to other fields after the end of inflation, thereby providing a natural mechanism to transfer the energy stored in the universe with $\varphi$ to the other fields which eventually thermalise, thereby providing the radiation dominated era we associate with the hot Big Bang phase. With this in mind we focus on one of the simplest incarnations of this decay,  where $\varphi$ is coupled to a massless scalar field $\chi$ via an interaction term ${\cal I}(\varphi,\chi)$. The system is described by the action
\begin{equation}
    \label{eq:Action_phi_chi}
    S[g_{\mu\nu},~\varphi,~\chi] = \int {\rm d}^4x~\sqrt{-g} \, \left[ \frac{m_p^2}{2} \, R - \frac{1}{2} \,  \partial_{\mu} \varphi \, \partial_{\nu}  \varphi \, g^{\mu\nu} - \frac{1}{2} \,  \partial_{\mu} \chi \, \partial_{\nu}  \chi \, g^{\mu\nu} - V(\varphi) -{\cal I}(\varphi,\chi)  \right] \, .
\end{equation}
The corresponding field equations are, 
\begin{gather}
    \label{eq:reh_EOM_phi}
    \ddot{\varphi} + 3 \, H \, \dot{\varphi} -\frac{\grad^2}{a^2} \varphi + V_{, \varphi} + {\cal I}_{, \varphi}= 0~, \\ 
    \label{eq:reh_EOM_chi}
    \ddot{\chi}+ 3 \, H \, \dot{\chi}  -\frac{\grad^2}{a^2}  \chi + {\cal I}_{, \chi} = 0 
\end{gather}
with the Friedmann constraint
\begin{equation}
    H^2 \, = \, \f{1}{3\, m_p^2} \, \l[ \l(  {\Blue \bm{ \f{1}{2} \, \dot{\varphi}^2} } + {\burntor \bm{ \f{1}{2} \,  \f{\nabla\varphi}{a} . \f{\nabla\varphi}{a}} } + {\violet \bm{ V(\varphi)} }  \r) +  \l(   {\fgreen \bm{ \f{1}{2} \, \dot{\chi}^2} } + {\red \bm{ \f{1}{2} \,  \f{\nabla\chi}{a} .  \f{\nabla\chi}{a} } } \r) +   {\Sepia \bm{ {\cal I} (\varphi, \, \chi)} } \r] \, .
    \label{eq:Hubble_fields}
\end{equation}
The massless offspring field $\chi$ is assumed to behave as  a spectator/test field during inflation which contributes negligibly to the total energy density and couples very weakly to the inflaton. Since the accelerated expansion during inflation rapidly dilutes the quanta of  the $\chi$ field,  we can assume $\chi$ to be in its vacuum state at the end of inflation.

\begin{figure}[htb]
\centering
\subfloat[]{
\includegraphics[width=0.5\textwidth]{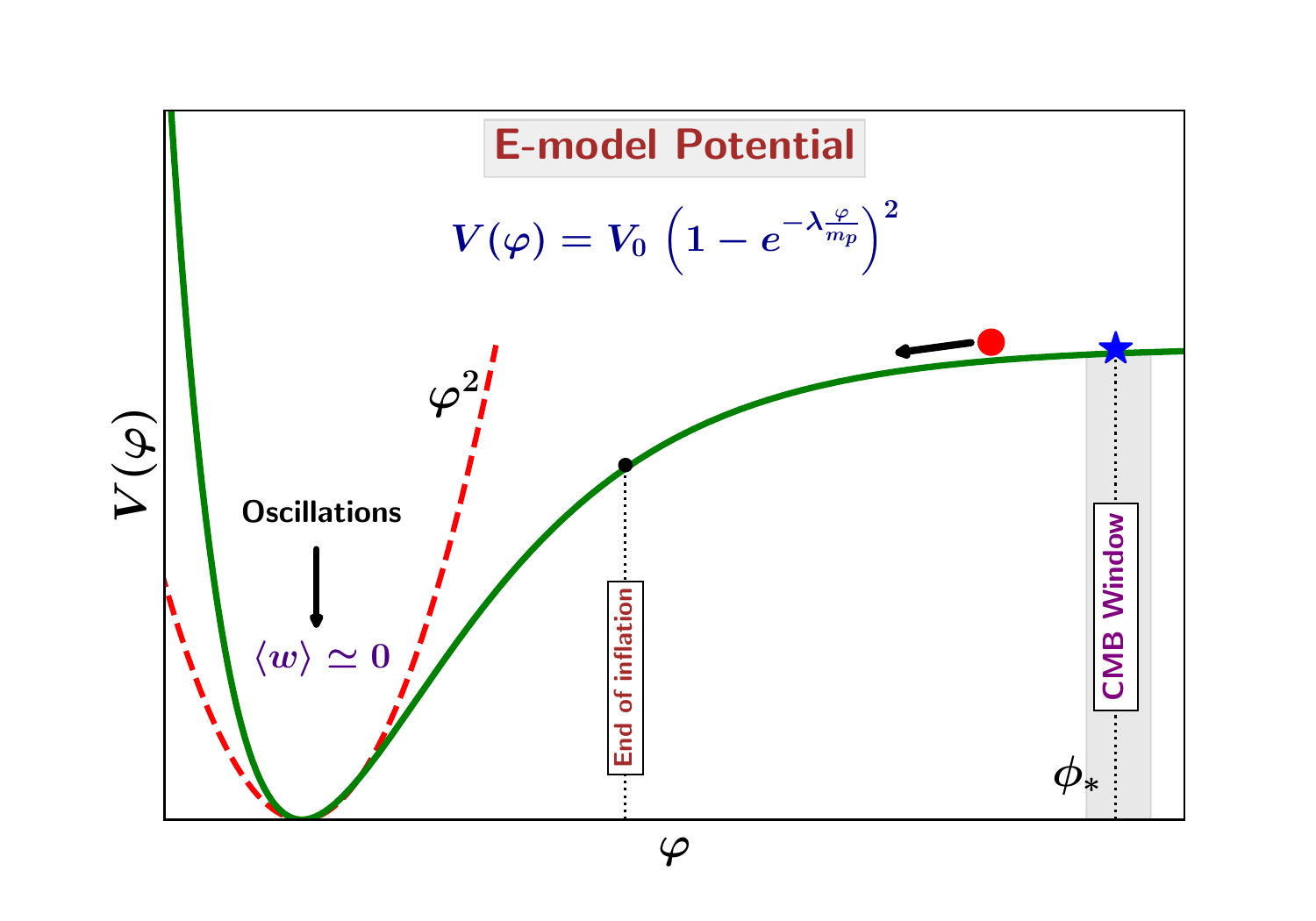}}
\subfloat[]{
\includegraphics[width=0.5\textwidth]{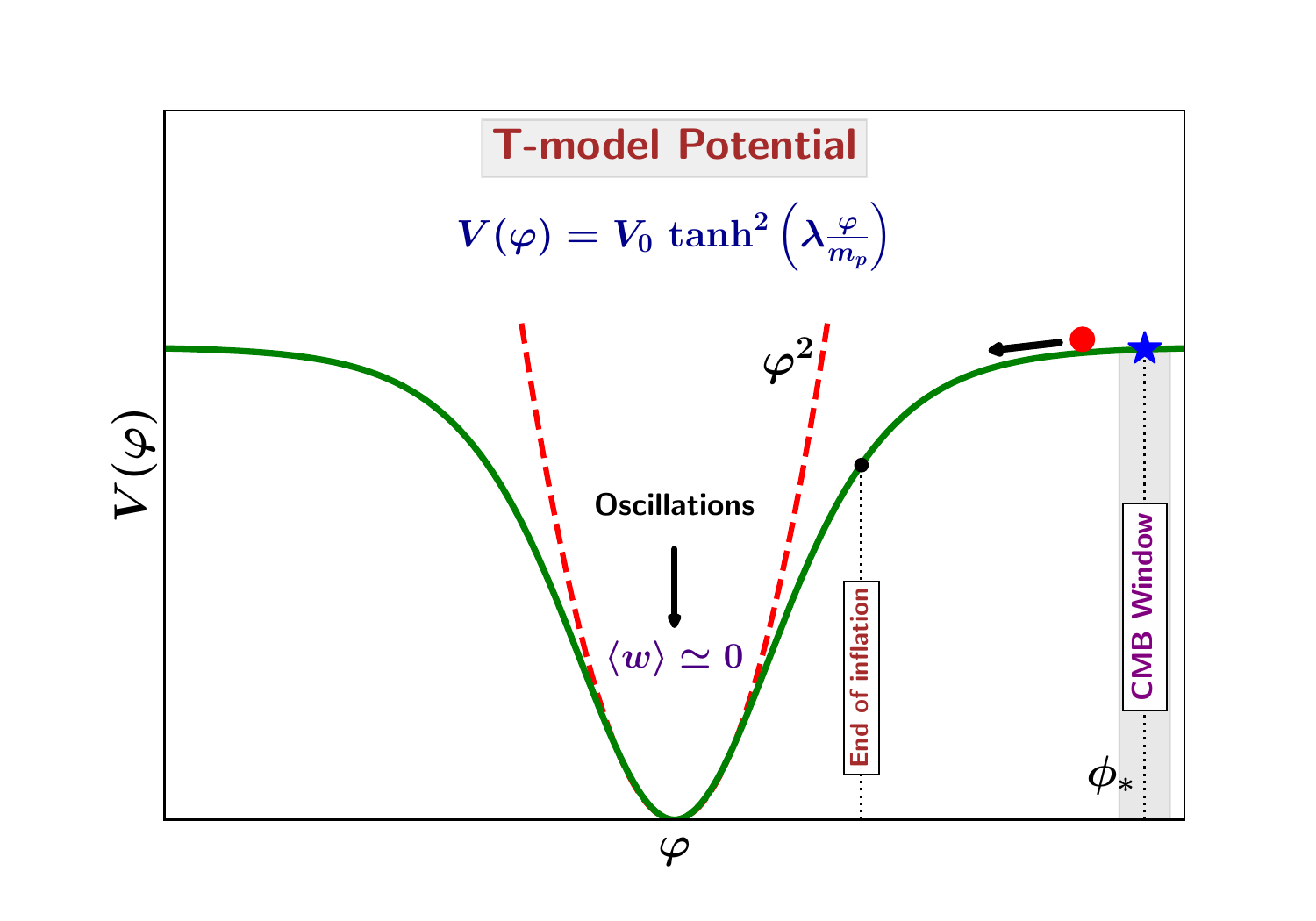}}
\caption{This figure schematically depicts asymptotically flat inflationary plateau potentials. The {\bf left panel} shows an asymmetric  potential featuring only a single plateau-like wing (E-model $\alpha$-attractor) which supports slow-roll inflation. While the {\bf right panel} shows a symmetric  potential possessing two plateau-like wings (T-model $\alpha$-attractor), both of which can support slow-roll inflation. During the post-inflationary oscillations, the attractive  non-linear self-interactions  result in the fragmentation of the oscillating inflaton condensate to form  oscillons.} 
\label{fig:pot_plateau_toy}
\end{figure}

As mentioned in Sec.~\ref{sec:intro},  we will be focusing on the formation  and lifetime of oscillons in both the asymmetric E-model and the symmetric T-model $\alpha$-attractor potentials,  which are shown in Fig.~\ref{fig:pot_plateau_toy}~\cite{Kallosh:2013hoa,Kallosh:2013yoa},  using parameters of the models which are consistent with CMB observations~\cite{Planck:2018jri, Planck:2018vyg}.  In particular we consider  the E-model potential of the form
\beq
    {\violet \bm{ V(\varphi)} } = V_0 \, \l[ 1 - e^{-\lambda_{_\text{E}} \, \f{\varphi}{m_p}}\r]^2  \, ; \label{eq:inf_pot_E-model} 
\eeq
 or, the T-model potential  of the form,
\beq
    {\violet \bm{ V(\varphi)} } = V_0 \, \tanh^2{\left(\lambda_{_\text{T}} \f{\varphi}{m_p}\right)} \, ,
    \label{eq:inf_pot_T-model}
\eeq
\noindent with a {\em quadratic-quadratic  coupling} to the offspring field  $\chi$ of the form
\begin{align}
    {\Sepia \bm{ {\cal I} (\varphi, \, \chi)} } &= \f{1}{2} \, g^2 \, \varphi^2 \, \chi^2  \, .
    \label{eq:Interact_term}
\end{align}
Given the forms of $V(\varphi)$ and ${\cal I(\varphi, \, \chi)}$, it follows that the post-inflationary dynamics of the system
is primarily dictated by the strength of the parameters $\lambda$ and $g^2$. Taking the importance of non-linear and non-thermal processes into account, the reheating dynamics can in general be divided into three distinct phases~\cite{Kofman:1996mv,Kofman:1997yn,Antusch:2021aiw} as described below. 

\subsubsection{Preheating}
\label{subsec:system_preheating}
As inflation ends with the breakdown of slow-roll, the (almost) homogeneous inflaton condensate $\phi$ rolls down the potential $V(\phi)$ and  begins to oscillate about its minimum. In the presence of couplings, both self and external, the fluctuations of $\varphi$ and $\chi$ are enhanced via \textit{parametric resonance}. In particular, using $f_k(t)$ to represent  the Fourier modes of the fluctuations of either $\varphi$ or $\chi$, it satisfies the equation of a damped parametric  oscillator of the form  
\begin{equation}  
    \label{eq:para_osc}
    \ddot{f}_k(t) + 3H \, \dot{f}_k(t) + \Omega_k^2(t) \, f_k(t) = 0 \, ; \quad  {\rm with} \quad \Omega_k(t+T_k) \simeq \Omega_k(t) \, ,
\end{equation}
  where $T_k$ is the periodicity of the  parametric frequency $\Omega_k(t)$
\cite{Kofman:1997yn,Greene:1997fu,Finelli:1998bu,Figueroa:2016wxr}.  During the coherent oscillations of the inflaton, it is possible for $\Omega_k(t)$ to change {\em non-adiabatically} for certain ranges of $k$ values, namely,
\beq
\bigg\vert \f{\dot{\Omega}_k(t)}{\Omega_k^2(t)} \bigg\vert \, \geq \, 1 \, ,
\label{eq:non-adiabaticity}
\eeq
thereby enabling efficient parametric resonance to occur during preheating which in turn results in the explosive (non-thermal) particle production  in discrete bands of the comoving wavenumber $k$ of the quanta of the field fluctuations~\cite{Kofman:1997yn}. As a result, the energy contained in the  oscillating condensate $\phi(t)$ is rapidly transferred to the fluctuations of $\varphi$ and $\chi$. The  precise nature of the periodicity relies on the functional form of the inflaton  self-interaction potential, $V(\varphi)$, and that of the external coupling, ${\cal I} (\varphi, \, \chi)$. In Sec.~\ref{sec:analyt}, we use  the approach of Floquet theory to analyse the nature of the growth of these fluctuations at the linear level  for the potentials and interaction terms given by Eqs.~(\ref{eq:inf_pot_E-model})-(\ref{eq:Interact_term}).

\subsubsection{Backreaction  and termination of resonance} 
\label{subsec:system_backreaction}
The second phase of reheating  can proceed along two different routes. The first occurs if the resonance is sufficiently strong and constitutes a significant  backreaction of the resonantly produced field quanta of $\varphi$ and $\chi$ on the homogeneous inflaton condensate $\phi(t)$. The presence of newly produced field quanta may change the effective mass of the oscillating inflaton. Moreover, when the energy densities of the fluctuations become comparable to that of the condensate, backreaction terminates the resonant production of the  field quanta mentioned in Sec.~\ref{subsec:system_preheating}, thereby quenching the growth of the fluctuations~\cite{Kofman:1996mv,Kofman:1997yn}. However, if the resonance is not strong enough  (\textit{i.e.} if fluctuations of $\varphi$ and $\chi$ remain small),  then the second route is followed, namely the  preheating phase terminates because of the redshifting of the  physical momenta $k_p = k/a$ (due to the expansion of the background space), in the absence of significant  backreaction.

\subsubsection{Thermalisation}
\label{subsec:system_thermalisation}

In the third and final phase, the remaining inflaton  condensate $\phi(t)$ decays perturbatively, leading to a large number of decay products. This in turn leads to the fluctuations of $\varphi$ and $\chi$ re-scattering off these new decay products and further decaying into the quanta of other fields that they are (very weakly) coupled to. Eventually the decay products thermalise resulting in the commencement of the familiar radiation-dominated  hot Big Bang phase.\\

\noindent  In Secs. \ref{sec:analyt} and \ref{sec:simulations}, we analyse in detail the first two phases of reheating: preheating and backreaction, from both a linear point of view (using linear Floquet theory) and a fully non-linear numerical perspective, in the latter case via detailed $3d$ lattice simulations using \cosmolattice~\cite{Figueroa:2020rrl, Figueroa:2021yhd}. 

\section{Linear parametric resonance and Floquet theory}
\label{sec:analyt}
\begin{figure}[htb]
    \centering
    \subfloat{\includegraphics[width = 0.5\textwidth]{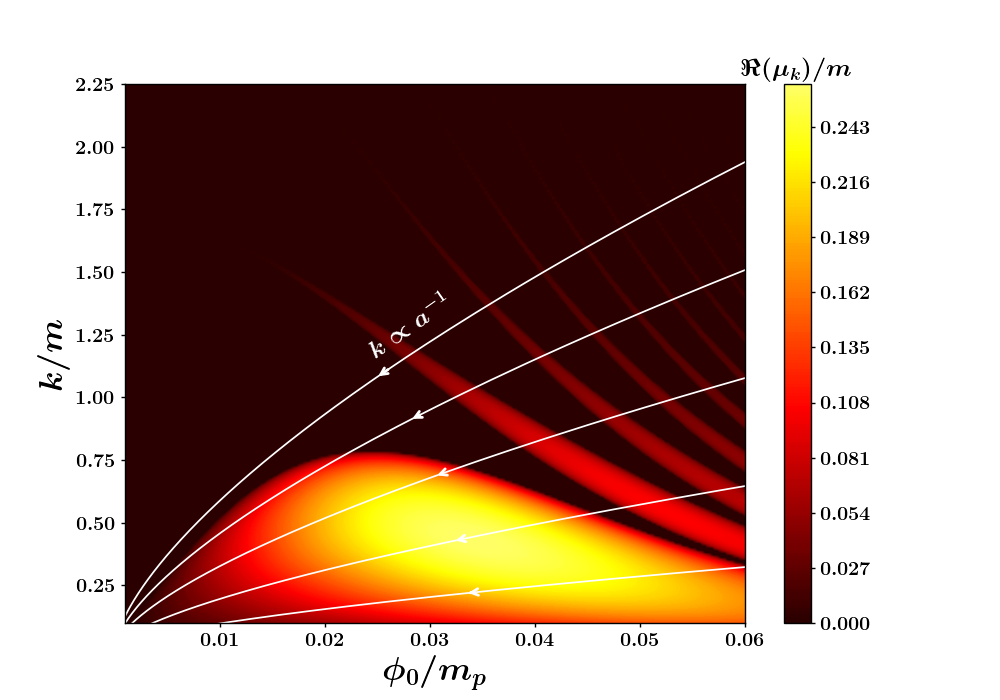}}
    \subfloat{\includegraphics[width = 0.5\textwidth]{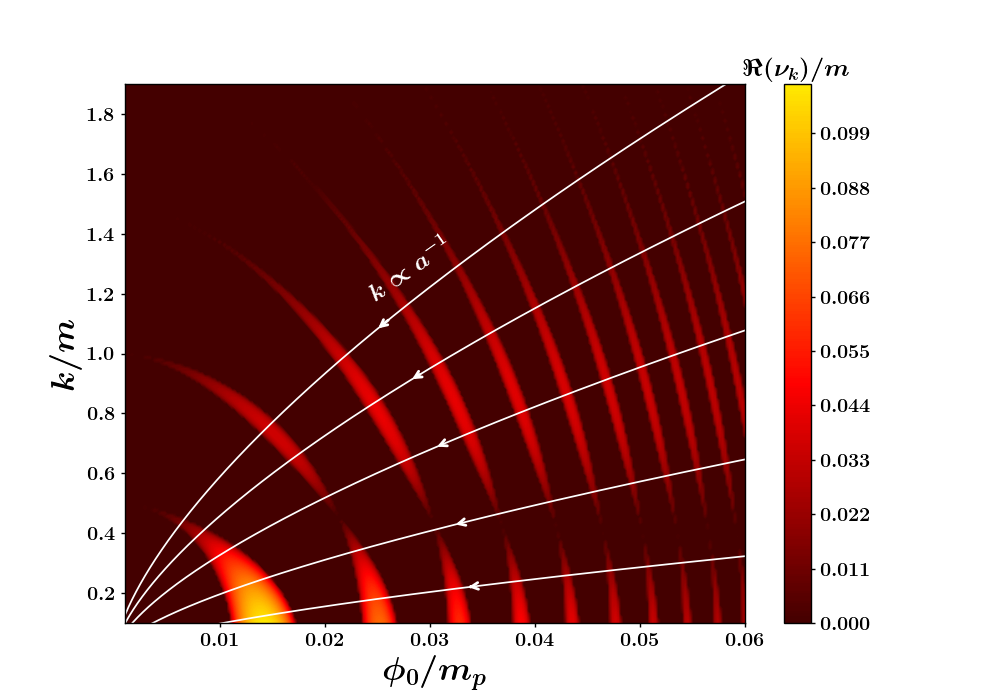}}
    \caption{The Floquet charts exhibiting the instability bands corresponding to the inflaton decay \textit{via} self-resonance (\textbf{left}) in the T-model with $\lambda_{_\text{T}}=50\sqrt{2/3}$ and external resonance (\textbf{right}) with $g^2=8\times 10^{-6}$ are shown, obtained by numerically solving Eqs.~\eqref{eq:delphi_k_linear} and \eqref{eq:chi_k_linear} ignoring the  expansion of the background space, \textit{i.e.}, setting $H = 0$ and $a = 1$. The darker regions correspond to stable solutions with $\lbrace {\mathfrak R}(\mu_k)$, ${\mathfrak R}(\nu_k) \rbrace \to 0$, while the brighter regions correspond to resonance (instability) bands where $\delta\varphi_k$, $\chi_k$ grow exponentially, namely $\delta\varphi_k \propto e^{{\mathfrak R}(\mu_k)t},~ \chi_k \propto e^{{\mathfrak R}(\nu_k)t}$. The white flow curves display the passage of the system through multiple resonance bands due to the expansion of the background space on longer time scales. The expansion of space reduces the amplitude of oscillations $\phi_0(t)$, and results in the redshifting of the physical momenta $k_{p}=k/a$.} 
    \label{fig:reh_resonance_self_ext}
\end{figure}

\begin{figure}[htb]
    \centering
    \subfloat{\includegraphics[width = 0.5\textwidth]{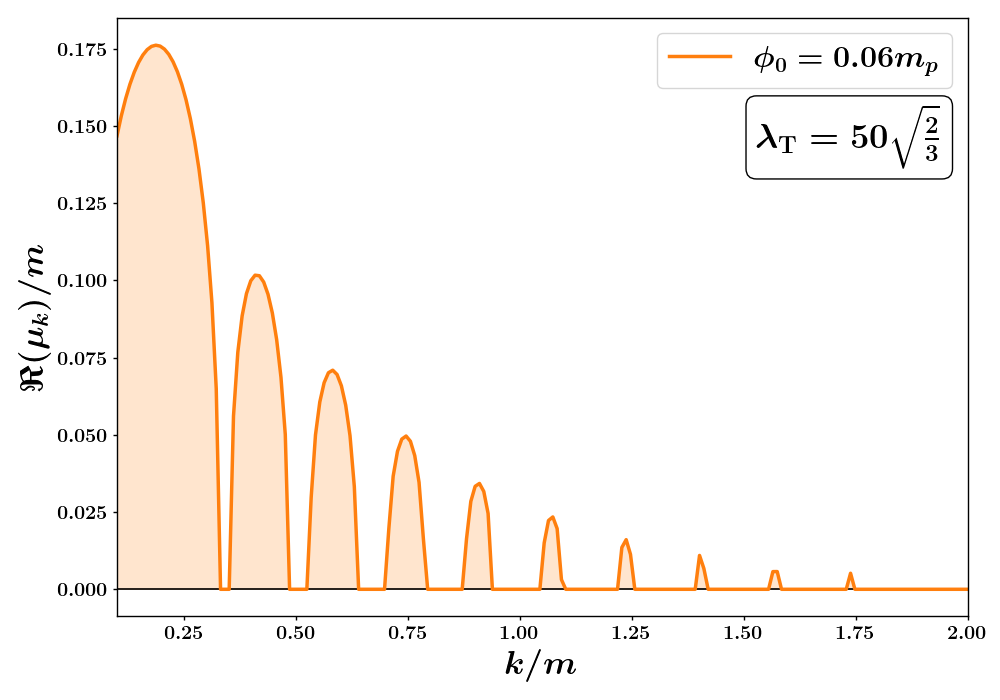}}
    \subfloat{\includegraphics[width = 0.5\textwidth]{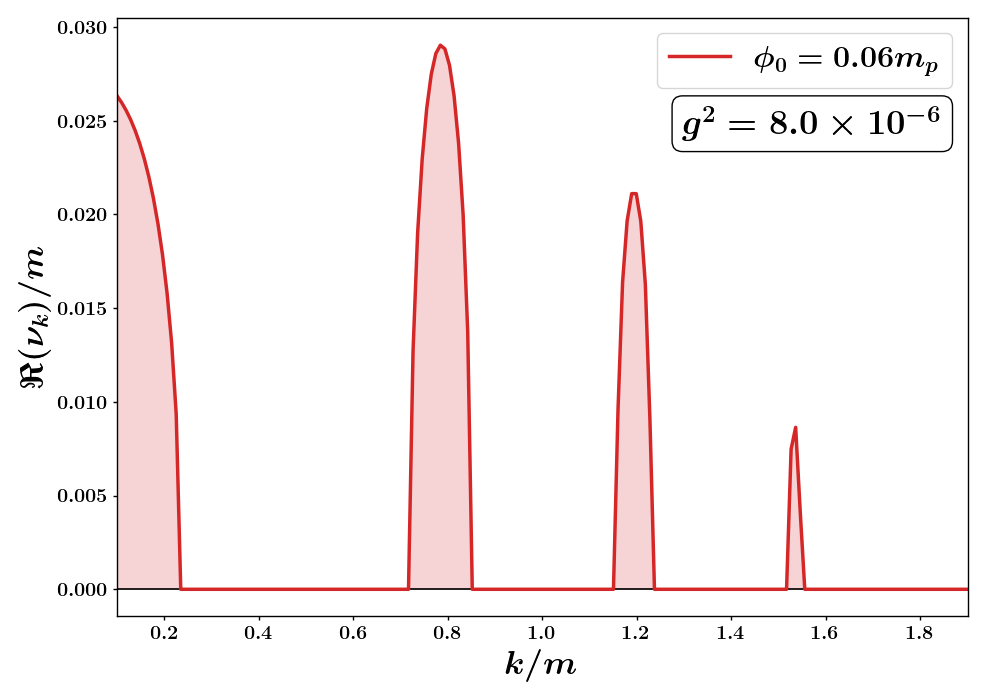}} \\
    \subfloat{\includegraphics[width = 0.5\textwidth]{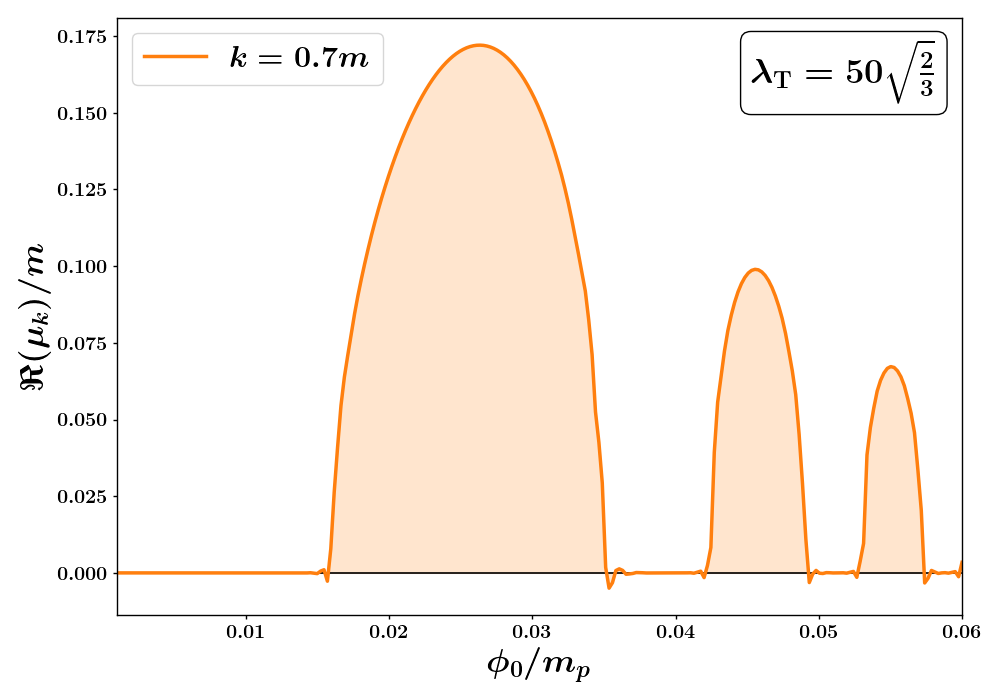}}
    \subfloat{\includegraphics[width = 0.5\textwidth]{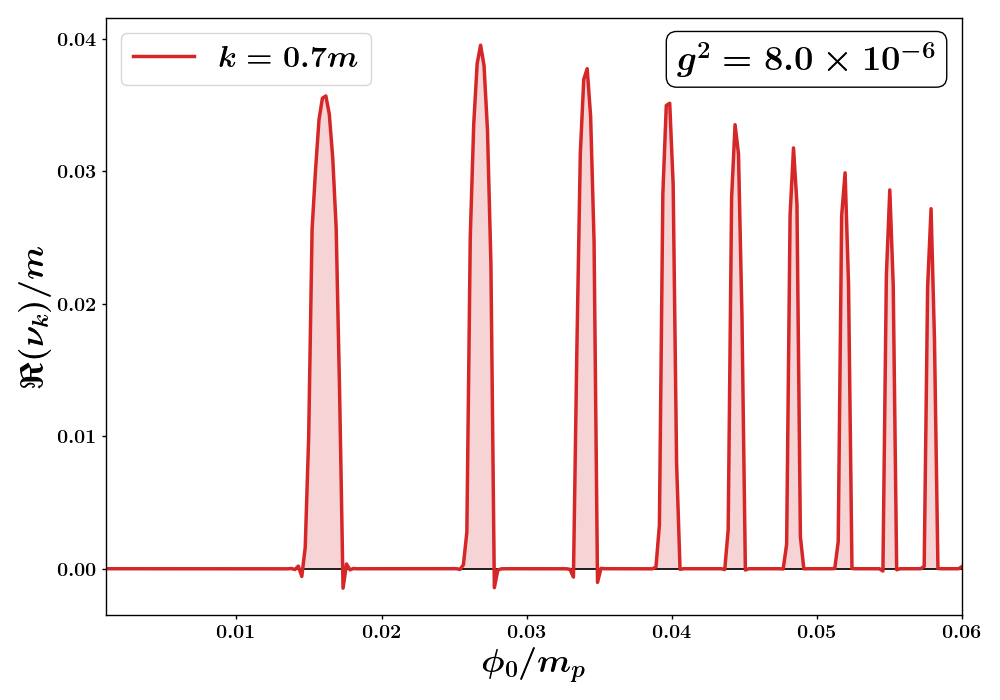}}
    \caption{Variations of the (real part of the) Floquet exponents, $ {\mathfrak R}(\mu_k)$ (orange), ${\mathfrak R}(\nu_k)$ (red), as a function of the comoving momenta $k$ for a fixed amplitude $\phi_0$ (\textbf{top row}), and as a function of $\phi_0$ for a fixed $k$ (\textbf{bottom row}) corresponding to self-resonance (\textbf{left}) and external resonance (\textbf{right}) in the T-model are shown here. Fluctuations $\delta\varphi_k$, $\chi_k$ experience resonant growth when ${\mathfrak R}(\mu_k),~{\mathfrak R}(\nu_k) > 0$ (shaded regions).} 
    \label{fig:FlqExp_self_ext}
\end{figure}

A convenient way to characterise the dynamics during the early stages of preheating is to linearize  the field fluctuations  in Eqs.~(\ref{eq:reh_EOM_phi})~and~(\ref{eq:reh_EOM_chi})  and  carry out  the standard  Floquet analysis~\cite{ASENS_1883_2_12__47_0,nwm,abramowitz+stegun,NIST:DLMF,Kofman:1997yn}. To begin with, we   split  each of the field contents in the Eqs.~(\ref{eq:reh_EOM_phi})-(\ref{eq:Hubble_fields}) into a  homogeneous background  and small fluctuations around that background, namely
\ber
\varphi(t,\bm{x}) &=& \phi(t) + \delta\varphi(t,\bm{x}) \, ; \label{eq:phi_split} \\
\chi(t,\bm{x}) &=& \bar{\chi}(t) + \delta\chi(t,\bm{x}) \, . \label{eq:chi_split}
\eer
As discussed in Sec.~\ref{sec:dyn_inf_post}, since the $\chi$ field is expected to be in its vacuum state at the end of inflation, we have $\bar{\chi}(t) \simeq 0$, leading to $\delta\chi(t,\bm{x}) = \chi(t,\bm{x})$. From hereon, we will simply denote the  fluctuations $\delta\chi$ as $\chi$.
Using Eqs.~\eqref{eq:reh_EOM_phi} and \eqref{eq:reh_EOM_chi}, one can arrive at  equations of the form of Eq.~\eqref{eq:para_osc} by making the following approximations to retain only terms that are linear in the fluctuations $\delta\varphi$, $\chi$.
\begin{enumerate}
\item The energy  density of the system is predominantly contained in the homogeneous  inflaton condensate $\phi(t)$, \textit{i.e.}
$$\phi_0(t) \, \gg \, |\delta\varphi(t,\bm{x})| , \, |\chi(t,\bm{x})|  \, ;$$
and
$$\rho_\phi \, \gg \, \rho_\chi, \, \rho_{\delta\varphi} \, ,$$
where $\phi_0(t)$ is the amplitude of oscillations of the inflaton.

\item Since $\bar{\chi}(t)=0$, the interaction term ${\cal I}(\varphi,\chi)$  given in Eq.~(\ref{eq:Interact_term}) becomes
$${\cal I}(\varphi,\chi) = {\cal I}(\phi+\delta\varphi,\chi) \simeq \f{1}{2} \, g^2 \, \phi^2(t) \, \chi^2 \, , $$
which can be dropped from the Friedmann constraint Eq.~(\ref{eq:Hubble_fields}) at linear order. Furthermore, we get 
$${\cal I}_{,\varphi} \equiv g^2 \, \varphi \, \chi^2 = g^2 \, \l( \phi(t)+\delta\varphi \r) \, \chi^2 \simeq g^2 \, \phi(t) \, \chi^2  \, , $$ which can be dropped from the evolution Eq.~(\ref{eq:reh_EOM_phi}) at linear order. Similarly, the last term of the left hand side of Eq.~(\ref{eq:reh_EOM_chi}) becomes
$${\cal I}_{,\chi} \equiv g^2 \, \varphi^2 \, \chi = g^2 \, \l( \phi(t)+ \delta\varphi \r)^2 \, \chi \simeq g^2 \, \phi^2(t) \, \chi $$
at linear order.
\item Under the linear approximation, the inflaton potential becomes 
$$ V(\varphi) = V(\phi+\delta\varphi) \simeq V(\phi) + V_{,\phi}(\phi) \, \delta\varphi$$
and its derivative becomes
$$V_{,\varphi}(\varphi) = V_{,\varphi}(\phi+\delta\varphi) \simeq V_{,\phi}(\phi)  +  V_{,\phi\phi}(\phi)  \, \delta\varphi \, .$$
\end{enumerate}
We again stress that these approximations are valid during the early stages of preheating, before backreactions from $\delta\varphi$ and $\chi$ become significant, \textit{i.e.} $\rho_{\delta \varphi}, \rho_{\chi} \sim \rho_\phi$, as discussed in Sec.~\ref{subsec:system_backreaction}. Under the aforementioned approximations, the field Eqs.~(\ref{eq:reh_EOM_phi}) and (\ref{eq:reh_EOM_chi}) reduce to 
\begin{align}
    \ddot{\phi} + 3H\dot{\phi} + V_{,\phi} &= 0  \, , \label{eq:phi_no_grad} \\
    \ddot{\delta\varphi}  + 3H\dot{\delta\varphi} +  \l[ - \frac{\grad^2}{a^2} + V_{,\phi\phi}(\phi) \r] \delta\varphi &= 0 \, , \label{eq:delphi_linear} \\
    \ddot{\chi}  + 3H\dot{\chi} + \l[ - \frac{\grad^2}{a^2} +  g^2 \, \phi^2 
    \r]  \, \chi &= 0 \, , \label{eq:chi_linear}
\end{align} 
and the Friedmann Eq.~(\ref{eq:Hubble_fields}) becomes  
\begin{gather}
    \label{eq:Hubble_no_grad}
    H^2 = \frac{1}{3\, m_p^2} \,  \left[  \frac{1}{2} \, \dot{\phi}^2 +  V(\phi) \right] \, .
\end{gather}
Note that in Eqs.~(\ref{eq:phi_no_grad}) and (\ref{eq:Hubble_no_grad}) we have dropped the $\delta\varphi$ terms in order to describe the background dynamics in terms of the purely homogeneous condensate at the end of inflation. Hence, they represent the background equations, with respect to which fluctuations are defined.  The corresponding equations for the evolution of the Fourier modes  $\delta\varphi_k$ and $\chi_k$, take the following form
\begin{gather}
    \ddot{\delta\varphi}_k+ 3H\dot{\delta\varphi}_k +  \left[ \frac{k^2}{a^2} + V_{,\phi\phi}(\phi)  \right] \delta\varphi_k = 0 \, ;  \label{eq:delphi_k_linear} \\ 
    \ddot{\chi}_k+ 3H\dot{\chi}_k +   \left[ \frac{k^2}{a^2} + g^2 \, \phi^2 \right]\chi_k = 0 \, . \label{eq:chi_k_linear}
\end{gather}
Eqs.~\eqref{eq:delphi_k_linear} and \eqref{eq:chi_k_linear} describe  two independent parametric oscillators with time-dependent damping terms  $3H\dot{\delta\varphi}_k$ and $3H\dot{\chi}_k$
and parametric frequencies of the form
\begin{equation}
    \label{eq:Resonance_Freq}
    \Omega^2_{k,\delta\varphi}(t) = \frac{k^2}{a^2} + V_{,\phi\phi}(\phi) \, ; \quad \Omega^2_{k,\chi}(t) = \frac{k^2}{a^2} + g^2 \, \phi^2(t) \, .   
\end{equation}
According to Floquet theory~\cite{ASENS_1883_2_12__47_0,nwm,abramowitz+stegun,NIST:DLMF}, the solutions to Eqs.~\eqref{eq:delphi_k_linear} and \eqref{eq:chi_k_linear} (which generally have to be obtained numerically) are of the form,
\begin{gather}
    \label{eq:sol_delphi_k_Floquet}
    \delta\varphi_k(t) = {\cal P}_{+}(t) \,  e^{\mu_k t} + {\cal P}_{-}(t) \, e^{-\mu_k t}~, \\
    \label{eq:sol_chi_k_Floquet}
    \chi_k(t) = {\cal Q}_{+}(t) \,  e^{\nu_k t} + {\cal Q}_{-}(t) \,  e^{-\nu_k t}
\end{gather}
where $\mu_k,~\nu_k$ are the corresponding \textit{Floquet exponents} and ${\cal P}_{\pm}(t),~{\cal Q}_{\pm}(t)$ are periodic functions with the same oscillation period as that of the background field. Floquet exponents are complex in general and those modes with ${\mathfrak R}(\mu_k),~{\mathfrak R}(\nu_k) \neq 0$ experience resonant growth with time, whereas for ${\mathfrak R}(\mu_k),~{\mathfrak R}(\nu_k) = 0$, the corresponding modes are oscillatory. One can map such regions onto a \textit{Floquet chart} as shown in Fig.~\ref{fig:reh_resonance_self_ext}, and further analyse the nature of the resonance (either narrow or broad) as a function of the oscillation amplitude and Fourier mode $k$ (see Fig.~\ref{fig:FlqExp_self_ext}).  Figures~\ref{fig:reh_resonance_self_ext} and \ref{fig:FlqExp_self_ext} are obtained numerically by solving Eqs.~\eqref{eq:delphi_k_linear} and \eqref{eq:chi_k_linear} ignoring the expansion of space by setting $H = 0$ and setting $a = 1$. The brighter (or shaded regions) correspond to resonance bands where $\delta \varphi_k$, $\chi_k$ experience exponential growth \textit{i.e.}  $\delta\varphi_k(t) \propto e^{{\mathfrak R}(\mu_k)t},~ \chi_k(t) \propto e^{{\mathfrak R}(\nu_k)t}$. Since the timescale of expansion of the background space is much longer  compared to that of the $\phi$-oscillations, its effect on the parameters, namely the redshifting of  physical momenta $k_p = k/a$ and the decrease in the amplitude of oscillations $\phi_0$, can  simply be represented by white flow curves through the resonance bands, as shown in Fig.~\ref{fig:reh_resonance_self_ext}.

As is clear from the evolution equations in Eqs.~(\ref{eq:phi_no_grad})~and~(\ref{eq:Hubble_no_grad}), making progress at the analytic level, even for the background evolution of $H(t)$ and $\phi(t)$ requires us to be able to handle the non-linear functional form of the potential $V(\varphi)$. In many cases, including the two potentials we are concentrating on, Eqs.~(\ref{eq:inf_pot_E-model})~and~(\ref{eq:inf_pot_T-model}), the functional form of $V(\varphi)$ is often quite complicated. However, we can still make progress by recalling that reheating in general takes place for small-amplitude regimes, where $\phi_0 \ll m_p$, in which case we can Taylor expand the potentials, leading to 
\begin{align}
    \label{eq:near_harmonic_approx_Emodel}
    V(\varphi)\big\vert_{\text{\tiny E-model}} &\simeq \f{1}{2} \, \l( 2\lambda_{_\text{E}}^2 \f{V_0}{m_p^2} \r) \, \varphi^2 - \f{1}{3} \, \l( 3\lambda_{_\text{E}}^3 \f{V_0}{m_p^3} \r) \, \varphi^3 + \f{1}{4} \, \l( \f{7}{3} \lambda_{_\text{E}}^4 \f{V_0}{m_p^4} \r) \, \varphi^4 + ... ~, \\
    \label{eq:near_harmonic_approx_Tmodel}
    V(\varphi)\big\vert_{\text{\tiny T-model}} &\simeq \f{1}{2} \, \l( 2\lambda_{_\text{T}}^2 \f{V_0}{m_p^2} \r) \, \varphi^2 - \f{1}{4} \, \l( \f{8}{3} \lambda_{_\text{T}}^4 \f{V_0}{m_p^4} \r) \, \varphi^4 + ... ~   
\end{align}
respectively. This allows us to evaluate the time-averaged equation of state (ignoring ${\cal O}(\varphi^3)$ and higher terms)  yielding $\left< w_\varphi \right> = \frac{n-1}{n+1},~{\rm for}~V(\varphi) \sim \varphi^{2n}$, and consequently, the 
 time-averaged expansion of the background space -- namely $a \sim t^{p}$, with $p=\frac{2}{3\left(1+\left< w_\phi \right>\right)}$ (refer to App.~\ref{app:N_osc_vs_DeltaN}).  From Eqs.~(\ref{eq:near_harmonic_approx_Emodel})~and~(\ref{eq:near_harmonic_approx_Tmodel}), it is easy to see that the mass of the inflaton is related to $V_0$ and $\lambda$ as
 \beq
m^2 = 2 \, \lambda^2 \, \f{V_0}{m_p^2} \, ,
\label{eq:V0_m_lambda_reln}
\eeq
where $\lambda = \lambda_{_\text{E}}, \, \lambda_{_\text{T}}$ for the E- and the T-models respectively.
 
It is worth noting that for a range of values of the parameter $\lambda$ (denoted as $\lambda_{_\text{T}}$ for the T-model and $\lambda_{_\text{E}}$ for the E-model), the first few oscillation of $\phi(t)$ about the  minimum of the potential  correspond to tachyonic regimes. Some details on tachyonic oscillations can be found in App.~\ref{app:tachyonicity}. As the inflaton evolves down its potential, since the amplitude of oscillations $\phi_0(t)$ decreases due to particle production (and due to expansion of the background space), the $\varphi^2$ term eventually becomes dominant in Eqs.~\eqref{eq:near_harmonic_approx_Emodel} and \eqref{eq:near_harmonic_approx_Tmodel} leading to Mathieu-like resonance. When this occurs, the structure of the resonance bands for self-resonance changes from the one shown in Fig.~\ref{fig:reh_resonance_self_ext} to that corresponding to the Mathieu equation.  We do not plot the Floquet-chart for the Mathieu equation here, as these are well known in the literature~\cite{Kofman:1994rk, Lozanov:2019jxc, Mishra:2024axb}. Hence, the Floquet chart corresponding to self-resonance in Fig.~\ref{fig:reh_resonance_self_ext} is valid only for the first few $\phi$-oscillations.

\medskip


Since our primary goal is to study the formation of oscillons, it is important to make the definition of oscillons clear. Oscillons are highly {\em non-linear} and {\em non-relativistic} localised field configurations where the spatial gradient (Laplacian) of the field profile is balanced by the field-derivative of the non-linear self-interaction potential. Therefore, both the attractive self-interaction  and the gradient terms play a crucial role in their field profile and energy density. In particular, their  formation  during preheating necessarily  requires a dynamical mechanism to facilitate the fragmentation of the homogeneous inflaton condensate, which can then clump together via attractive self-interaction to form oscillons. In summary,  the possibility of oscillon formation during preheating requires two {\em necessary} conditions: (\textit{i}) the presence of an attractive non-linear self-interaction term in the potential \textit{i.e.} $V(\varphi) = \frac{1}{2} m^2 \varphi^2 - |U(\varphi)|\, ,$  and (\textit{ii}) an exponential  growth of the inflaton inhomogeneities $\delta \varphi_k$ \textit{via} broad parametric resonance.  Eqs.~\eqref{eq:near_harmonic_approx_Emodel} and \eqref{eq:near_harmonic_approx_Tmodel} show that in the small-amplitude regime the potentials contain the desired attractive self-interaction terms  suitable for scalar field fragmentation, and oscillon formation, which can be seen from  Fig.~\ref{fig:pot_plateau_toy}.  Our analysis in this section, as plotted in Figs.~\ref{fig:reh_resonance_self_ext} and \ref{fig:FlqExp_self_ext}, demonstrates that the second condition is also easily satisfied  in the linear regime for a range of values of $\lbrace \phi_0, \, k \rbrace$, even in the presence of an external coupling.


Nevertheless, it is important to remember that while the aforementioned conditions are necessary for oscillon formation, they are not sufficient. If the resonant growth of the  inflaton and $\chi$ fluctuations continues for long enough, then  the energy densities of the fluctuations  eventually become comparable to that of the homogeneous condensate and our linear treatment based on parametric resonance breaks down. In general, the growing fluctuations are expected to  backreact on the oscillating  condensate and  break it into inflaton fragments, which can then  clump together due to the non-linear self-interaction to form oscillons.  Since oscillons are necessarily non-linear field configurations,  in order to firmly establish the formation (and later, the lifetime) of  oscillons, one needs to carry out detailed  numerical lattice simulations in order to capture the non-perturbative dynamics. 


In fact, lattice simulations carried out in Refs.~\cite{Mahbub:2023faw} and~\cite{Lozanov:2017hjm} show that oscillons do form during the inflaton oscillations around the E- and the T-model potentials, in the absence of any external coupling. However, the question of oscillon formation in the presence of a coupled offspring field has not been investigated systematically in the current literature, which is the primary focus of this work. In  Sec.~\ref{sec:simulations}, we  carry out detailed  numerical lattice simulations of the non-linear field  Eqs.~(\ref{eq:reh_EOM_phi})-(\ref{eq:Hubble_fields}) in the presence of the external coupling~(\ref{eq:Interact_term}) as discussed below.

\section{Non-linear dynamics: Lattice simulations}
\label{sec:simulations}
\subsection{$\mathcal{C}$osmo$\mathcal{L}$attice setup and parameters}
\label{subsec:sim_setup}
For the  $3d$ lattice simulations of the field equations, we use the publicly available lattice code \cosmolattice ~\cite{Figueroa:2020rrl, Figueroa:2021yhd}. In order to carry out numerical simulations conveniently,  we define the following dimensionless variables,
\beq
{\tilde t} = m \, t ~ ; ~~ {\widetilde x} = m \, x ~; ~~ {\tilde f} = \f{1}{\beta} \, \frac{f}{m_p} \, ,
\label{eq:dimless_t_x_phi}
\eeq
where $f = \lbrace \varphi,~\chi \rbrace$, $ m^2 = \frac{2 \, V_0\lambda^2}{m_p^2}$ from Eq.~(\ref{eq:V0_m_lambda_reln}), and   $\beta \in {\mathbb R}^+$ is a parameter that we have introduced for numerical convenience. In general $\beta$ should be of the order $\phi_{\rm end}/m_p$, where recall that $\phi_{\rm end}$ marks the end of inflation. In our lattice simulations, we fix $\beta=10$ throughout. In terms of these variables, we convert the kinetic ($K$), the gradient ($G$), the potential ($V$) and the interaction ($\mathcal{I}$) terms in the action in Eq.~\eqref{eq:Action_phi_chi} to their corresponding dimensionless counterparts by dividing each term with $\beta^2 m^2 m_p^2$ as follows- 
\begin{equation}
    \label{eq:Conversion_Lagrangian_terms_dimensionless}
    {\widetilde F}({\tilde f}) = \l( \frac{1}{\beta} \, \f{m_p}{m} \r)^2 \, \f{F(f)}{m_p^4} \, ,
\end{equation}
where $F = \lbrace K,~G,~V,~\mathcal{I} \rbrace$ (and $f = \lbrace \varphi,~\chi \rbrace$). As a result, the kinetic and gradient terms in Eq.~\eqref{eq:Action_phi_chi} become,
\begin{align}
    \label{eq:kinetic_gradient_phi}
    {\Blue {\widetilde K}_{\tilde \varphi}} &= \frac{1}{2} \l( \frac{\partial{\widetilde \varphi}}{\partial{\tilde t}}\r)^2 \, , & {\burntor {\widetilde G}_{\tilde \varphi}} &= \frac{1}{2}\f{1}{a^2({\tilde t})} \left[ \left(\frac{\partial{\widetilde \varphi}}{\partial{\widetilde x}}\right)^2 + \left(\frac{\partial{\widetilde \varphi}}{\partial{\widetilde y}}\right)^2 + \left(\frac{\partial{\widetilde \varphi}}{\partial{\widetilde z}}\right)^2 \right] \, , \\
    {\fgreen {\widetilde K}_{\tilde \chi}} &= \frac{1}{2}  \left(\frac{\partial{\widetilde \chi}}{\partial{\tilde t}}\right)^2 \, , & {\red {\widetilde G}_{\widetilde \chi}} &= \frac{1}{2}\f{1}{a^2({\tilde t})}   \left[ \left(\frac{\partial{\widetilde \chi}}{\partial{\widetilde x}}\right)^2 + \left(\frac{\partial{\widetilde \chi}}{\partial{\widetilde y}}\right)^2 + \left(\frac{\partial{\widetilde \chi}}{\partial{\widetilde z}}\right)^2 \right] \, ;
\end{align}
where we note that the scale factor $a$ is dimensionless. The inflationary potentials and the interaction term Eqs.~\eqref{eq:inf_pot_E-model}, \eqref{eq:inf_pot_T-model} and \eqref{eq:Interact_term} get re-scaled as, 
\begin{eqnarray}
    {\violet \widetilde{V}(\widetilde{\varphi}) } &=& \f{1}{2} \, \f{1}{\beta^2 \lambda_{_\text{E}}^2}~\left(1 - e^{-\lambda_{_\text{E}} \beta \Tilde{\varphi}} \right)^2  \, ,  \label{eq:sim_rescaled_potentials-E} \\
   {\violet \widetilde{V}(\widetilde{\varphi}) } &=& \f{1}{2} \, \f{1}{\beta^2 \lambda_{_\text{T}}^2}~\tanh^2({\lambda_{_\text{T}} \beta \widetilde{\varphi}})\, , \\
    \label{eq:sim_rescaled_potentials-T}
    {\Sepia \widetilde{\cal I}(\widetilde{\varphi}, \widetilde{\chi}) } &=& \frac{1}{2} \, q  \, \widetilde{\varphi}^2 \widetilde{\chi}^2 \, ; \quad \quad q = g^2 \, \beta^2 \, \f{m^2_p}{m^2} \, ,
    \label{eq:sim_rescaled_coupling}
\end{eqnarray}
where we used Eq.~(\ref{eq:V0_m_lambda_reln}) to replace $V_0$ in terms of $m$ and $\lambda$ in Eqs.~(\ref{eq:sim_rescaled_potentials-E})~and~(\ref{eq:sim_rescaled_potentials-T}). The field Eqs.~\eqref{eq:reh_EOM_phi} and \eqref{eq:reh_EOM_chi} thus take the form
\begin{align}
    \label{eq:sim_rescaled_EOM_phi}
    &\frac{\partial^2 {\widetilde \varphi}}{\partial {\tilde t}^2}  + 3{\widetilde H} \, \frac{\partial {\widetilde \varphi}}{\partial {\tilde t}} - \frac{{\widetilde \grad}^2}{a^2} \, {\widetilde \varphi} + {\widetilde V}_{,{\tilde \varphi}} + {\widetilde {\cal I}}_{,{\tilde \varphi}} = 0 \, , \\
    \label{eq:sim_rescaled_EOM_chi}
    &\frac{\partial^2 {\widetilde \chi}}{\partial {\tilde t}^2} + 3{\widetilde H}\, \frac{\partial {\widetilde \chi}}{\partial {\tilde t}} - \frac{{\widetilde \grad}^2}{a^2} \, {\widetilde \chi}  + {\widetilde {\cal I}}_{,{\tilde \chi}} = 0 \, , 
\end{align}
with  
\begin{equation}
    \label{eq:sim_rescaled_Hubble fields}
    {\widetilde H}^2 \equiv \left(\frac{{\rm d} \ln{a}}{{\rm d} {\tilde t}}\right)^2 = \frac{\beta^2}{3} \left( {\Blue {\widetilde K}_{\tilde \varphi} } + {\burntor {\widetilde G}_{\widetilde \varphi} } + {\violet {\widetilde V}(\tilde \varphi)} + {\fgreen {\widetilde K}_{\tilde \chi} } + {\red {\widetilde G}_{\tilde \chi}} + {\Sepia {\widetilde {\cal I}} ({\widetilde \varphi},~{\widetilde \chi})}  \right) \, .
\end{equation}
We define the re-scaled energy densities and pressures for ${\widetilde \varphi}$ and ${\widetilde \chi}$ as\footnote{We note that due to the way in which {\cosmolattice} is configured, interaction terms such as ${\widetilde{\cal I}}({\widetilde \varphi}, {\widetilde \chi})$ have to be absorbed within the definition of a potential in the output files, which we do above by including it in the potential of the inflaton, \textit{i.e.} as   ${\widetilde V}(\tilde \varphi) + {\widetilde{\cal I}({\tilde \varphi}, {\tilde \chi})}$.}
\begin{align} 
    \label{eq:rescaled_en_pressure_phi}
    {\widetilde \rho}_{\tilde \varphi} &= {\Blue {\widetilde K}_{\tilde \varphi}} + {\burntor {\widetilde G}_{\tilde \varphi}} + {\violet \widetilde{V}(\widetilde{\varphi}) } + {\Sepia \widetilde{\cal I}(\widetilde{\varphi}, \widetilde{\chi}) }, & 
    {\widetilde p}_{\tilde \varphi} &= {\Blue {\widetilde K}_{\tilde \varphi}} - \frac{1}{3} {\burntor {\widetilde G}_{\tilde \varphi}} - {\violet \widetilde{V}(\widetilde{\varphi}) } - {\Sepia \widetilde{\cal I}(\widetilde{\varphi}, \widetilde{\chi}) } \\
    \label{eq:rescaled_en_pressure_chi}
    {\widetilde \rho}_{\tilde \chi} &= {\fgreen {\widetilde K}_{\tilde \chi}} + {\red {\widetilde G}_{\tilde \chi}} , & 
    {\widetilde p}_{\tilde \chi} &= {\fgreen {\widetilde K}_{\tilde \chi}} - \frac{1}{3} {\red {\widetilde G}_{\tilde \chi}} 
\end{align}
with their respective equation of state (EoS) parameters 
\beq
{\widetilde w}_{\tilde \varphi} = \f{{\widetilde p}_{\tilde \varphi}}{{\widetilde \rho}_{\tilde \varphi}} \, , \quad {\widetilde w}_{\tilde \chi} = \f{{\widetilde p}_{\tilde \chi}}{{\widetilde \rho}_{\tilde \chi}} \, .
\label{eq:rescaled_EoS}
\eeq
Furthermore, we have defined the dimensionless total energy density and pressure of the system as,
\begin{align}
    \label{eq:rescaled_en_pressure_total}
    {\widetilde \rho} &= {\Blue {\widetilde K}_{\tilde \varphi}} + {\fgreen {\widetilde K}_{\tilde \chi}} + {\burntor {\widetilde G}_{\tilde \varphi}}  + {\red {\widetilde G}_{\tilde \chi}} + {\violet \widetilde{V}(\widetilde{\varphi}) } + {\Sepia \widetilde{\cal I}(\widetilde{\varphi}, \widetilde{\chi}) }, \\ 
    {\widetilde p} &= {\Blue {\widetilde K}_{\tilde \varphi}} + {\fgreen {\widetilde K}_{\tilde \chi}} - \frac{1}{3} \left({\burntor {\widetilde G}_{\tilde \varphi}} + {\red {\widetilde G}_{\tilde \chi}} \right) - {\violet \widetilde{V}(\widetilde{\varphi}) } - {\Sepia \widetilde{\cal I}(\widetilde{\varphi}, \widetilde{\chi}) }
\end{align}
with the associated total EoS parameter  
\beq
{\widetilde w} = \f{{\widetilde p}}{{\widetilde \rho}}  \, .
\label{eq:rescaled_EoS_total}
\eeq
Finally we have defined the \textit{fractional energy density} of different components $F = \lbrace K,~G,~V,~\mathcal{I} \rbrace$, with $f = \lbrace \varphi,~\chi \rbrace $ as,
\begin{equation}
    \varepsilon^{\tilde{f}}_{F} = \frac{{\widetilde F}({\tilde f})}{\widetilde \rho}\, .
    \label{eq:rescaled_frac_rho}
\end{equation}

We note that the dynamics of inflaton decay in the simulations depend primarily on the values of the  two  key parameters, $\lbrace \lambda, \, g^2 \rbrace$, denoting the strengths of the self and external interactions respectively. In Tables~\ref{tab:Sim_Parameters_emodel}~and~\ref{tab:Sim_Parameters_tmodel}, we present the values of the self-coupling $\lambda$ (denoted as $\lambda_{\text{\tiny E}}$ and $\lambda_{\text{\tiny T}}$) and the external coupling $g^2$,  along with the corresponding values of $V_0$ and initial conditions $\lbrace \phi_{\rm in},\, \dot{\phi}_{\rm in} \rbrace$, for the E- and the T-model potentials that are used in our lattice simulations.  In our simulations,  we consider values of $\lambda$ that are large enough to ensure that the predicted levels of the tensor-to-scalar ratio satisfy the latest CMB bound, namely $r \leq 0.036$~\cite{BICEP:2021xfz,Mishra:2022ijb}. The value of $V_0$ for a given $\lambda$ has been fixed by the  CMB normalisation  $\mathcal{P}_\zeta=2.1\times10^{-9}$~\cite{Planck:2018jri},  the details of which can be found in App.~\ref{app:CMB_T_E_models}. The corresponding inflaton mass $m$ is then determined from $V_0$ and $\lambda$ using Eq.~(\ref{eq:V0_m_lambda_reln}). 

\begin{figure}[hbt]
    \centering
    \subfloat{\includegraphics[width = 0.48\textwidth]{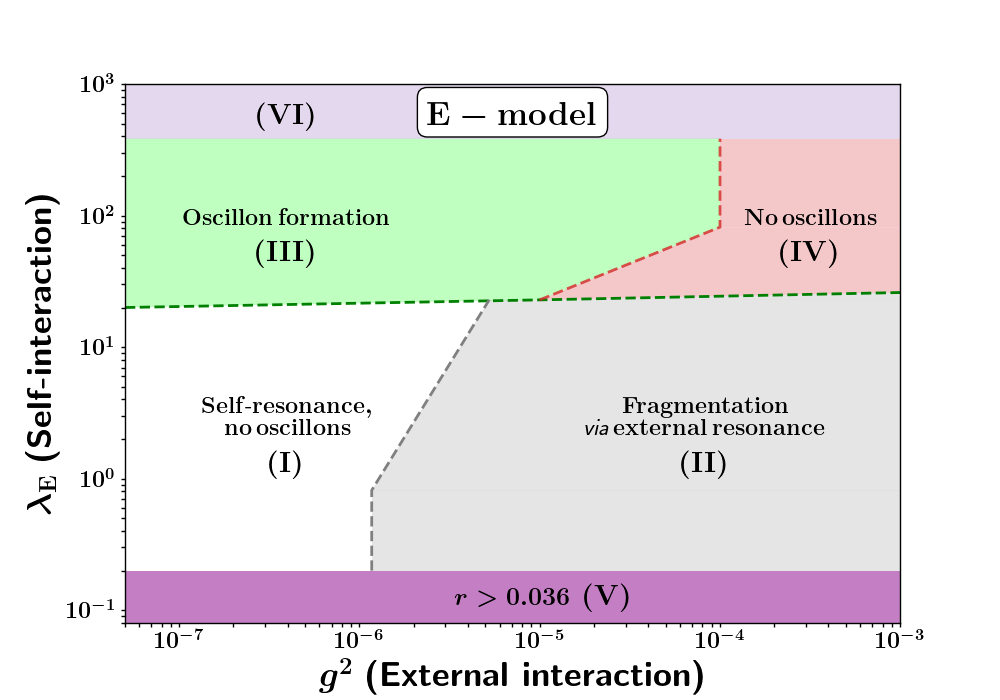}}
    \subfloat{\includegraphics[width = 0.48\textwidth]{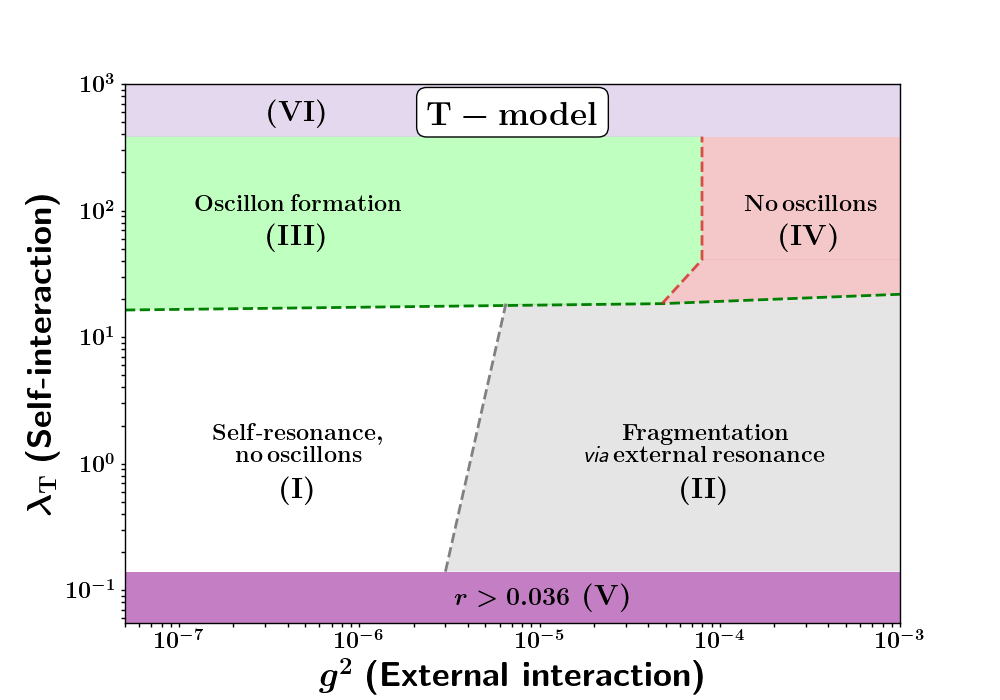}}
    \caption{The parameter space, $\lbrace \lambda, \, g^2 \rbrace$, of the self and the external coupling constants explored in our lattice simulations for the E-model ({\bf left}) and T-model ({\bf right}) are shown here.  Region\,(I) (shaded in white) corresponds to the parameter space where scalar field fragmentation is negligible, while the inflaton condensate  is guaranteed to fragment \textit{via} external resonance,  without the formation of oscillons, in region\,(II) (shaded in grey). The green-shaded region\,(III)  corresponds to the parameter space for which self-interaction is high enough and external coupling is low enough for oscillon formation to take place. However, in region\,(IV) (red-shaded),  oscillon formation is disrupted due to sufficiently high $g^2$. Region\,(V) corresponds to  $r > 0.036$, which violates the tensor-to-scalar ratio bound. Region\,(VI) does not host oscillons since the time-averaged EoS of the inflaton is negative and the inflaton mass term is negligible compared to higher-order interactions, as discussed in the text below.}
    \label{fig:par_space_lambda_g^2}
\end{figure}

The primary results of this work were obtained by carrying out simulations with a lattice size of $N = 128^3$ and ${\tilde k}_\text{\tiny IR} = 0.05$ \footnote{The infrared cut-off ${\tilde k}_\text{\tiny IR}$ is defined as, 
\begin{gather}
    {\tilde k}_\text{\tiny IR} = \frac{2 \pi}{\widetilde L} \nonumber
\end{gather}
where ${\widetilde L} = L m$, with $L$ being the comoving length of the cubic lattice. For ${\tilde k}_\text{\tiny IR} = 0.05$, the lattice has sides of comoving length $L = 40\pi \, m^{-1}$.} (which is the minimum cut-off for the reciprocal lattice ${\tilde k} = \frac{k}{am}$). The system was evolved using the second order Velocity-Verlet (VV2) algorithm available in \cosmolattice. In order to ensure inflaton fragmentation  and oscillon formation in a shallow potential,  approximate analytical bounds on the parameters of the potential were obtained in Refs.~\cite{Kim:2017duj, Kim:2021ipz}. For example, in the  case of  a symmetric potential of the form $V(\varphi) = \frac{1}{2} \, m^2 \, \varphi^2 - A^{(4)} \, \varphi^4$, the bound on $\lambda_\text{\tiny T}$ was obtained to be 
\begin{equation}
    \label{eq:sim_sym_lambda_frag_bound_Tmodel}
    \lambda_\text{\tiny T} \gtrsim 54.77 \,  \sqrt{\frac{0.1}{\gamma_\text{\tiny T}}}
\end{equation}
where $\gamma_\text{\tiny T} \equiv 2A^{(4)}\phi^2_\text{\tiny in}/m^2 \sim {\cal O}(1)$.  Similarly, for an asymmetric potential of the form $V(\phi) = \frac{1}{2}m^2\phi^2 - A^{(3)}\phi^3$, the corresponding  bound on $\lambda_\text{\tiny E}$ was found to be 
\begin{equation}
    \label{eq:sim_sym_lambda_frag_bound_Emodel}
    \lambda_\text{\tiny E} \gtrsim 11.55 \left(\frac{0.1}{\gamma_\text{\tiny E}}\right) 
\end{equation}
where $\gamma_\text{\tiny E} \equiv 2A^{(3)}\phi_\text{\tiny in}/m^2 \sim {\cal O}(1)$.  These analytic arguments strongly suggest that the inflaton condensate is  expected to fragment with our choice of parameters $\lbrace \lambda, \, g^2 \rbrace$ given in  Tables~\ref{tab:Sim_Parameters_emodel}~and~\ref{tab:Sim_Parameters_tmodel},  corresponding to four representative values of $g^2$ used in our lattice simulations. However, while indicative, the analytical bounds do not completely guarantee the formation of oscillons, for which we need to take into account the non-linear effects. Hence, in order to investigate the formation of oscillons in our set up,  we carry out fully non-linear numerical lattice simulations of the scalar field dynamics.  Figure~\ref{fig:par_space_lambda_g^2} provides an illustrative summary of the key  results of our numerical investigations in the form of a $\lbrace \lambda, \, g^2 \rbrace$ parameter space plot of the E- and the T-model potentials, highlighting regions of inflaton fragmentation and oscillon formation, distinguishing them from regions where oscillon formation is disrupted due to the presence of large external coupling.
\begin{table}[htb]
    \centering
    \caption{Values of various parameters used in the simulations with the E-model  inflaton potential.}
    \label{tab:Sim_Parameters_emodel}
    \begin{tabular}{|c c c c c c|}
        \hline
        \Tstrut
        $\bm{\lambda_{\text{\tiny E}}}$ & $\bm{V_0}~[m^4_p]$ & $\bm{\phi_{\rm in}}~[m_p]$ & $\bm{\dot{\phi}_{\rm in}}~[m^2_p]$ & $ \bm{m}~[m_p]$ & $\bm{g^2}$\\ [1.2ex]
        \hline \Tstrut
        \multirow{4}{4em}{$50 \sqrt{\frac{2}{3}}$} & \multirow{3}{5em}{$4.8 \times 10^{-13}$} & \multirow{3}{3em}{0.038} & \multirow{3}{6em}{$-5.49 \times 10^{-7}$} & \multirow{3}{5em}{$4.00 \times 10^{-5}$} & 0 \\ 
        & & & & & $1.6 \times 10^{-6}$ \\
        & & & & & $8.0 \times 10^{-6}$ \\
        & & & & & $4.0 \times 10^{-5}$ \\
        \hline \Tstrut
        \multirow{4}{4em}{$100 \sqrt{\frac{2}{3}}$} &\multirow{3}{5em}{$8.5 \times 10^{-14}$} & \multirow{3}{3em}{0.020} & \multirow{3}{6em}{$-2.28 \times 10^{-7}$} & \multirow{3}{5em}{$3.37 \times 10^{-5}$} & 0 \\
        & & & & & $1.6 \times 10^{-6}$ \\
        & & & & & $8.0 \times 10^{-6}$ \\
        & & & & & $4.0 \times 10^{-5}$ \\
        \hline 
    \end{tabular}
\end{table}
\begin{table}[htb]
    \centering
    \caption{Values of various parameters used for the simulations with the T-model inflaton potential.}
    \label{tab:Sim_Parameters_tmodel}
    \begin{tabular}{|c c c c c c|}
        \hline\Tstrut
         $\bm{\lambda_{\text{\tiny T}}}$ & $\bm{V_0}~[m^4_p]$ & $\bm{\phi_{\rm in}}~[m_p]$ & $\bm{\dot{\phi}_{\rm in}}~[m^2_p]$ & $ \bm{m}~[m_p]$ & $\bm{g^2}$ \\ [1.2ex]
        \hline\Tstrut
        \multirow{4}{4em}{$50 \sqrt{\frac{2}{3}}$} & \multirow{3}{5em}{$8.5 \times 10^{-13}$} & \multirow{3}{3em}{0.027} & \multirow{3}{6em}{$-7.41 \times 10^{-7}$} & \multirow{3}{5em}{$5.32 \times 10^{-5}$} & 0 \\
        & & & & & $1.6 \times 10^{-6}$ \\
        & & & & & $8.0 \times 10^{-6}$ \\
        & & & & & $4.0 \times 10^{-5}$ \\
        \hline\Tstrut
        \multirow{4}{4em}{$100 \sqrt{\frac{2}{3}}$} & \multirow{3}{5em}{$6.5 \times 10^{-14}$} & \multirow{3}{3em}{0.014} & \multirow{3}{6em}{$-2.06 \times 10^{-7}$} & \multirow{3}{5em}{$2.94 \times 10^{-5}$} & 0 \\
        & & & & & $1.6 \times 10^{-6}$ \\
        & & & & & $8.0 \times 10^{-6}$ \\
        & & & & & $4.0 \times 10^{-5}$ \\
        \hline
    \end{tabular}
\end{table}
 
\medskip

Before proceeding further to discuss the results from our lattice simulations, let us recall that the  inflaton oscillations around the minimum of the aforementioned asymptotically flat potentials may be tachyonic in nature. In fact, our analysis in App.~\ref{app:tachyonicity} shows that the inflaton oscillations are tachyonic  for  $\lambda_{\text{\tiny T}}\geq 0.28$ in the T-model, and for  $\lambda_{\text{\tiny E}} \geq 1.44$  in the E-model, where the effective inflaton mass-squared term $m^2_{\text{eff}} \equiv V_{,\phi\phi}$ is negative~\cite{Felder:2000hj, Felder:2001kt}. This indeed is the case for the chosen values of $\lambda_\text{\tiny E}, \, \lambda_\text{\tiny T}$ in our simulations, as given  in  Tables~\ref{tab:Sim_Parameters_emodel}~and~\ref{tab:Sim_Parameters_tmodel}. For example,  Fig.~\ref{fig:phi_field_tachy} in App.~\ref{app:tachyonicity} demonstrates that  the inflaton condensate explores the tachyonic regime for the first few oscillations before cosmic expansion dampens out the field amplitude (see App.~\ref{app:tachyonicity} for details). Furthermore, for ultra-large values of the self-couplings, namely $\lambda_{\text{\tiny E}}, \, \lambda_{\text{\tiny T}} \gg {\cal O}(10^2)$, the mass terms in the Taylor expansion of the potentials given in Eqs.~(\ref{eq:near_harmonic_approx_Emodel}) and (\ref{eq:near_harmonic_approx_Tmodel}) become negligible compared to the self-interaction terms. In this case inflaton fragmentation does not lead to oscillon formation, since oscillons are predominantly non-relativistic field configurations~\cite{Kim:2021ipz}. This is shown by the shaded purple region of the parameter space plot in Fig.~\ref{fig:par_space_lambda_g^2}. We will further explore scalar field fragmentation in the presence of such large self-couplings analytically in our upcoming paper~\cite{Mishra:2024Part2}.  

\subsection{Oscillon formation}
\label{subsec:sim_oscillon_form}
\begin{figure}[t]
    \centering
    \subfloat{\includegraphics[width = 0.48\textwidth]{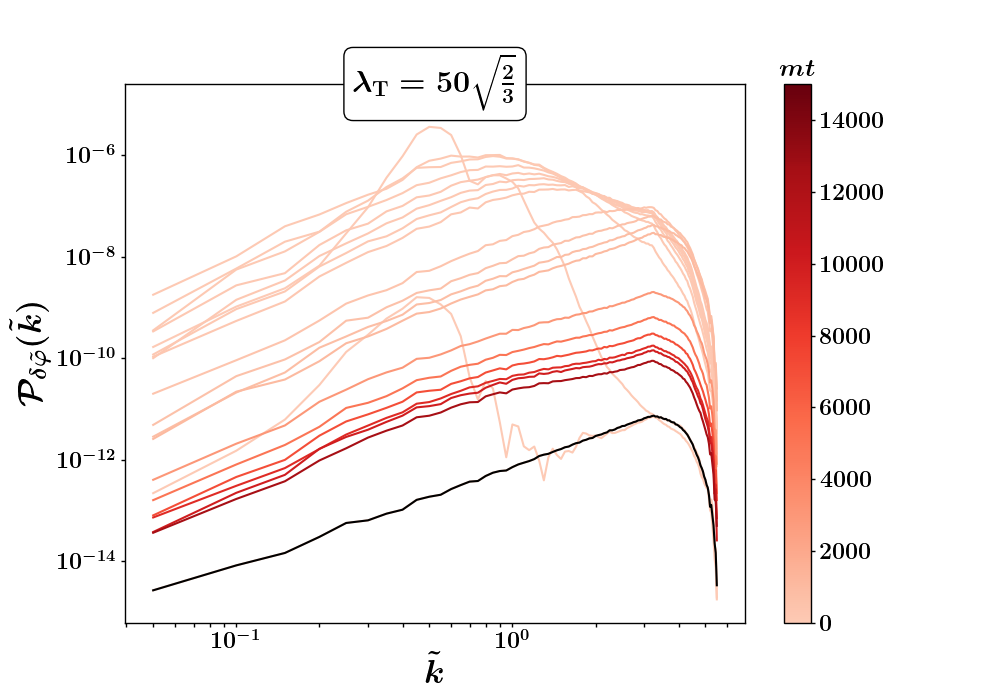}} 
    \subfloat{\includegraphics[width = 0.48\textwidth]{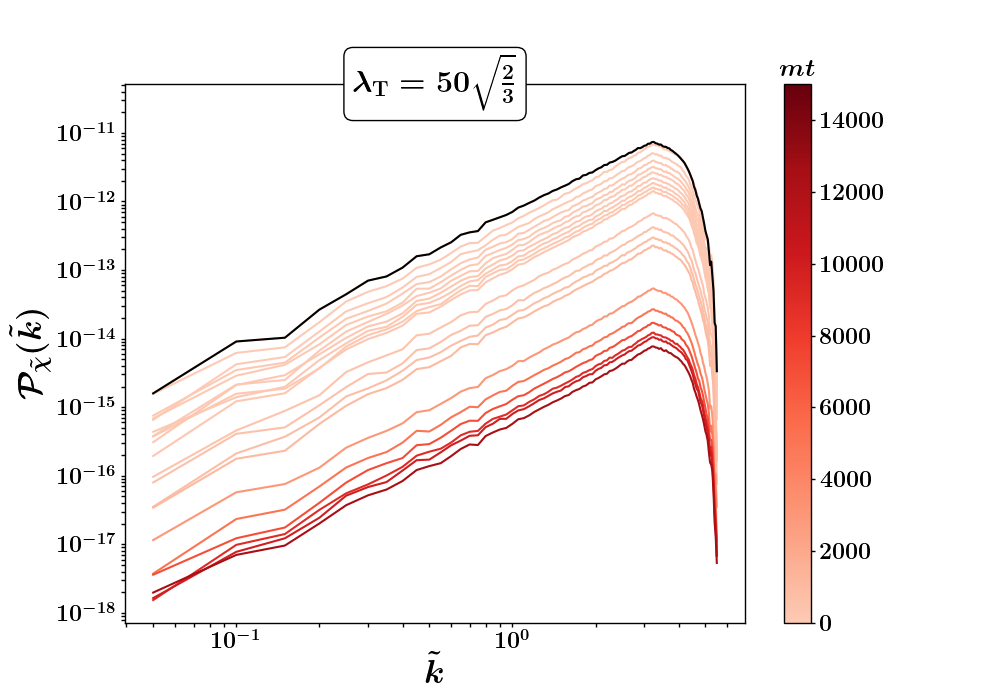}} \\
    \subfloat{\includegraphics[width = 0.48\textwidth]{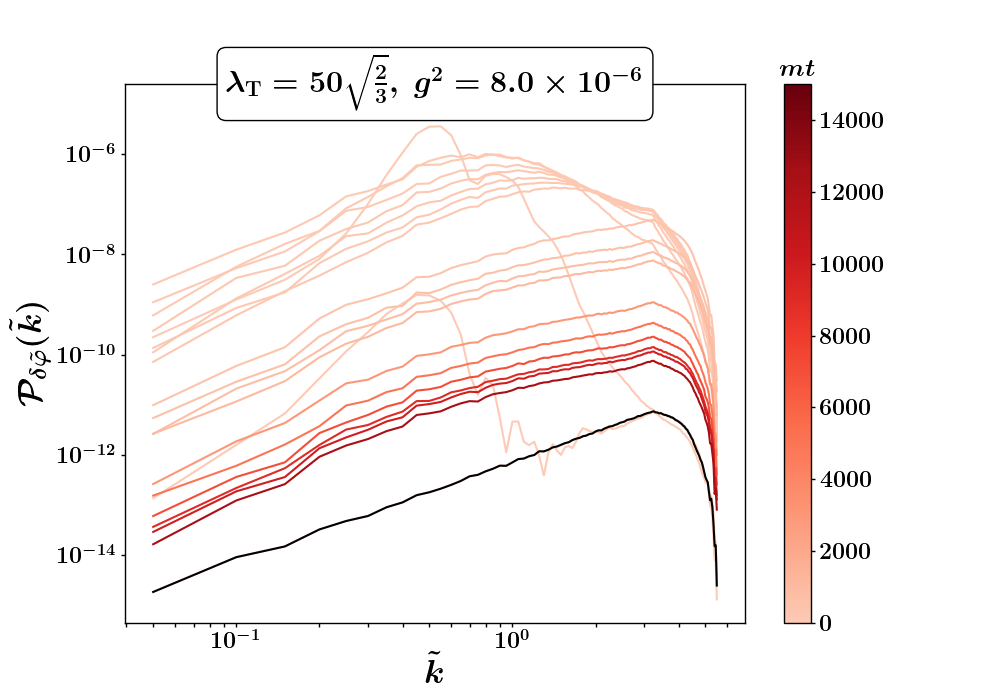}}
    \subfloat{\includegraphics[width = 0.48\textwidth]{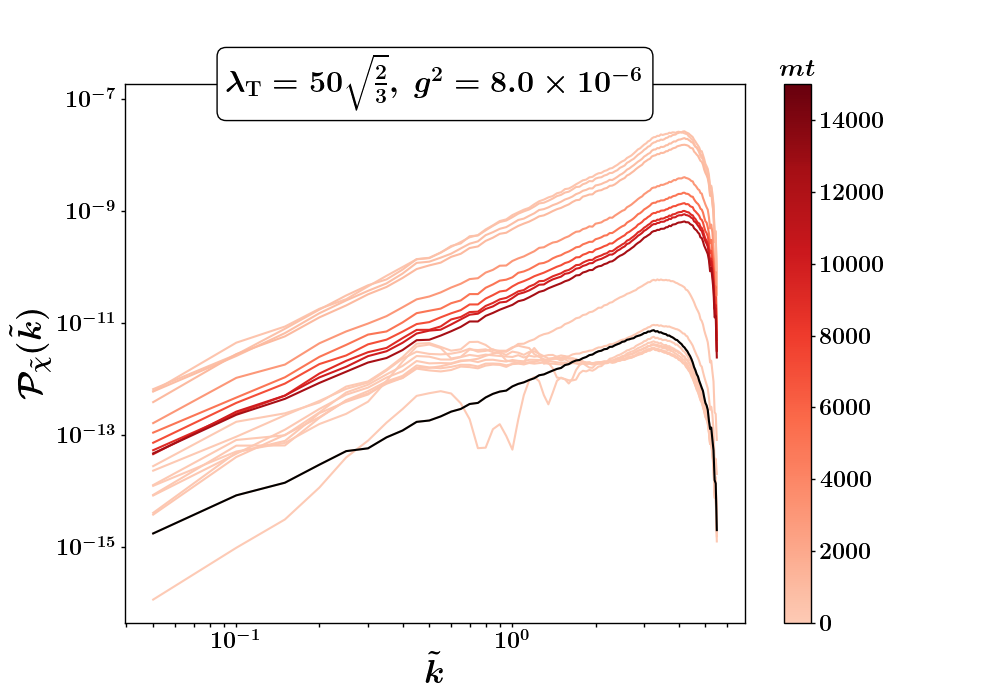}} \\
    \caption{Power spectra of field fluctuations in the T-model, for $\lambda_\text{\tiny T} = 50\sqrt{2/3}$, in the absence (\textbf{top}) and presence (\textbf{bottom}) of external coupling  obtained from lattice simulations are plotted with the black lines depicting the initial power spectra.  The results from lattice simulations are in excellent agreement with the  linear analyses  in Sec.~\ref{sec:analyt} until $t \sim 100\,m^{-1}$, after which  significant backreaction of the field fluctuations onto the homogeneous condensate disrupts the resonant processes.  For inflaton fluctuations shown in the left panels, the peaks in the power spectra occurring for  $\tilde{k}\lesssim 0.75$ correspond to particle production from the low-momentum modes that traverse the broad-resonance band.}
    \label{fig:Power_Spectrum_sim}
\end{figure}

Our numerical simulations demonstrate that resonant processes during preheating  amplify certain ${\tilde k}$  modes within the first few oscillations, $t \sim {\cal O}(10^2)\:m^{-1}$. For example, this can be seen for the power spectrum of  inflaton fluctuations ${\cal P}_{\delta\tilde{\varphi}}(\tilde{k})$  shown in the left panels of  Fig.~\ref{fig:Power_Spectrum_sim}  which is  qualitatively different from the initial one (shown in black). This is in excellent agreement with the  linear analyses made in Sec.~\ref{sec:analyt}  using Floquet theory, where in the left panel of Fig.~\ref{fig:reh_resonance_self_ext}, we observe that the Fourier modes of inflaton fluctuations with $\tilde{k}\lesssim 0.75$ pass through the broad resonance band in the Floquet chart, which are indicative of the amplifications occurring in $\mathcal{P}_{\delta\tilde{\varphi}}(\tilde{k})$. Larger $\tilde{k}$ modes also experience amplification, albeit at slightly later times, when they pass through  the  first resonance band in the narrow regime. On the other hand,  in the top right panel of Fig.~\ref{fig:Power_Spectrum_sim}, we see that, in the absence of $g^2$, the power spectrum of the offspring field $\mathcal{P}_{\delta\tilde{\chi}}(\tilde{k})$ steadily decreases in amplitude with the expansion of the universe. With the interaction term turned on, we observe, in the bottom right panel of Fig.~\ref{fig:Power_Spectrum_sim}, a gradual amplification in the power spectrum, reflected by the fact that multiple modes cross the various narrow-resonance bands, although the $\chi$-particle production is not as significant as that of the inflaton fluctuations (for the given value of $g^2$).

\begin{figure}[htb]
    \centering
    \subfloat{\includegraphics[width = 0.47\textwidth]{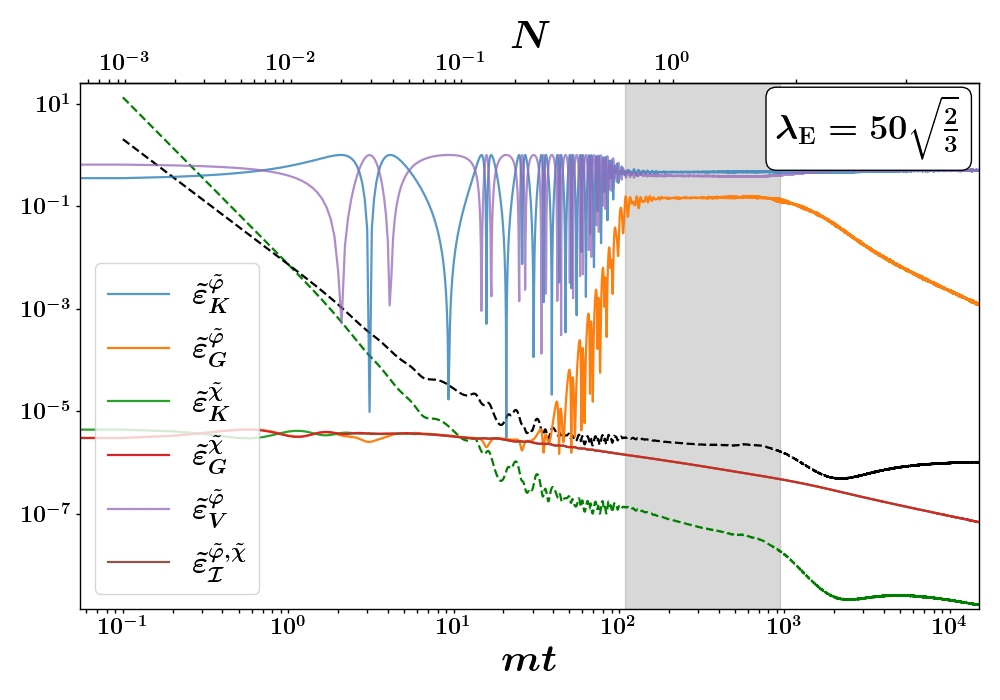}} 
    \subfloat{\includegraphics[width = 0.47\textwidth]{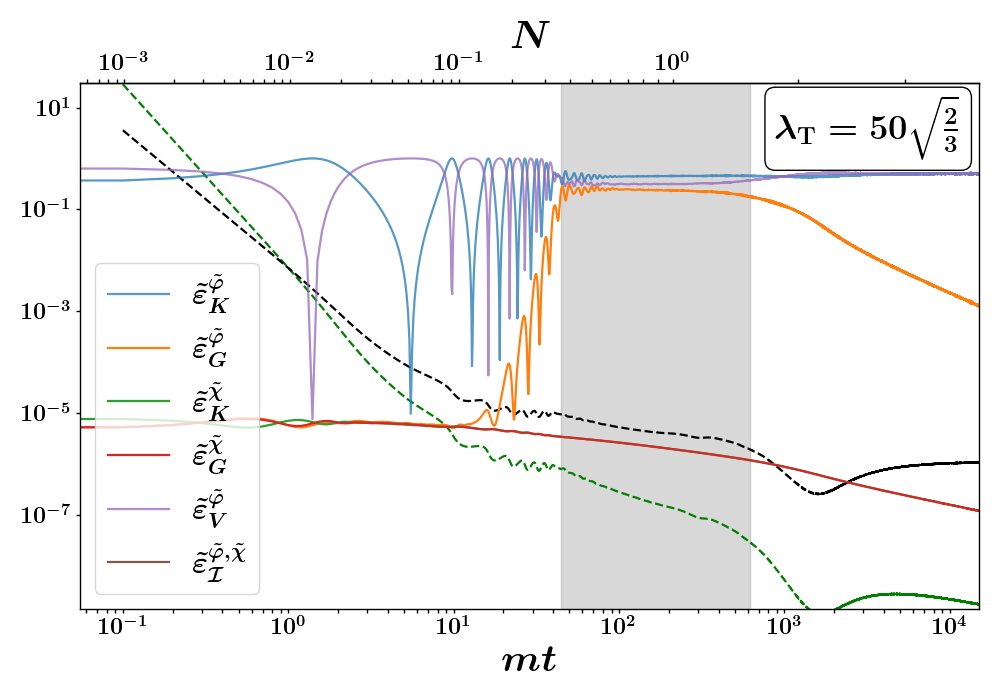}} \\
    \subfloat{\includegraphics[width = 0.47\textwidth]{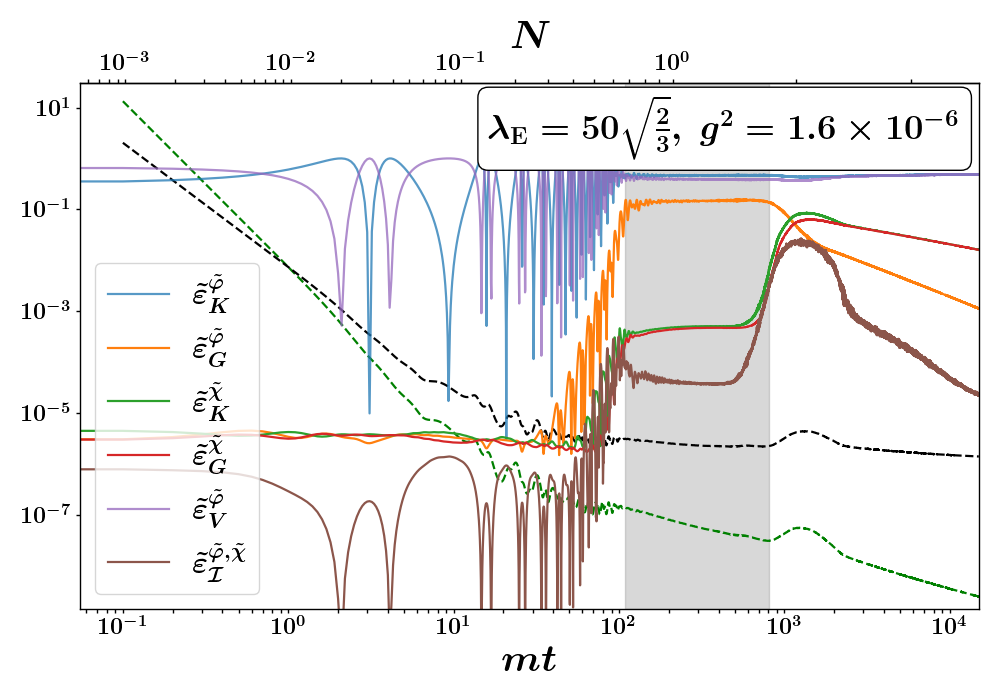}}
    \subfloat{\includegraphics[width = 0.47\textwidth]{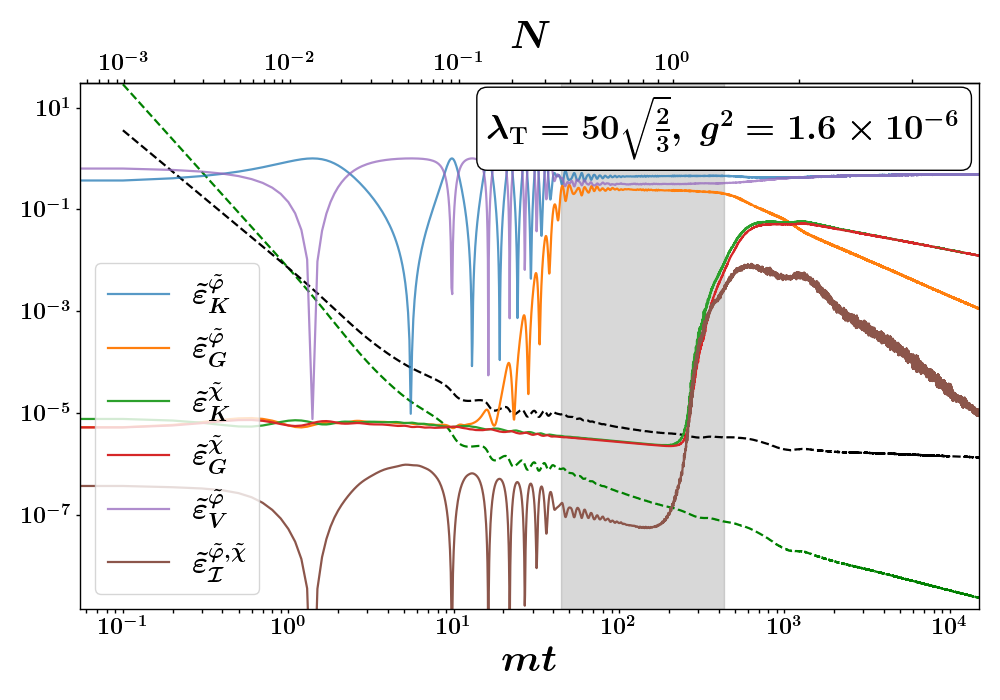}} \\
    \subfloat{\includegraphics[width = 0.47\textwidth]{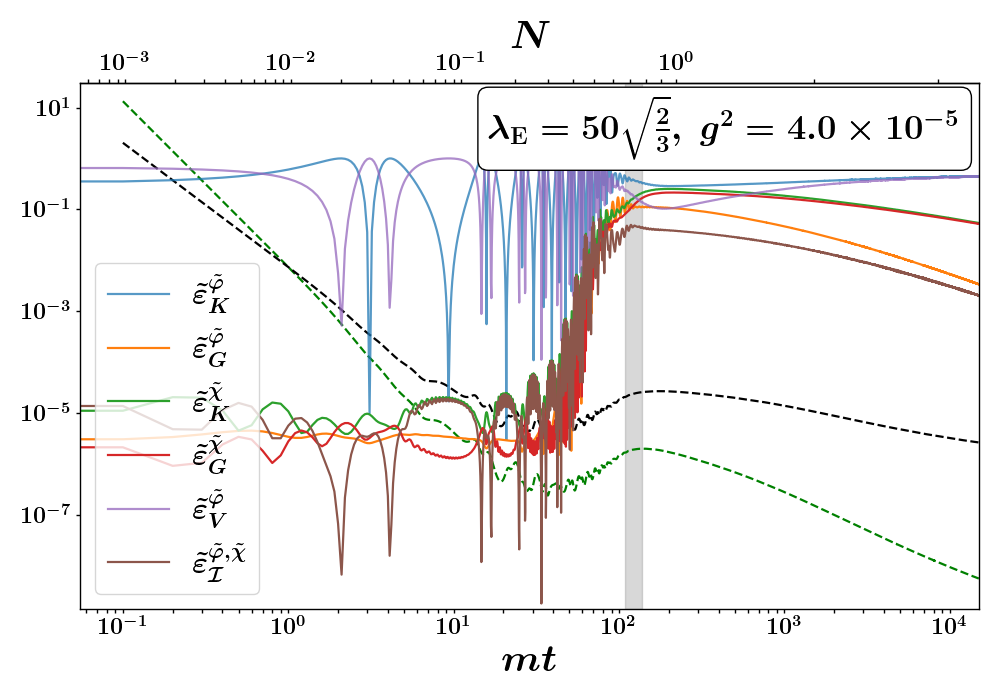}}
    \subfloat{\includegraphics[width = 0.47\textwidth]{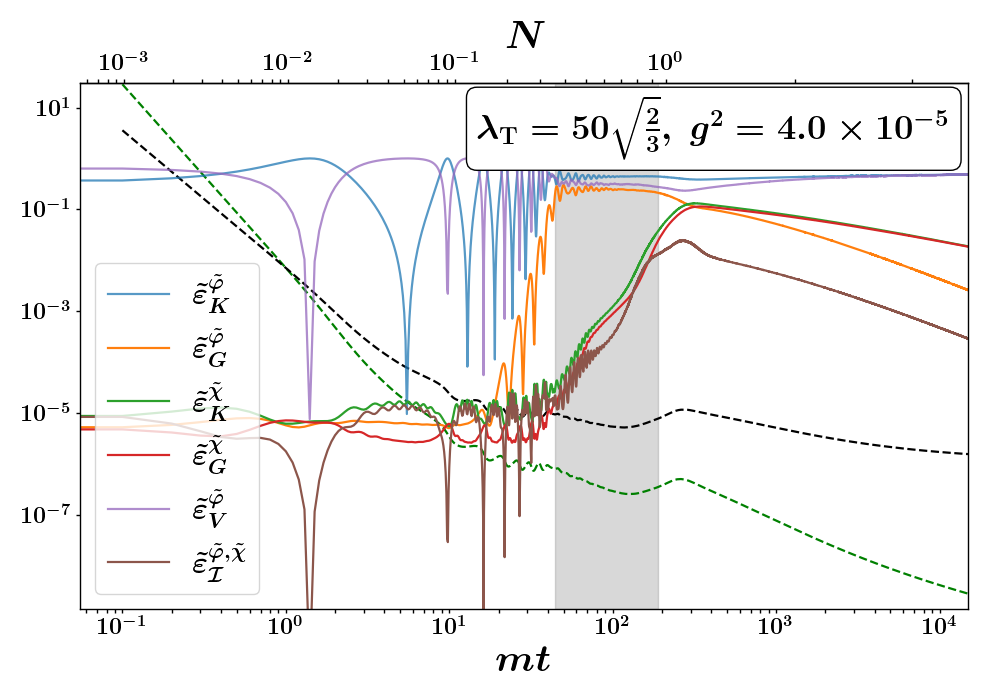}}
    \caption{Evolution of the volume-averaged  fractional energy density components for $g^2 = 0$ ({\bf top row}), $g^2 =1.6 \times 10^{-6}$ ({\bf middle row}) and, $g^2 =4.0 \times 10^{-5}$ ({\bf bottom row})  for the E-model (\textbf{left column}) and T-model (\textbf{right  column}) potentials with $\lambda_\text{\tiny E}, \, \lambda_\text{\tiny T} = 50\sqrt{2/3}$ are shown. The shaded regions  represent the duration during which $\widetilde{G}_{\tilde{\varphi}} \propto \widetilde{K}_{\tilde{\varphi}} \, (\simeq \widetilde{V}_{\tilde{\varphi}})$. We observe that the gradient energy density of the inflaton ${\widetilde G}_{\varphi}$ grows considerably during  preheating due to parametric resonance, which results in the formation of long-lived oscillons in the absence of external coupling.  However, although non-linear inflaton lumps (proto-oscillons) do form initially, the presence of a strong enough external coupling,  such as  $g^2 \gtrsim 4 \times 10^{-5}$, quickly disrupts the formation of robust oscillons, as the newly formed dense  fragmented transients of the inflaton  quickly dissipate into the $\chi$-particles. From the dashed black and green lines, acting as proxies for the evolution of test matter and radiation fields respectively, we observe that the gradient energy density of the inflaton ${\widetilde G}_{\tilde \varphi}$ falls roughly as non-relativistic matter after the formation of oscillons, and  as radiation at asymptotically late times.}
    \label{fig:energy_den_comp_smaller_lambda}
\end{figure}

We search for the formation of oscillon-like field configurations by analysing the behaviour of the gradient  energy density $\widetilde{G}_{\tilde{\varphi}}$  of the inflaton field, as defined in Eq.~(\ref{eq:kinetic_gradient_phi}). At the onset of preheating, we expect the gradient term to provide a subdominant contribution to the total  energy density of the system, owing to the fact that significant nonlinearities do not tend to develop towards the end of inflation for smooth asymptotically flat potentials. However, after the first $\sim {\cal O}(10^1)$ oscillations, nonlinearities begin to develop   on sub-Hubble scales, inducing a rather sharp increase  in $\widetilde{G}_{\tilde{\varphi}}$. In Fig.~\ref{fig:energy_den_comp_smaller_lambda}, we showcase the  evolution of the  volume-averaged fractional energy densities of the fields  $\varphi$ and $\chi$, defined in Eq.~(\ref{eq:rescaled_frac_rho}), with time (number of $e$-folds are shown on the upper horizontal axis). For the given values of  $\lambda_{\text{\tiny E}}, \, \lambda_{\text{\tiny T}}$, we observe that the gradient term ${\widetilde G}_{\varphi}$ of the inflaton field grows considerably within $t\sim 100\,m^{-1}$,  contributing an appreciable fraction of the total energy density of the system. 
 
\begin{figure}[hbt]
    \centering
    \subfloat{\includegraphics[width = 0.75\textwidth]{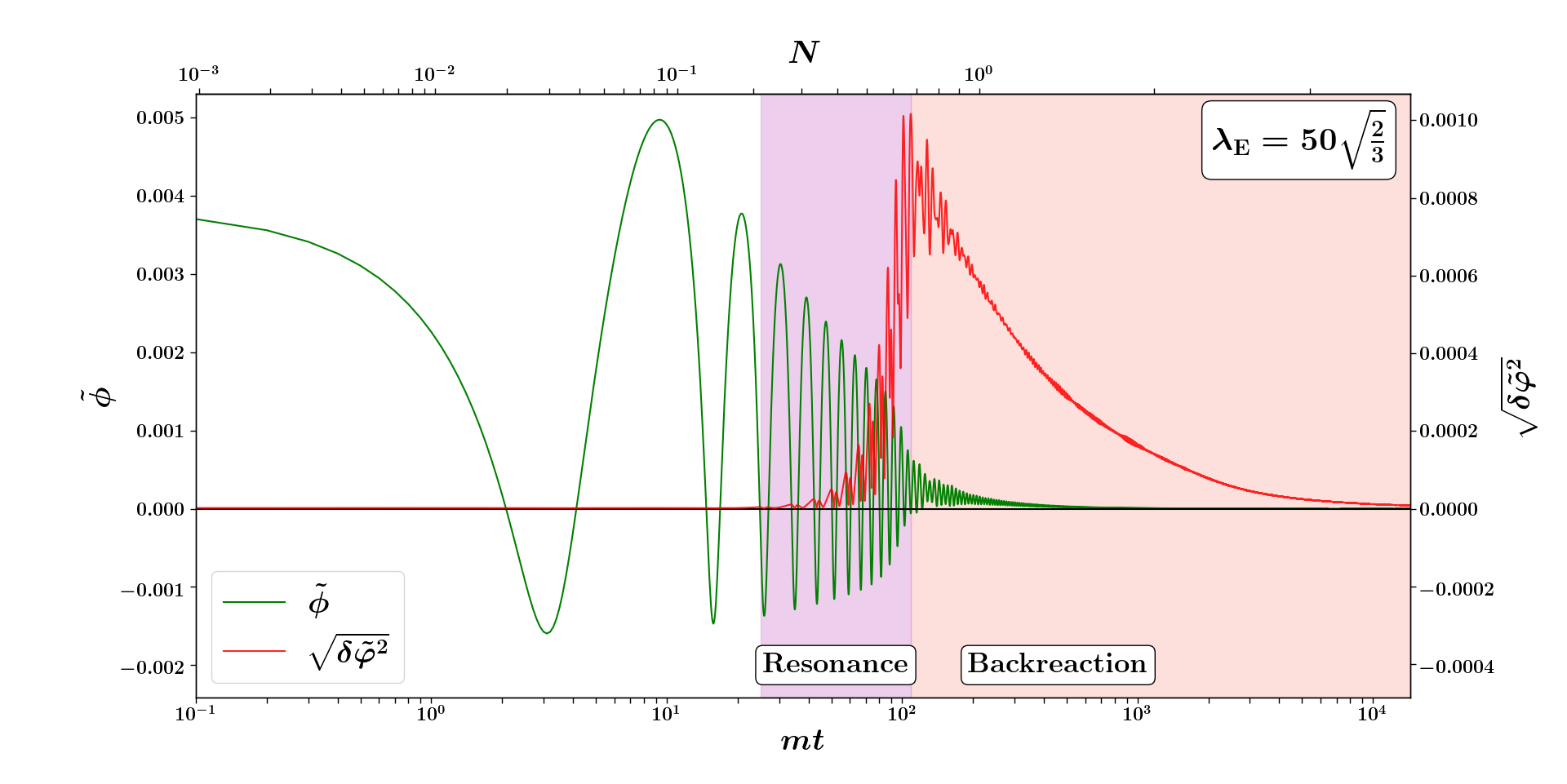}} \\
    \subfloat{\includegraphics[width = 0.75\textwidth]{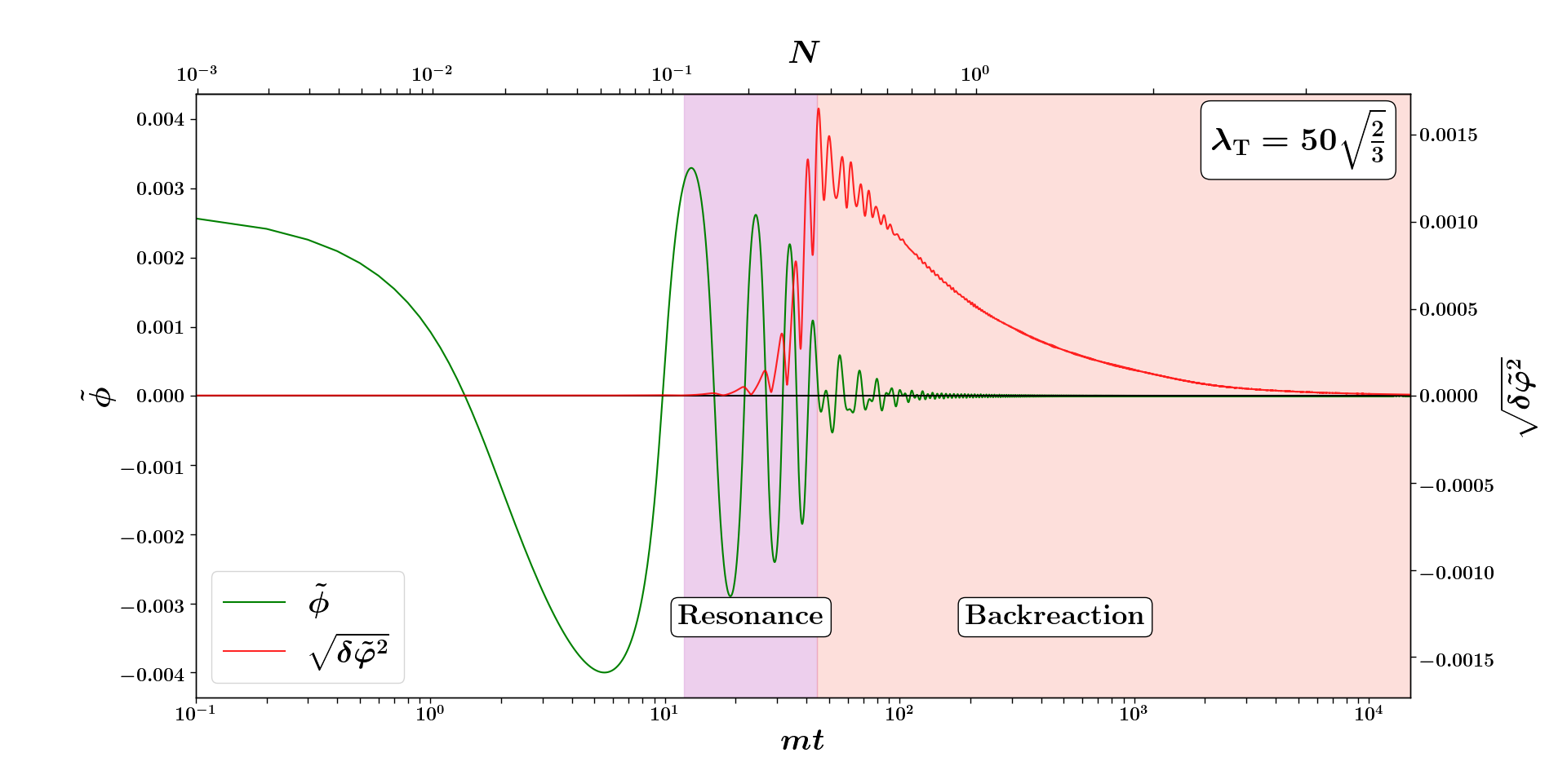}}
    \caption{ Evolution of the volume-averaged inflaton configurations are plotted for the case $g^2 = 0$ of the E-model (\textbf{top panel}) and T-model (\textbf{bottom panel}) with $\lambda_\text{\tiny E}, \, \lambda_\text{\tiny T} = 50\sqrt{2/3}$, with the upper horizontal axis being the number of $e$-folds $N$ elapsed  since the end of inflation.  Successful self-resonance results in the amplification of $\delta {\widetilde \varphi}$, shown here as a solid red line. As a result,  the homogeneous inflaton condensate (shown in solid green) fragments within $t \lesssim 100\,m^{-1}$, forming lumps in the process, leading to the formation of oscillons.  We note that most oscillons form at the onset of the backreaction phase (the transition from resonance to backreaction phases) near the peak of $\delta {\tilde \varphi}$.}
    \label{fig:field_dynamics_no_coupling}
\end{figure}

The sharp rise in the gradient term can be understood from Fig.~\ref{fig:field_dynamics_no_coupling}, where the  time evolution of the volume-averaged inflaton fluctuation $\sqrt{ \delta\widetilde{\varphi}^2 }$ is plotted in red.  Note that the volume-averaged fluctuation is calculated by first determining the time evolution of the inflaton fluctuations on the $3d$ grid as $\delta\widetilde{\varphi}\left(\tilde{t},\widetilde{\bm{x}}\right)=\widetilde{\varphi}\left(\tilde{t},\widetilde{\bm{x}}\right)-\widetilde{\phi}\left(\tilde{t}\right)$, and then using the definition
 \beq
\langle\delta\widetilde{\varphi}\left(\tilde{t},\widetilde{\bm{x}}\right)^2\rangle_{_\text{\tiny V}} =\frac{1}{\text{vol}}\int_{\text{vol}}\d^3 \widetilde{x}\:\left[ \widetilde{\varphi}\left(\tilde{t},\widetilde{\bm{x}}\right)-\widetilde{\phi}\left(\tilde{t}\right) \right]^2 \, .
 \label{Eq:vol_avg_inflaton}
 \eeq 

\begin{figure}[hbt]
    \centering
    \subfloat{\includegraphics[width = 0.33\textwidth]{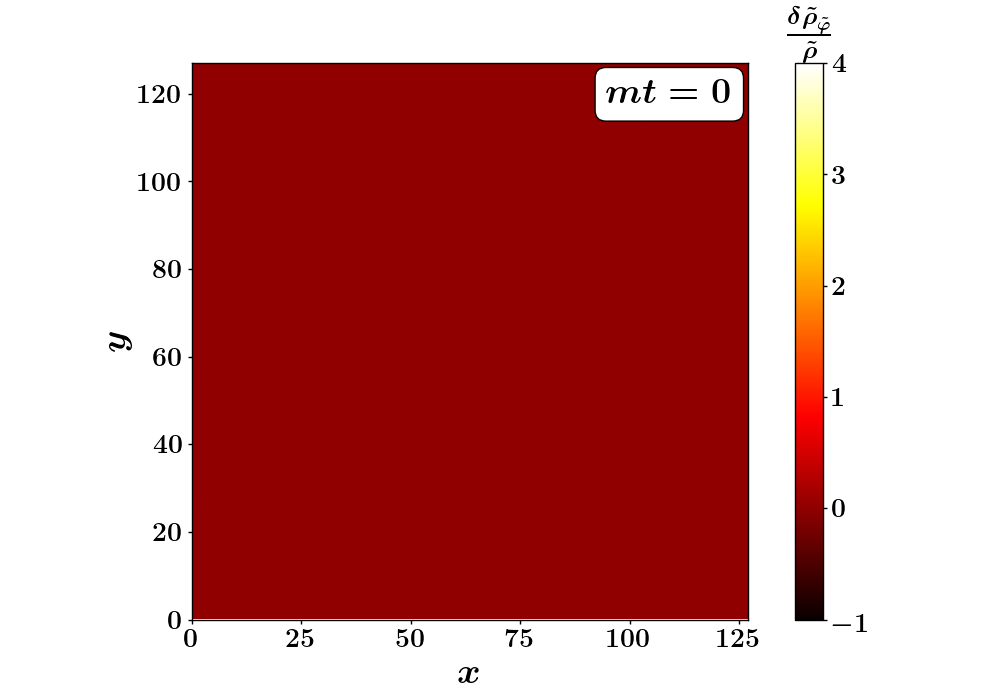}} 
    \subfloat{\includegraphics[width = 0.33\textwidth]{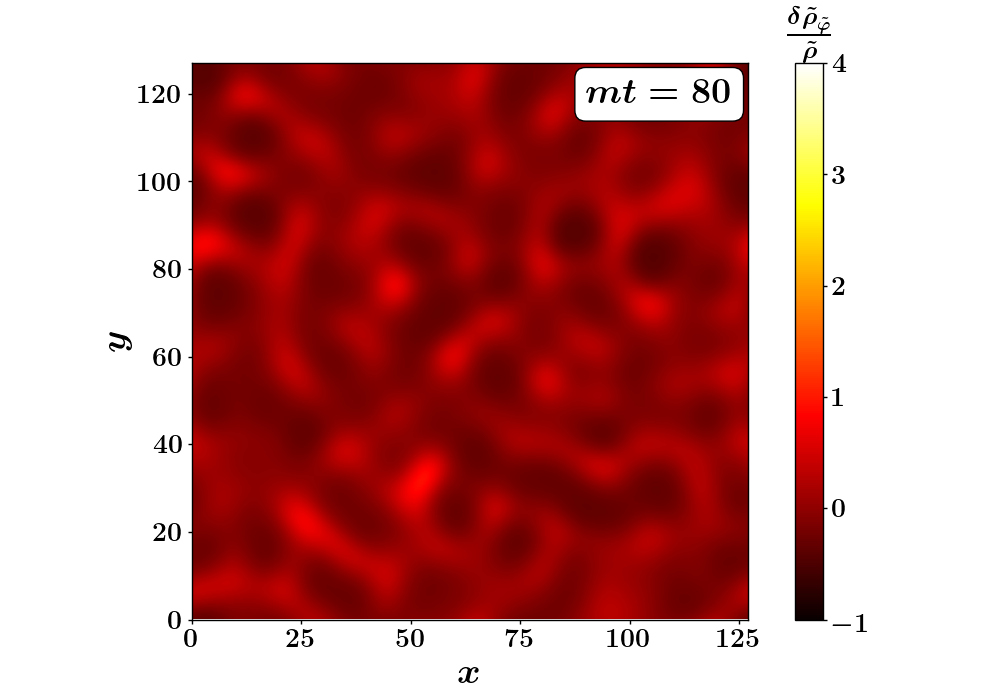}}  
    \subfloat{\includegraphics[width = 0.33\textwidth]{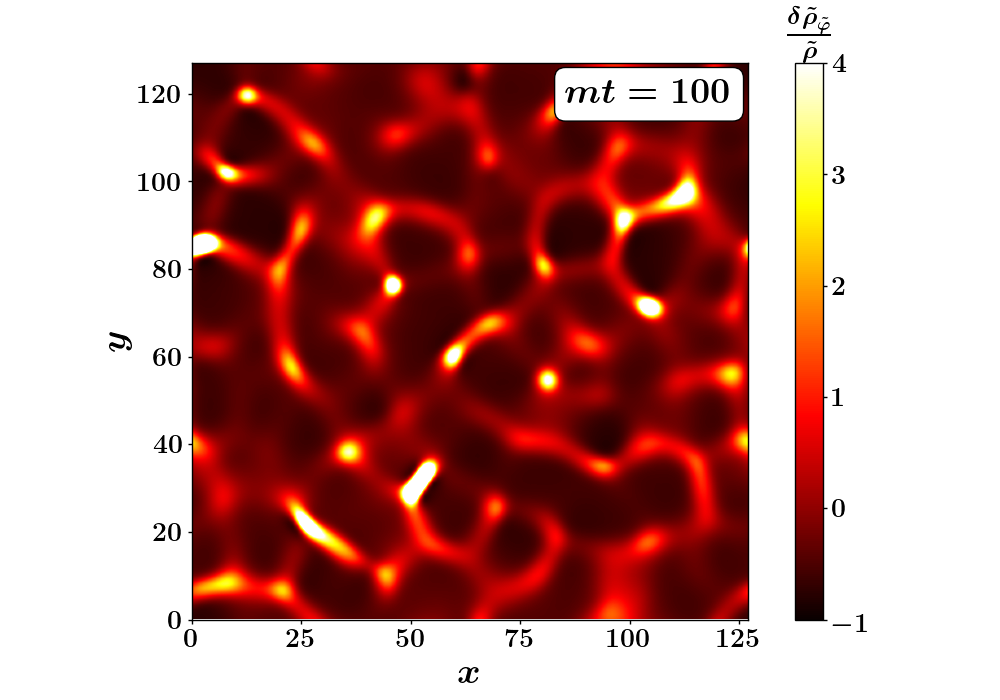}} \\
    \subfloat{\includegraphics[width = 0.33\textwidth]{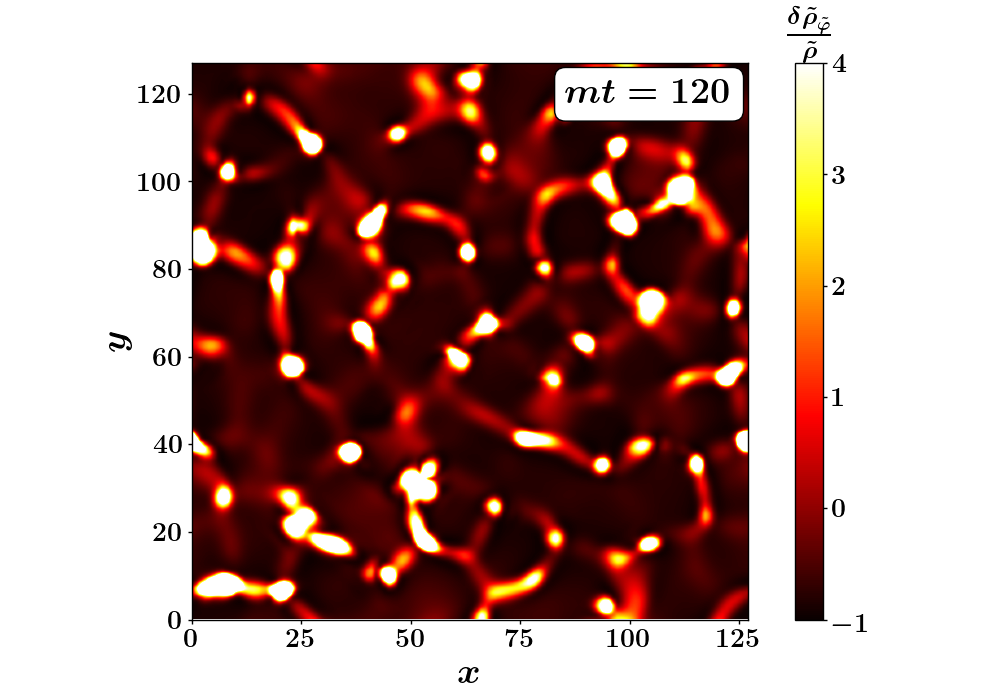}} 
    \subfloat{\includegraphics[width = 0.33\textwidth]{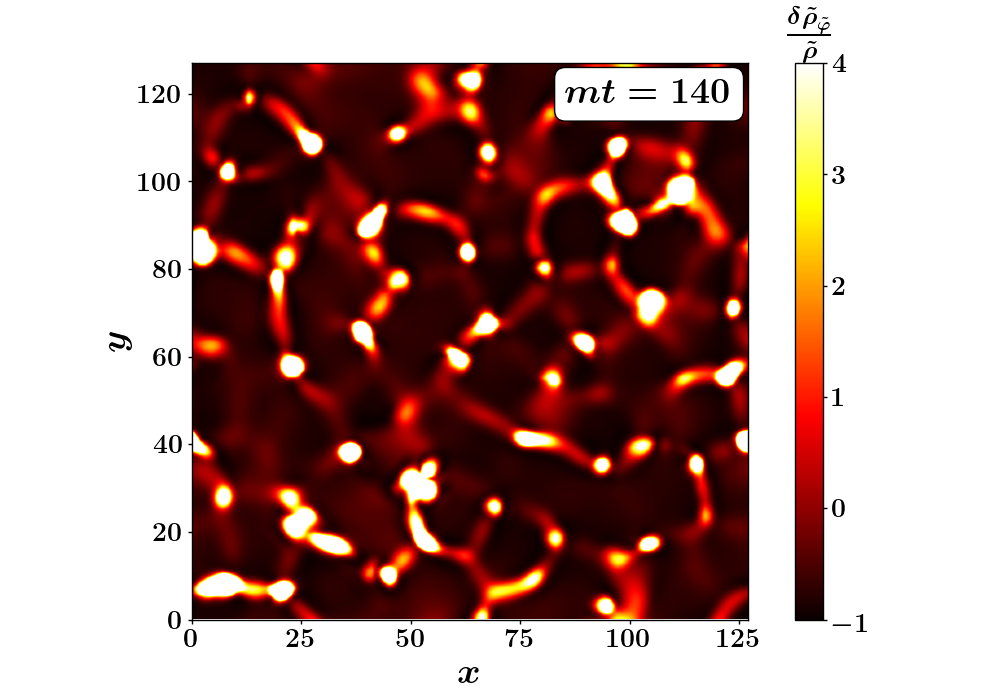}} 
    \subfloat{\includegraphics[width = 0.33\textwidth]{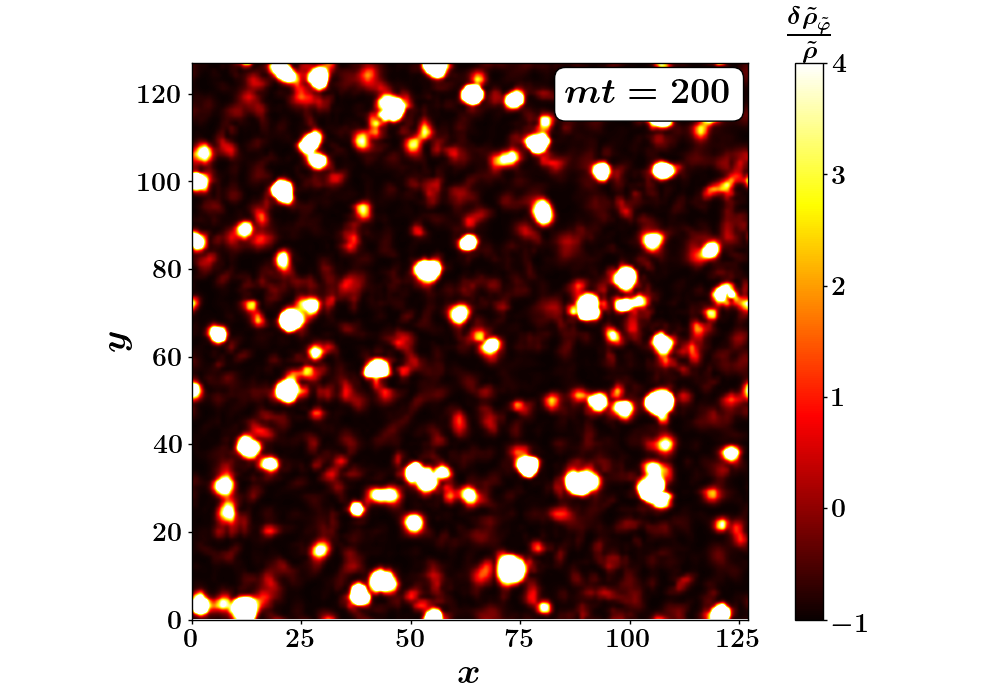}} \\
    \subfloat{\includegraphics[width = 0.33\textwidth]{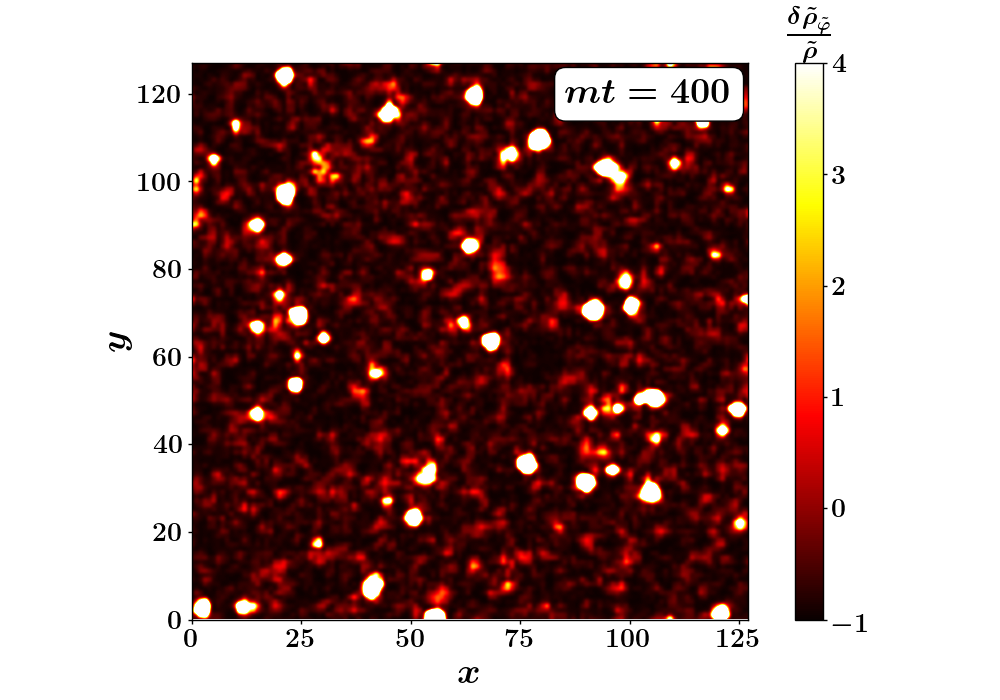}} 
    \subfloat{\includegraphics[width = 0.33\textwidth]{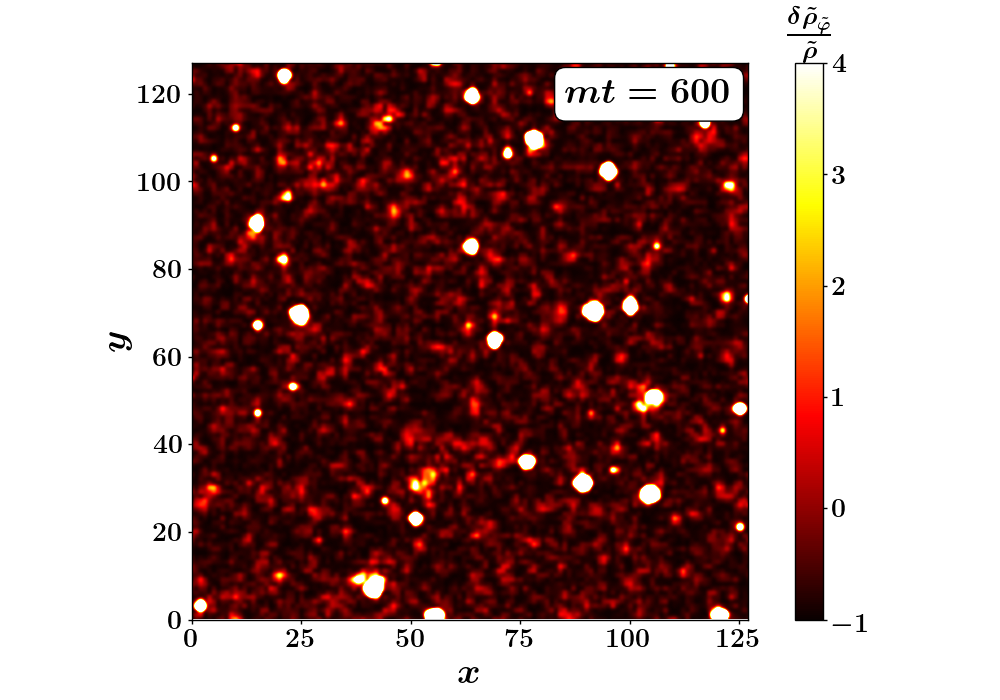}} 
    \subfloat{\includegraphics[width = 0.33\textwidth]{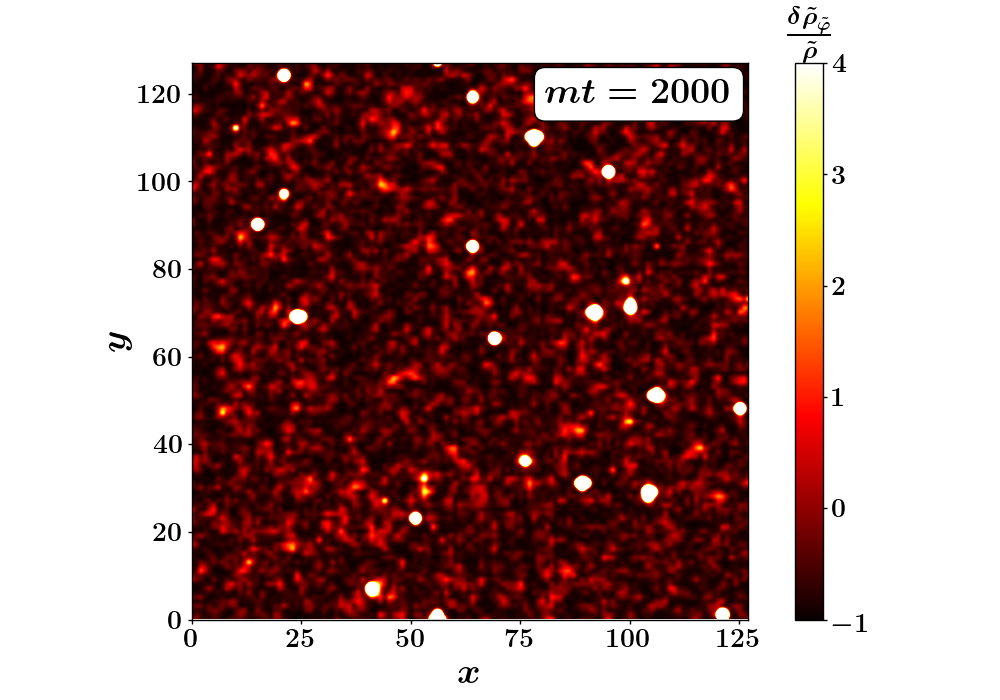}} 
    \caption{A sequence depicting the formation of oscillons from  fragmented inflaton condensate after self-resonance in the  E-model, with self-coupling $\lambda_\text{\tiny E} = 50 \sqrt{2/3}$ and external coupling $g^2 = 0$. As can be observed,  there is significant  inflaton fragmentation within the first $t \sim 100\,m^{-1}$, followed by the appearance  of oscillon-like localised non-linear structures by $t \sim 120\,m^{-1}$. Hence, oscillons form rather quickly after the inflaton fragmentation, \textit{i.e.} within a short duration  $\Delta t \sim 20\,m^{-1}$.   Also notice that most of the oscillons are long-lived, persisting beyond $t \gtrsim 10^3 \,m^{-1}$.}
\label{fig:oscillon_formation_progression_success}
\end{figure}

\begin{figure}[hbt]
    \centering
    \subfloat{\includegraphics[width = 0.33\textwidth]{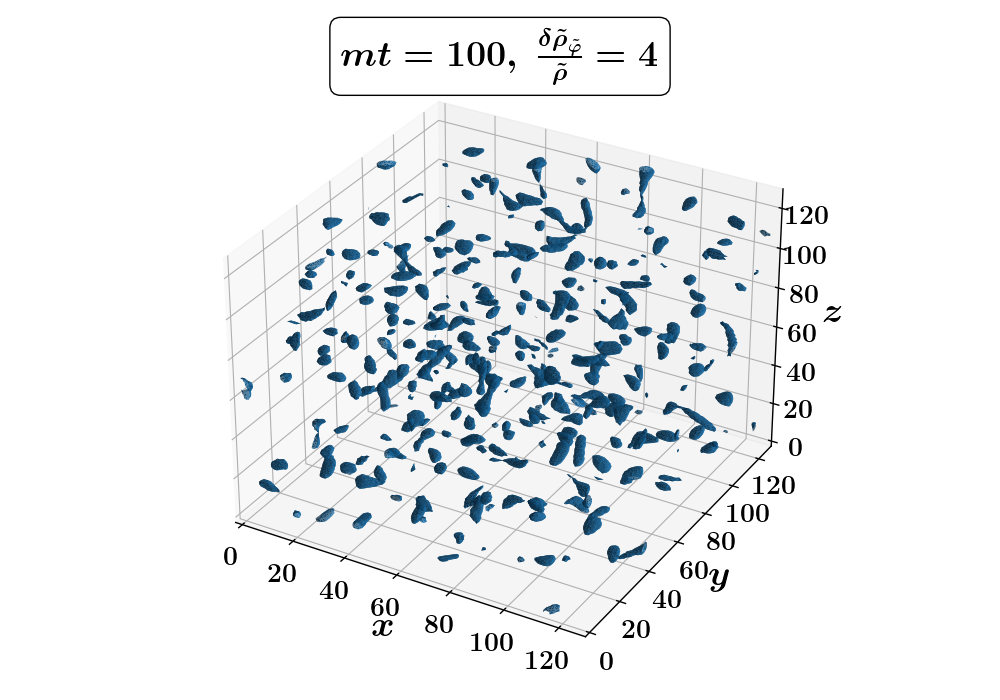}} 
    \subfloat{\includegraphics[width = 0.33\textwidth]{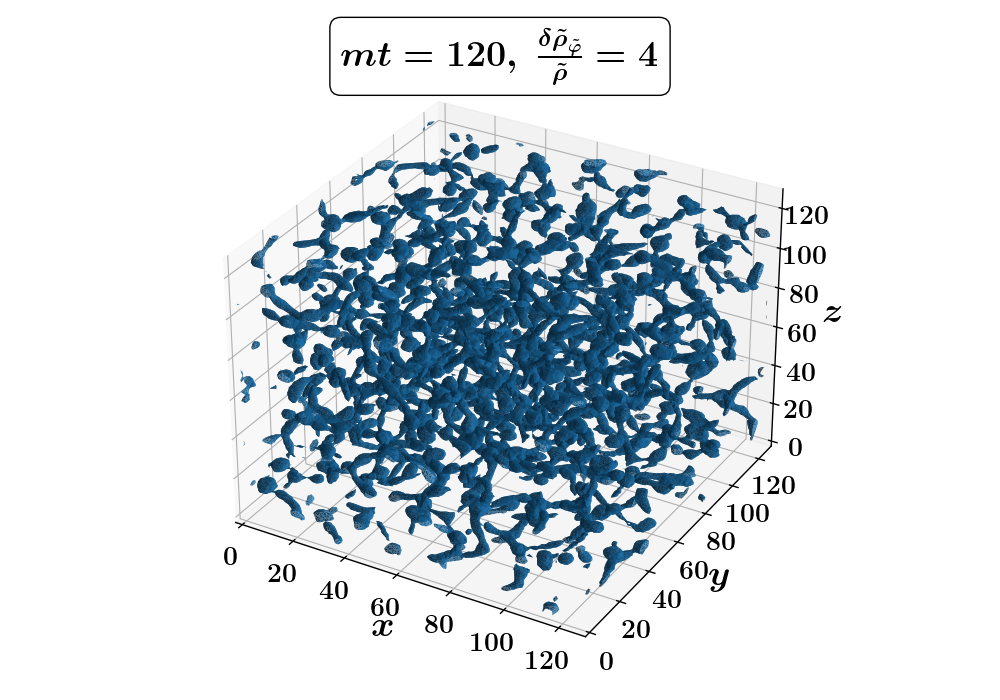}} 
    \subfloat{\includegraphics[width = 0.33\textwidth]{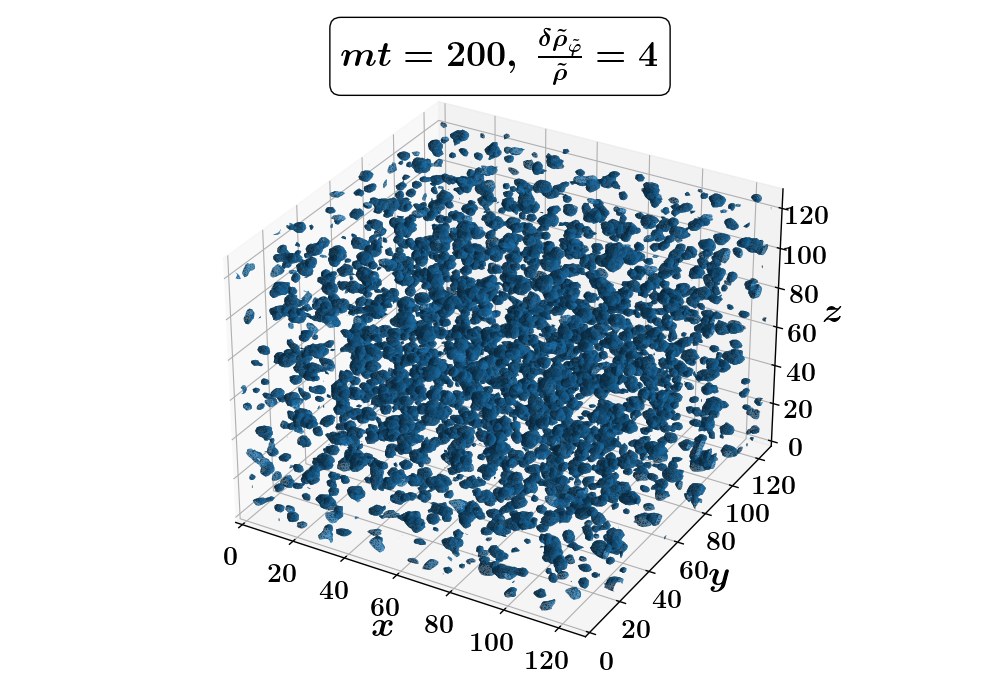}} \\
    \subfloat{\includegraphics[width = 0.33\textwidth]{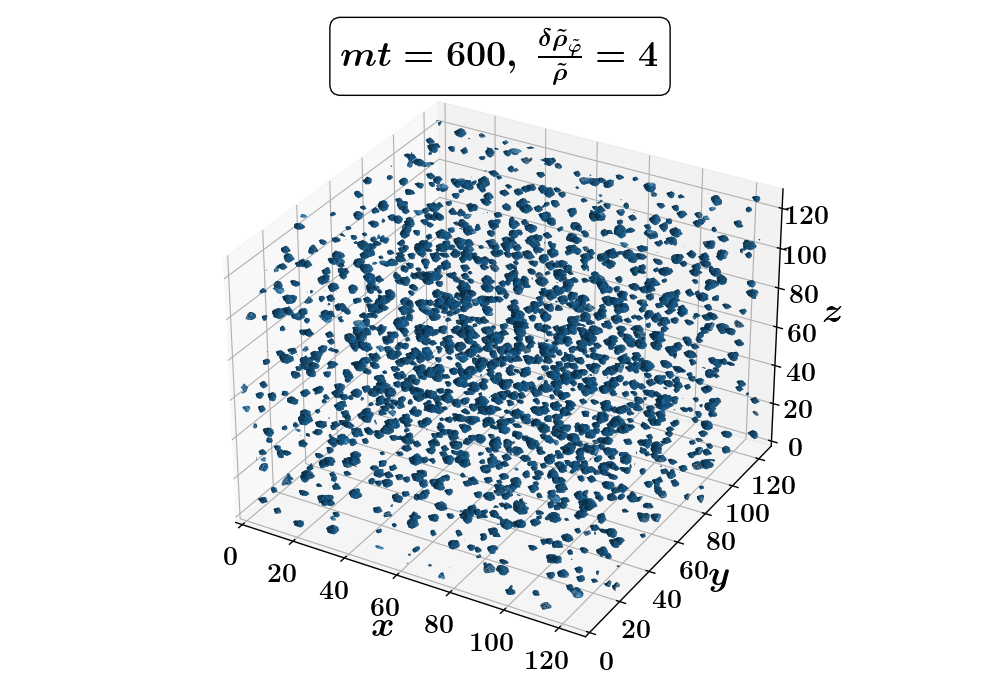}} 
    \subfloat{\includegraphics[width = 0.33\textwidth]{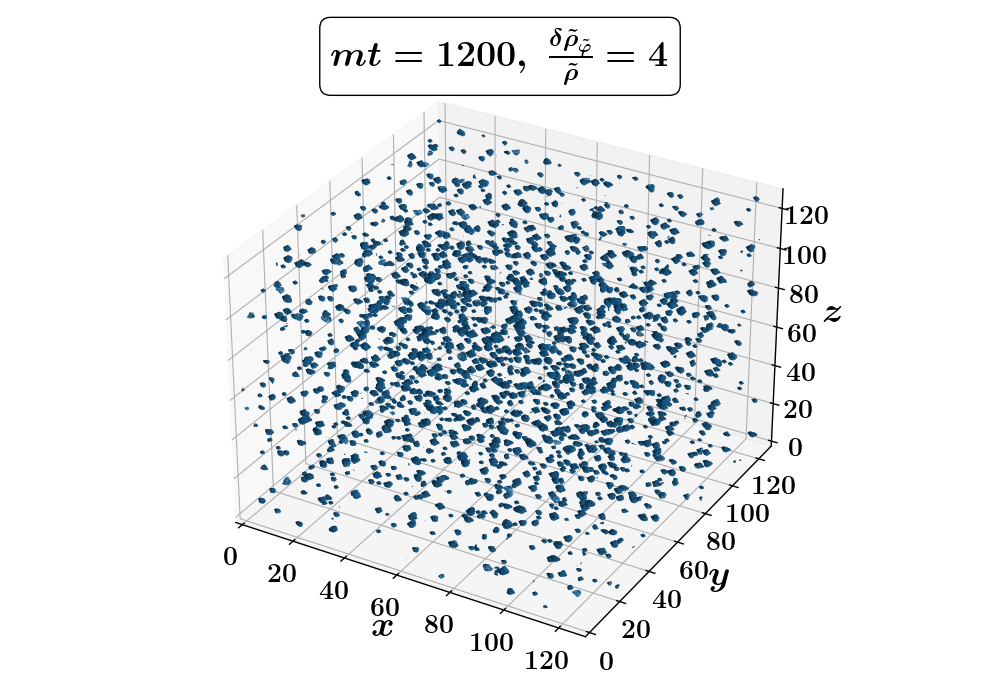}} 
    \subfloat{\includegraphics[width = 0.33\textwidth]{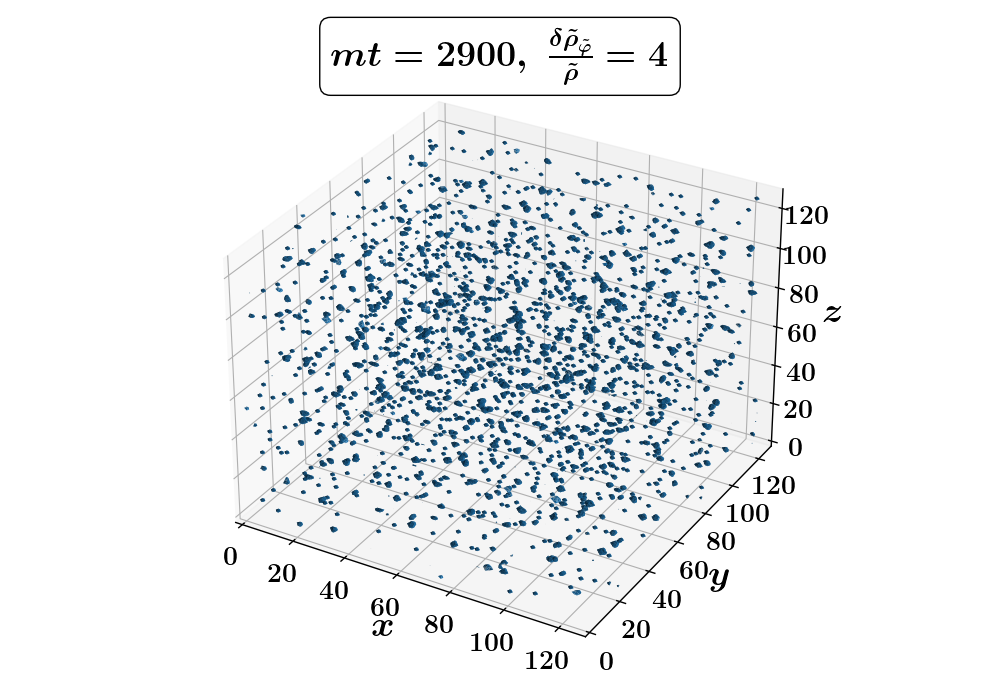}} 
    \caption{ Evolution of the constant density ($3d$) isocontours  showing the formation (and  relatively slower decay) of oscillons in the absence of an external coupling ($g^2 = 0$) in the E-model  ($\lambda_{\text{\tiny E}} = 50\sqrt{2/3}$) are shown.  Note that the sides of the  boxes are in the comoving scale. So, oscillons with fixed physical  sizes correspond to structures with decreasing comoving size in the figure.}
    \label{fig:oscillon_decay_progression_Emodel_nocoupling}
\end{figure}

\begin{figure}[hbt]
    \centering
    \subfloat{\includegraphics[width = 0.33\textwidth]{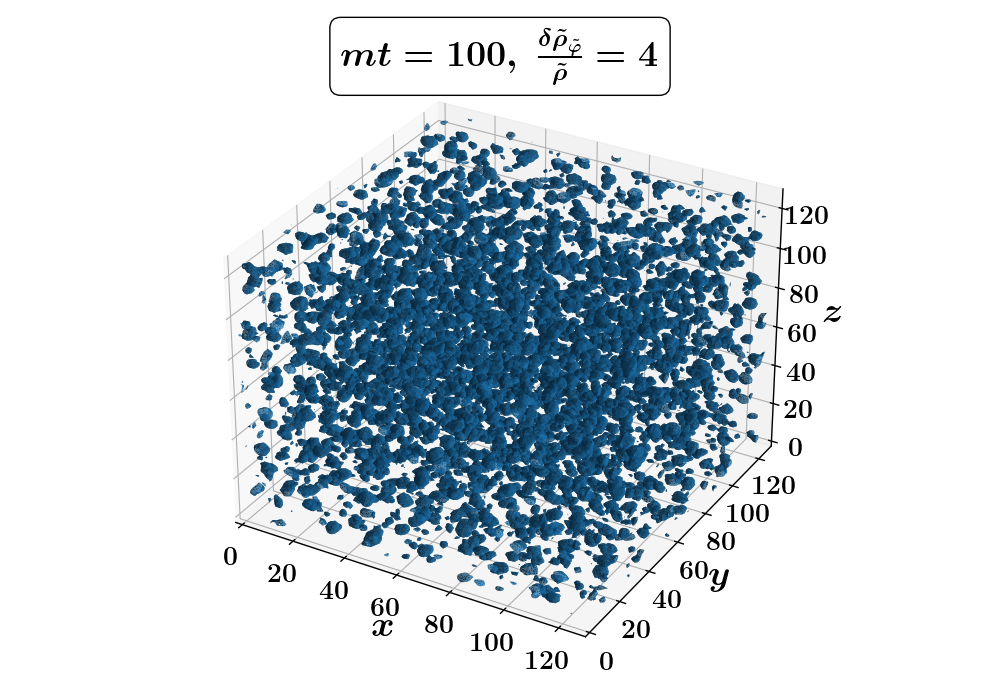}}
    \subfloat{\includegraphics[width = 0.33\textwidth]{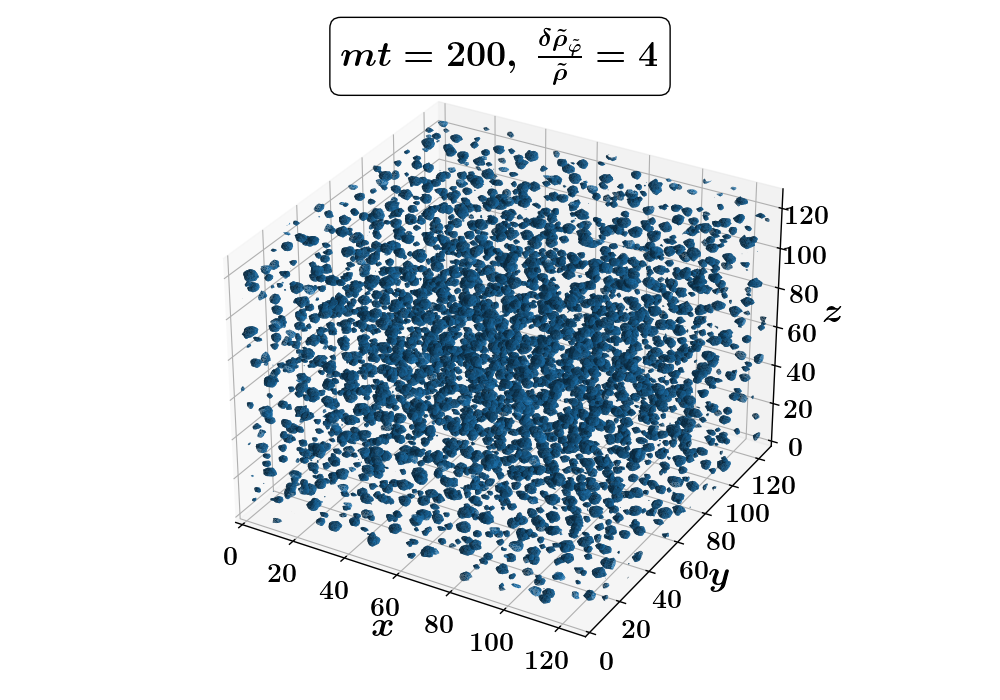}} 
    \subfloat{\includegraphics[width = 0.33\textwidth]{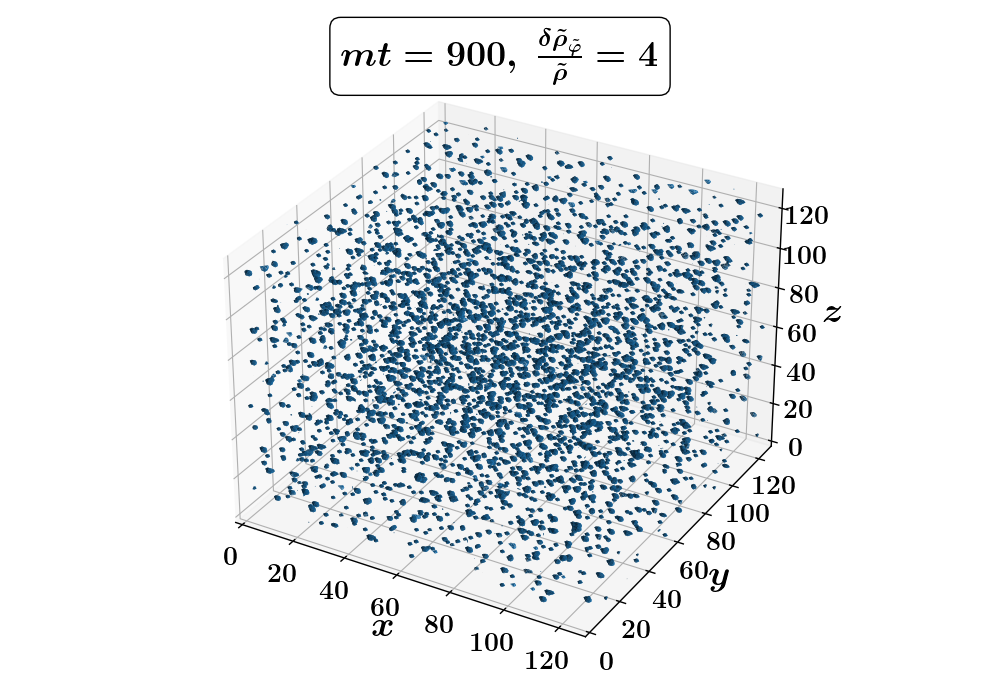}} \\
    \subfloat{\includegraphics[width = 0.33\textwidth]{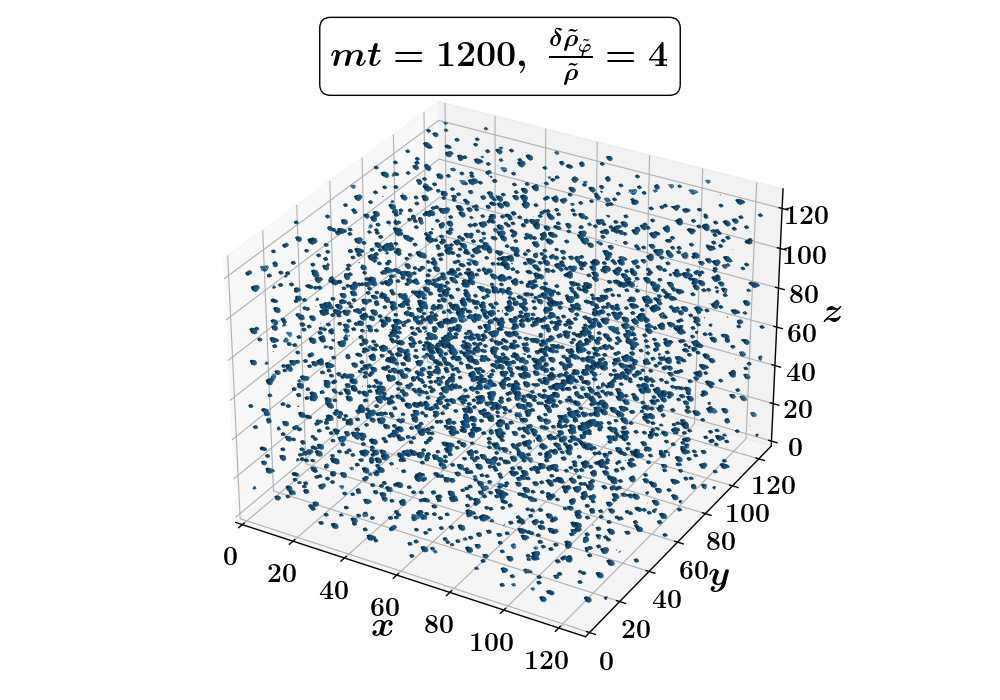}}
    \subfloat{\includegraphics[width = 0.33\textwidth]{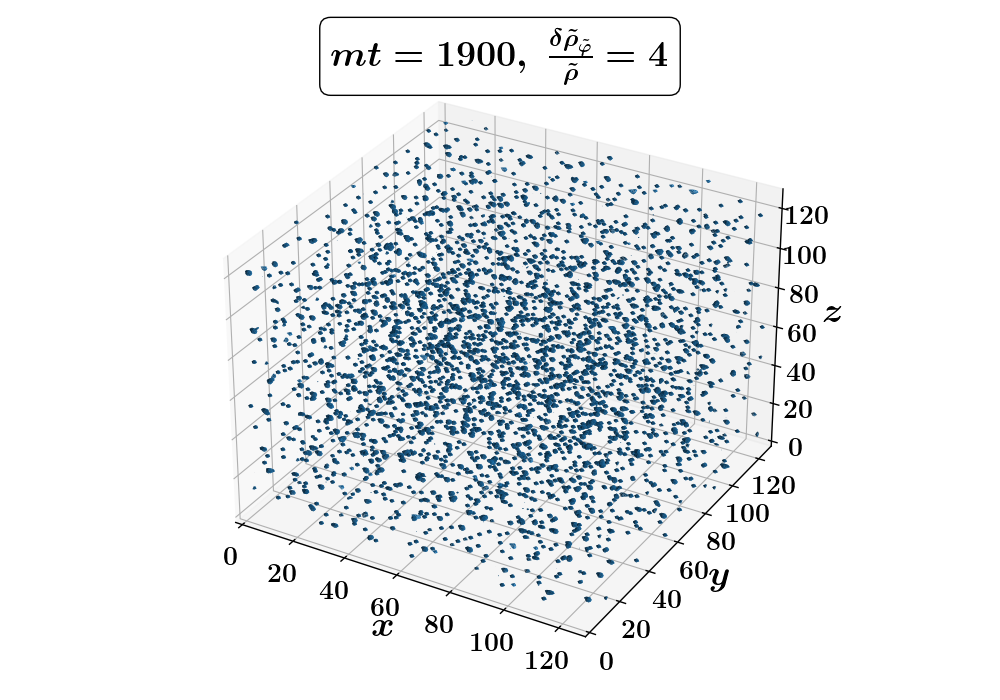}}
    \subfloat{\includegraphics[width = 0.33\textwidth]{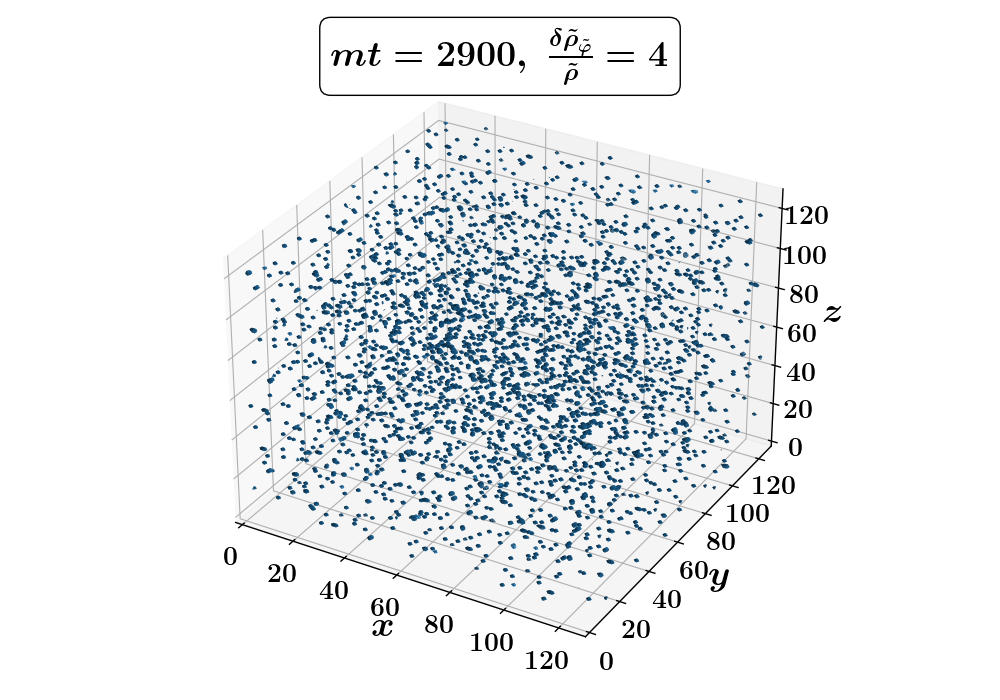}}
    \caption{ Constant density ($3d$) isocontours showing the formation (and  relatively slower decay) of oscillons for $g^2 = 0$ in the T-model with $\lambda_{\text{\tiny T}} = 50\sqrt{2/3}$.}
    \label{fig:oscillon_decay_progression_Tmodel_nocoupling}
\end{figure}

The top panel of Fig.~\ref{fig:field_dynamics_no_coupling} shows that  within $t \lesssim 100\:m^{-1}$ after the end of inflation, a significant amount of inflaton inhomogeneities develop in the  E-model, directly translating to the precipitous increase in the  gradient term observed in the  top left panel of  Fig.~\ref{fig:energy_den_comp_smaller_lambda}. We also observe,  in the bottom panel of  Fig.~\ref{fig:field_dynamics_no_coupling}, that the amplification  of inhomogeneities  in the T-model is very similar, although  slightly quicker, where maximum resonant growth is achieved around $t\sim 50\:m^{-1}$. 
The enhanced fluctuations then strongly backreact onto the condensate, fragmenting it in the process, with the fragmentation occurring  near the peaks of $\delta {\widetilde \varphi}$ and ${\widetilde G}_{\tilde \varphi}$.  This marks the onset of the backreaction phase,  where the oscillations of the inflaton condensate  become incoherent, as can be seen in Fig.~\ref{fig:field_dynamics_no_coupling}. Upon fragmentation, the strong attractive self-coupling results in the formation of dense and localised  (proto-oscillon) lumps. The time evolution of the  lumps is illustrated for the case of no external coupling $g^2=0$ in Fig.~\ref{fig:oscillon_formation_progression_success} in the form of $2d$ spatial slices.  Similarly, Figs.~\ref{fig:oscillon_decay_progression_Emodel_nocoupling} and \ref{fig:oscillon_decay_progression_Tmodel_nocoupling} show the evolution of such lumps as $3d$ constant energy isocontours.  Oscillons form from the fragmented inflaton lumps very quickly, in fact,  within a duration $\Delta t  \sim20\,m^{-1}$ (in between the inflaton fragmentation and oscillon formation) to be specific. Moreover, they are  long-lived --  most oscillons in our simulations  exhibit a lifetime $\tau_{\rm osc} \sim {\cal O}(10^3) \, m^{-1}$ in the absence of any external coupling,  while a few continuing to persist  even beyond $t \sim 15000\,m^{-1}$. Additionally, they contribute significantly to the the energy budget of the system,  as will be discussed in Sec.~\ref{subsec:sim_lifetime_decay}.

\begin{figure}[hbt]
    \centering
    \subfloat{\includegraphics[width = 0.33\textwidth]{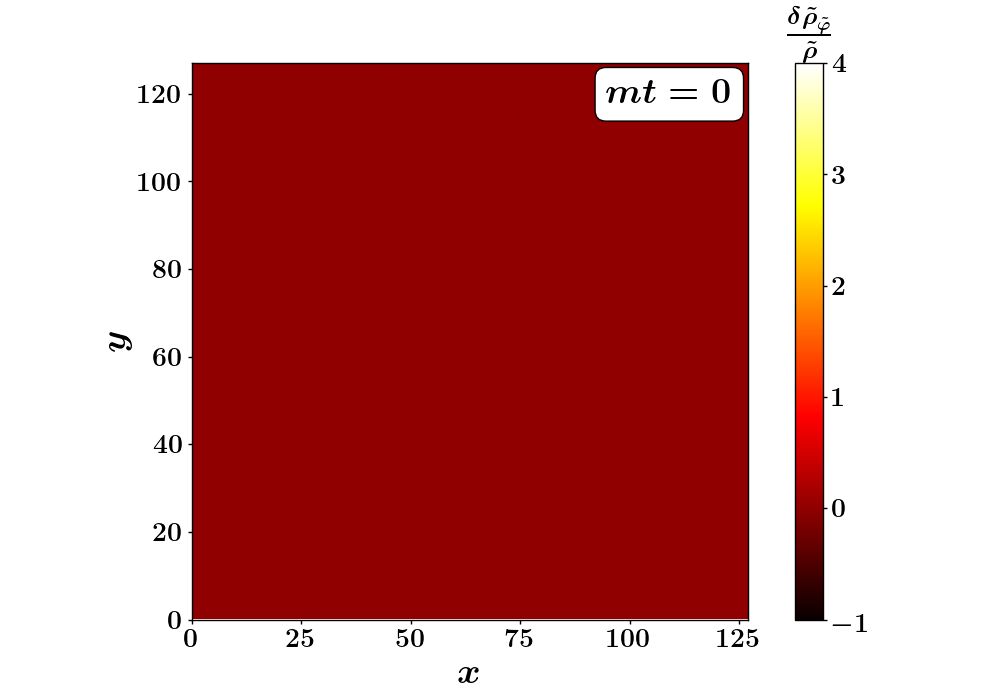}} 
    \subfloat{\includegraphics[width = 0.33\textwidth]{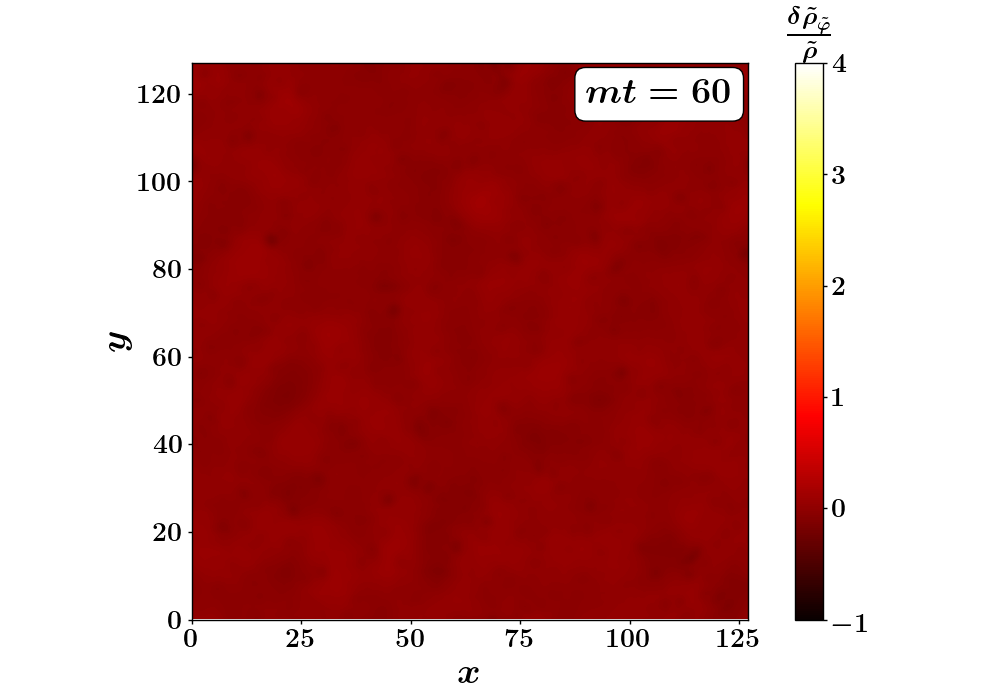}}
    \subfloat{\includegraphics[width = 0.33\textwidth]{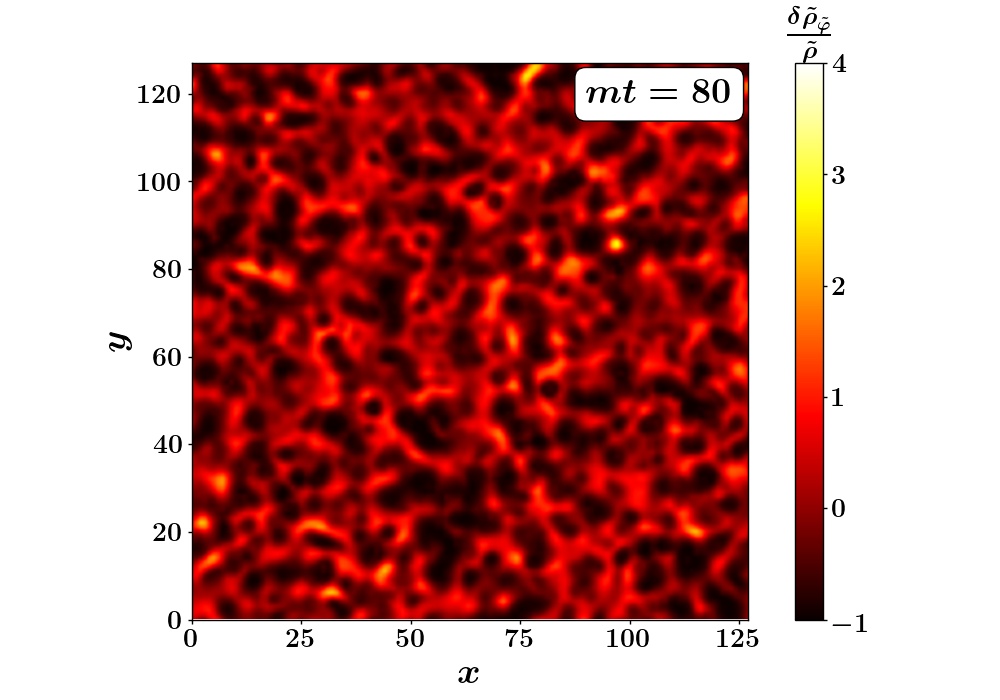}} \\
    \subfloat{\includegraphics[width = 0.33\textwidth]{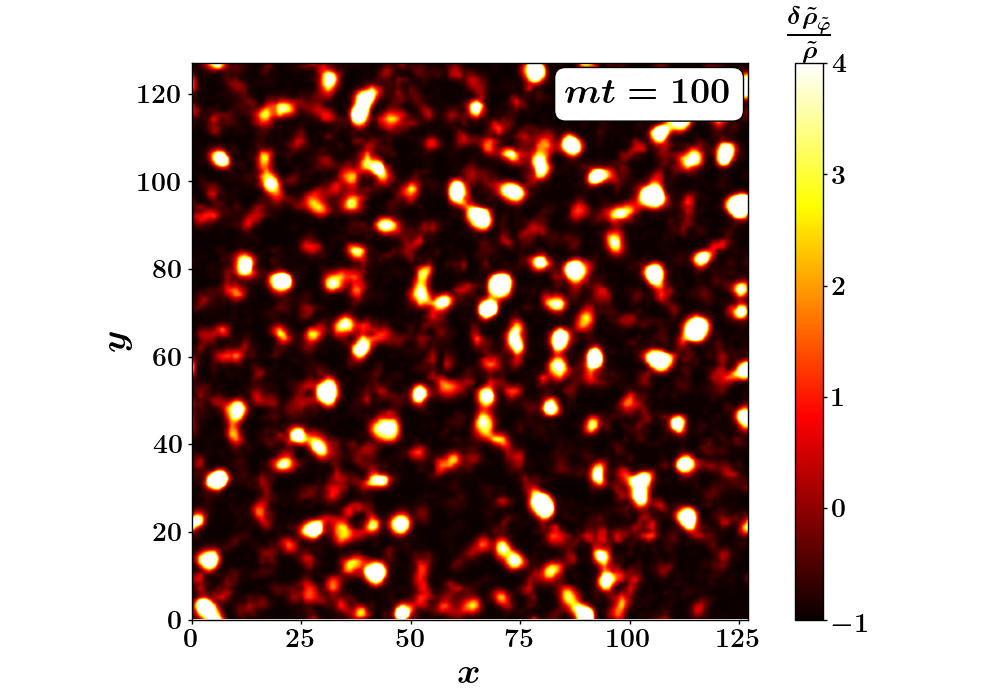}}
    \subfloat{\includegraphics[width = 0.33\textwidth]{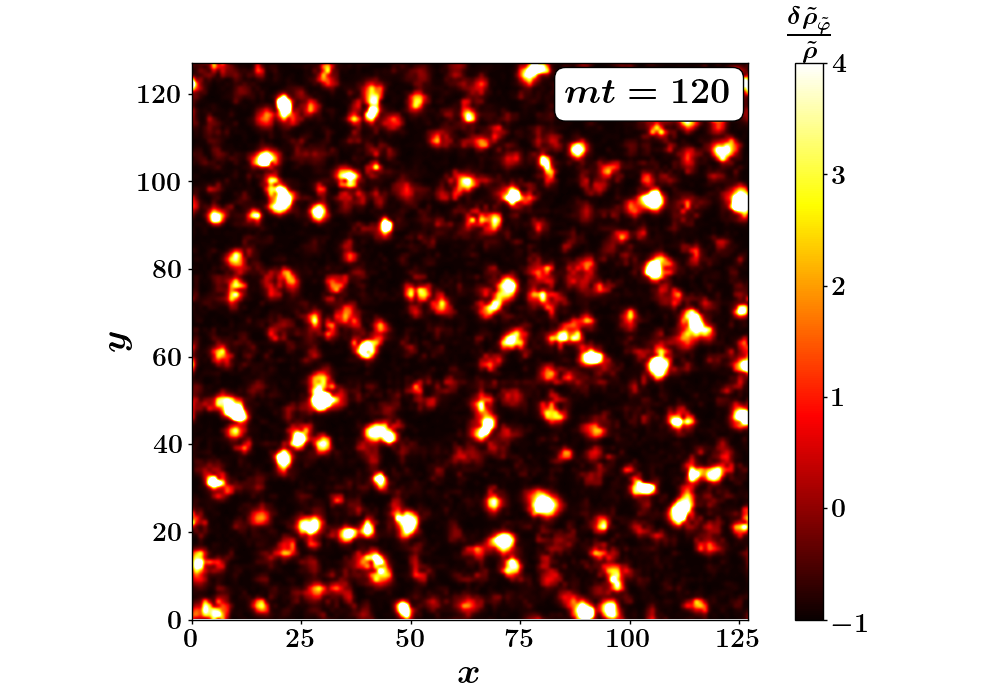}}
    \subfloat{\includegraphics[width = 0.33\textwidth]{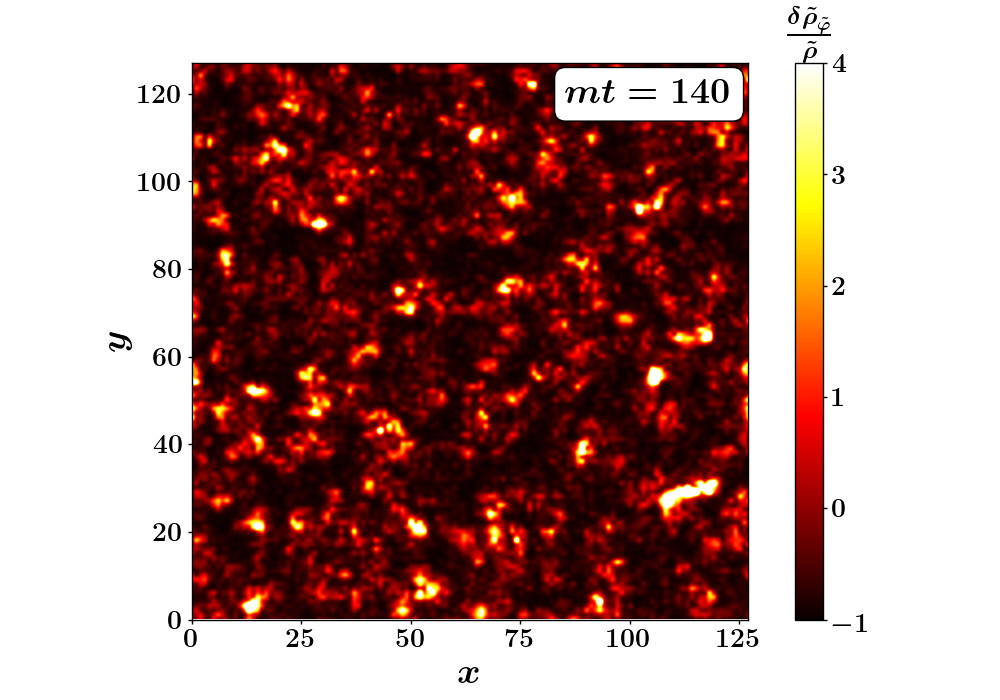}} \\
    \subfloat{\includegraphics[width = 0.33\textwidth]{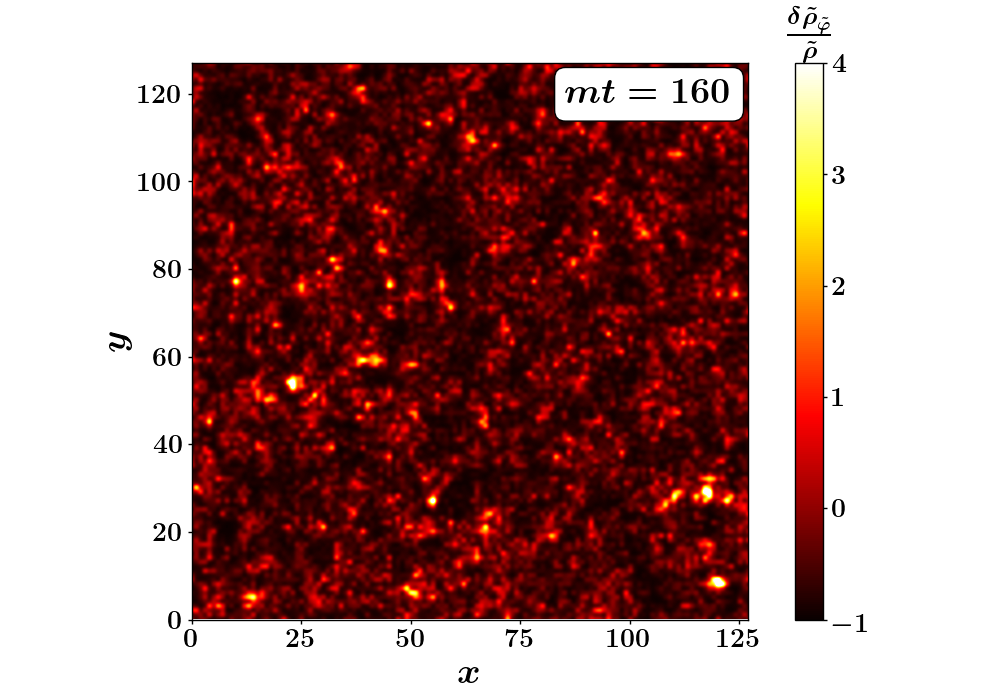}}
    \subfloat{\includegraphics[width = 0.33\textwidth]{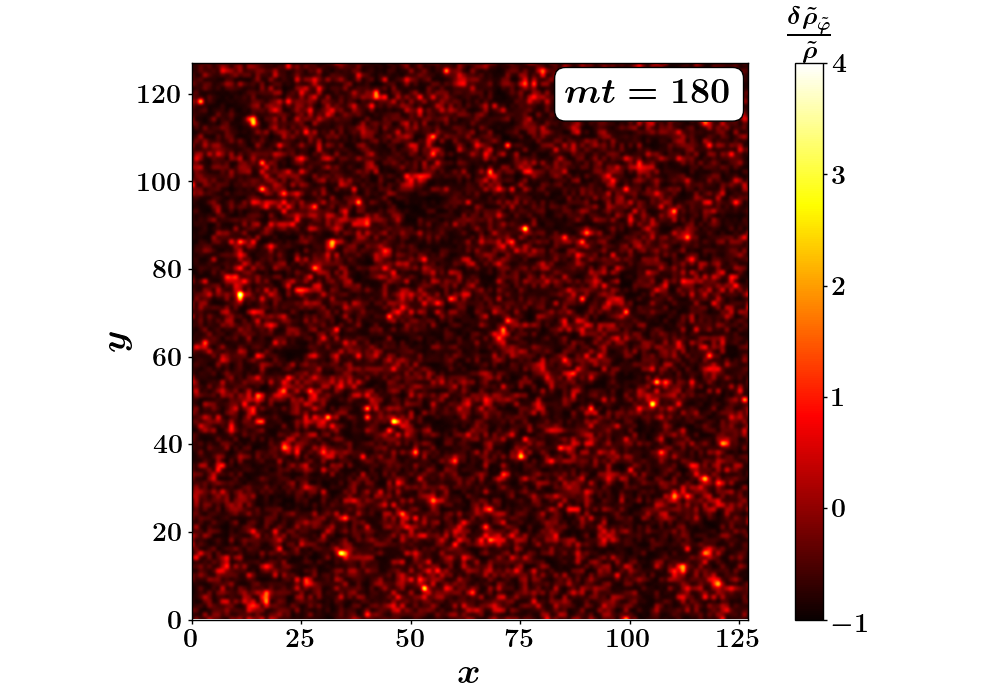}}
    \subfloat{\includegraphics[width = 0.33\textwidth]{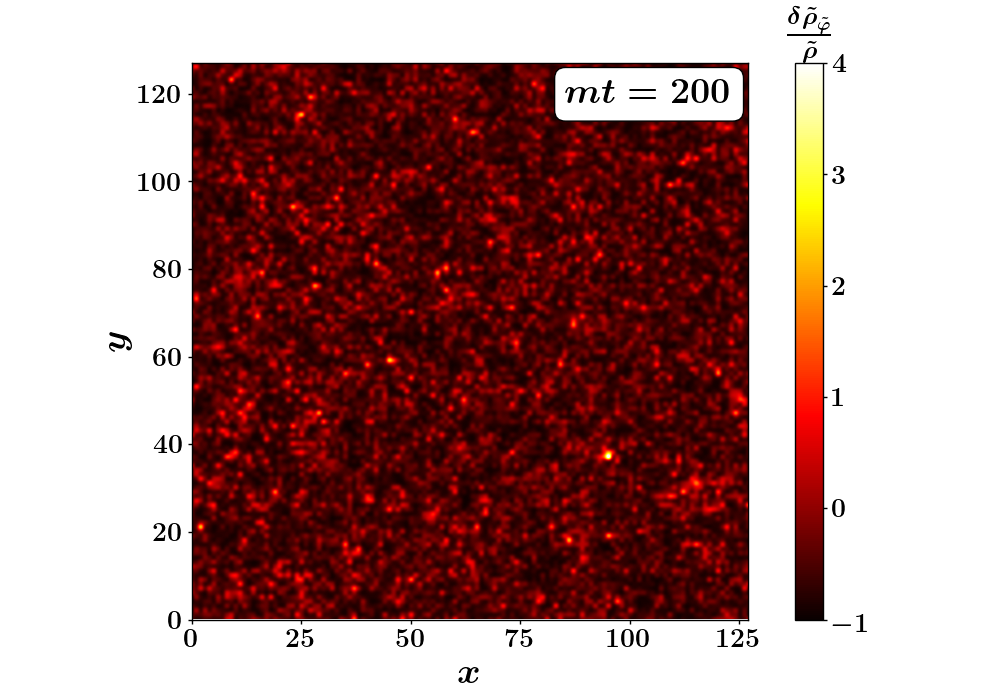}}
    \caption{Evolution of the inhomogeneities of the fragmented inflaton condensate after self-resonance in the  E-model, with self-coupling $\lambda_\text{\tiny E} = 50 \sqrt{2/3}$, and a relatively \textit{large} external coupling $g^2=4.0\times 10^{-5}$ is shown. Note how it now prevents the formation of  robust long-lived oscillons, even after a successful period of self-resonance  producing large inhomogeneities (proto-oscillons) around $t \sim 100\,m^{-1}$. As a result, any transient dense configuration that forms, quickly dissipates into $\chi$ particles, leaving  behind a fragmented inflaton field which dominates the energy density of the system.}
    \label{fig:oscillon_formation_progression_failure}
\end{figure}

On the other end, in the presence of strong external couplings $g^2\gtrsim 4\times 10^{-4}$, we initially observe the formation of  inflaton nonlinearities. However, as can be seen in Fig.~\ref{fig:oscillon_formation_progression_failure}, these  initial nonlinearities are transients, and they  quickly dissipate into $\chi$-particles due to the increased coupling strength. Consequently, the system is composed of a bath of fragmented inflaton inhomogeneities and $\chi$-particles. Nevertheless, for intermediate values of the external coupling, \textit{i.e.} in the range $10^{-6} \lesssim g^2 \lesssim 10^{-4}$, we still observe the formation of nonlinearities  within a duration of $\Delta t\sim 20\:m^{-1}$ which clump together to form robust oscillons. Although, such intermediate couplings appear to reduce the longevity  and energy fraction of oscillons, which will be discussed in Sec.~\ref{subsec:sim_lifetime_decay}. We further note that the values of $g^2$ quoted above are valid for $\lambda_{\text{\tiny E}}, \, \lambda_{\text{\tiny T}} \simeq {\cal O}(10^2) \times \sqrt{2/3}$. For lower values of $\lambda_{\text{\tiny E}}, \, \lambda_{\text{\tiny T}}$ in the parameter space, the corresponding values  $g^2$ for which oscillon formation is disrupted  will be further lowered accordingly. 

\subsection{The lifetime and decay of oscillons}
\label{subsec:sim_lifetime_decay}
Oscillons are quasi-stable nonlinear configurations that can be exceptionally long-lived in the absence of an external coupling,  (\textit{i.e.} $g^2 =0$).  There have been numerous studies, both analytical and numerical,  on the lifetime and decay of oscillons, some of which  can be found in Refs.~\cite{Hertzberg:2010yz,Ibe:2019vyo,Zhang:2020bec,Zhang:2020ntm,Cyncynates:2021rtf}. It was confirmed in Ref.~\cite{Zhang:2020bec} that  the lifetime of an individual oscillon can be extremely long ($10^8 \, m^{-1}$),  although not \textit{infinite}, based on their analysis with symmetric potentials $V(\varphi) = V(-\varphi)$  using a single-frequency oscillon profile. Furthermore, a class of axion-inspired symmetric potentials was constructed in Ref.~\cite{Cyncynates:2021rtf} using  a newly developed \textit{physical quasibreather} technique  that can sustain the so-called \textit{ultra long-lived oscillons} with lifetimes  up to $t \gtrsim 10^{11} \, m^{-1}$ for some specific range of the parameter space of the potentials.  The aforementioned analyses were carried out (semi-) analytically to determine the lifetime of a single oscillon.

In this work,  since we  numerically deal with a population of oscillons, we determine the oscillon lifetime using the time evolution of  the gradient and the kinetic (as well as the potential) energy densities of the inflaton field.  More specifically,  given that oscillons are non-relativistic objects, the  lifetime ($\tau_{\rm osc}$) of a population of oscillons can be estimated by comparing the duration for which $\widetilde{G}_{\tilde{\varphi}} \,\propto \, \widetilde{K}_{\tilde{\varphi}} \, (\simeq \widetilde{V}_{\tilde{\varphi}}) \propto a^{-3}$. The oscillon lifetime in this work can be categorized based on the type of potential used -- symmetric or asymmetric. Using results from the lattice simulations, in particular observing how the gradient energies $\widetilde{G}_{\tilde{\varphi}}$ evolve compared to the energy densities of matter and radiation test fields in an expanding universe, we find that lifetime of oscillons formed in the E-model is relatively longer than their T-model counterparts in the absence of an external coupling. This can be attributed to a stronger self-interaction due to the presence of the cubic term in the Taylor expansion of the E-model potential in Eq.~\eqref{eq:near_harmonic_approx_Emodel}, which is absent in the T-model potential.  We intend to carry out a thorough analytical study of the difference between the oscillon lifetimes for the E- and the T-model potentials in an upcoming paper~\cite{Mishra:2024Part2}.

\begin{figure}[hbt]
    \centering
    \subfloat{\includegraphics[width = 0.33\textwidth]{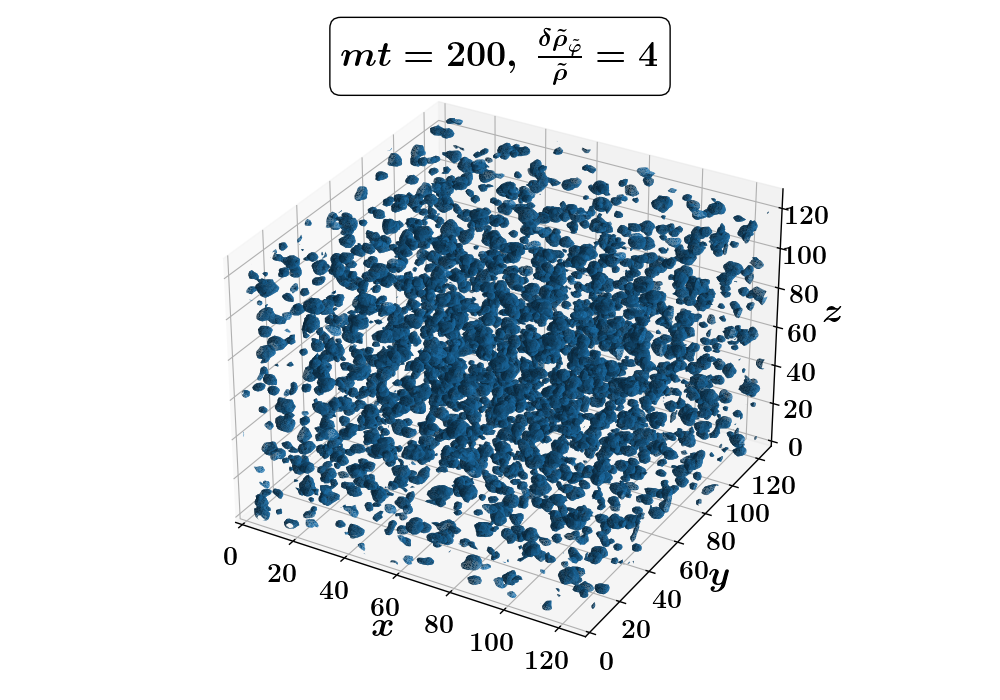}} 
    \subfloat{\includegraphics[width = 0.33\textwidth]{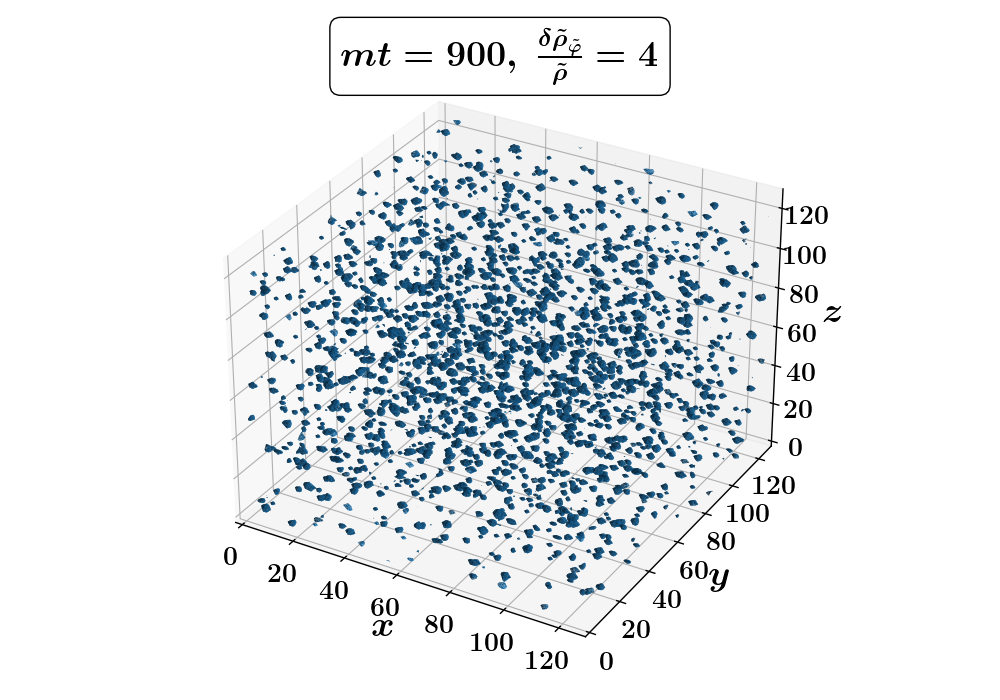}} 
    \subfloat{\includegraphics[width = 0.33\textwidth]{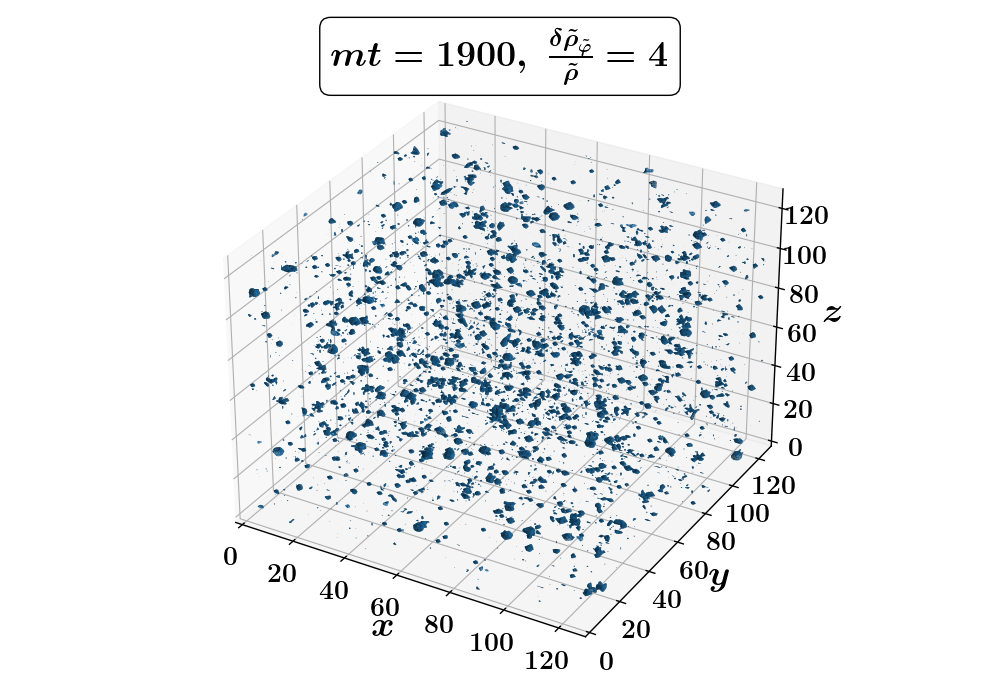}} \\
    \subfloat{\includegraphics[width = 0.33\textwidth]{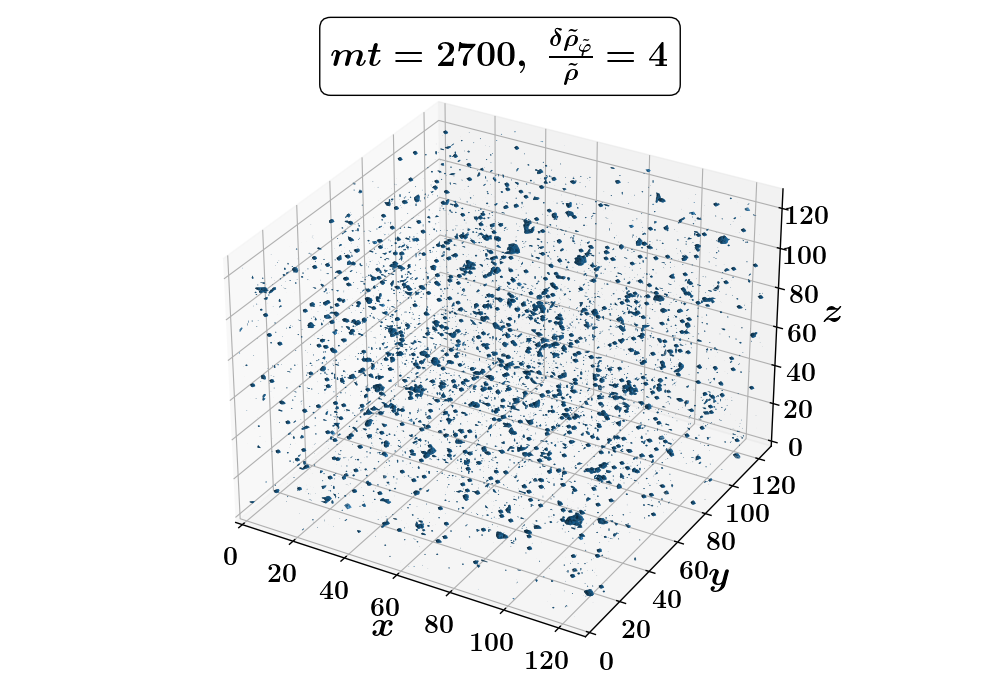}} 
    \subfloat{\includegraphics[width = 0.33\textwidth]{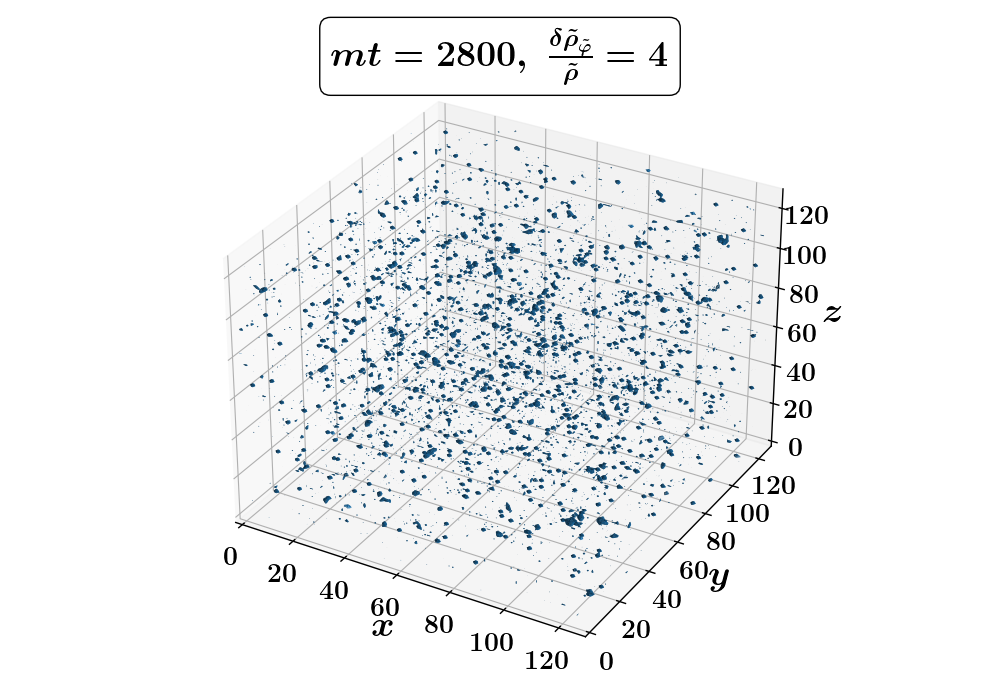}} 
    \subfloat{\includegraphics[width = 0.33\textwidth]{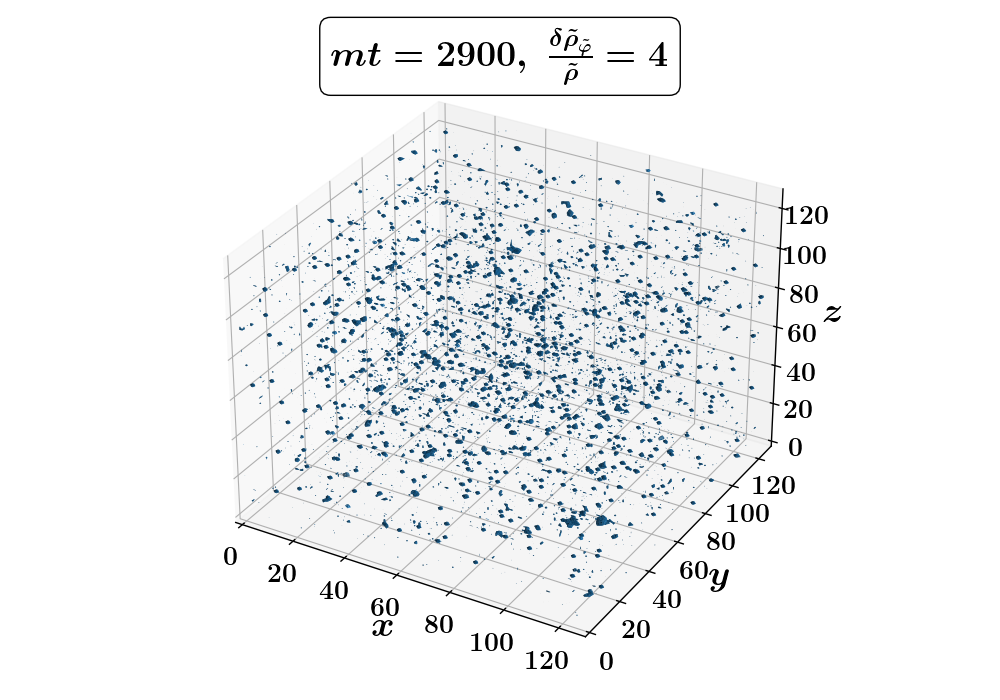}} 
    \caption{Evolution of the constant density ($3d$) isocontours for sufficiently high external coupling (here $g^2 = 1.6 \times 10^{-6}$ for the E-model $\lambda_{\text{\tiny E}} = 50 \sqrt{2/3}$) shows that oscillons decay to the coupled offspring field $\chi$ at a rate greater than that in the absence of the coupling. The rapid decay of oscillons into $\chi$ particles is transient and the system is eventually dominated by the slowly decaying inflaton fragments (including oscillons),  see the middle row of the left panel of Fig.~\ref{fig:energy_den_comp_smaller_lambda}.}
    \label{fig:oscillon_decay_progression_Emodel_coupling}
\end{figure}

\begin{figure}[hbt]
    \centering
    \subfloat{\includegraphics[width = 0.33\textwidth]{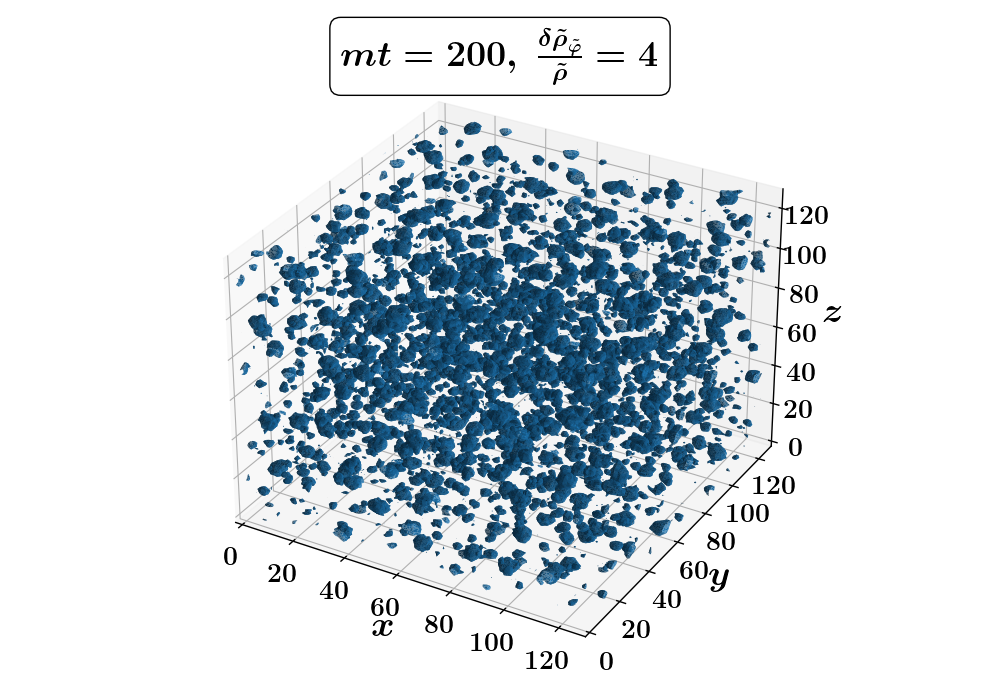}}
    \subfloat{\includegraphics[width = 0.33\textwidth]{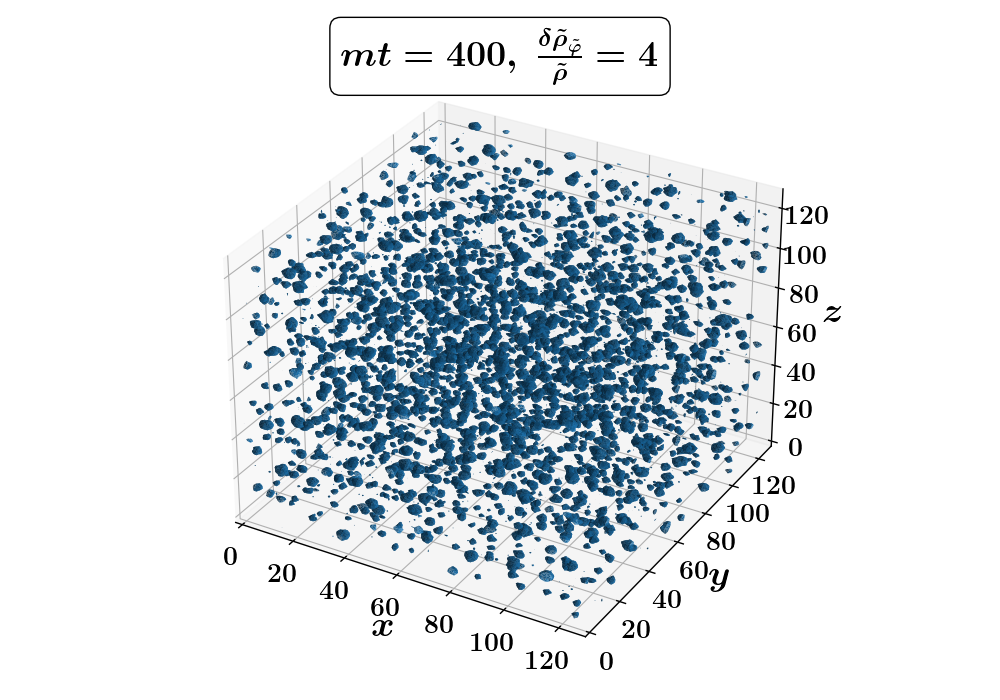}} 
    \subfloat{\includegraphics[width = 0.33\textwidth]{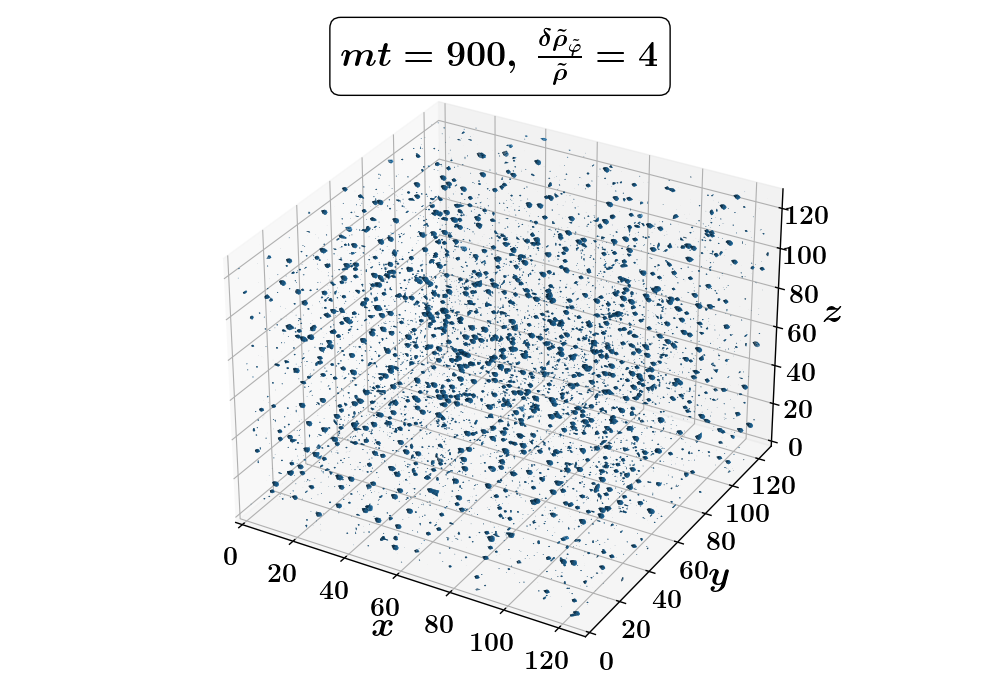}} \\
    \subfloat{\includegraphics[width = 0.33\textwidth]{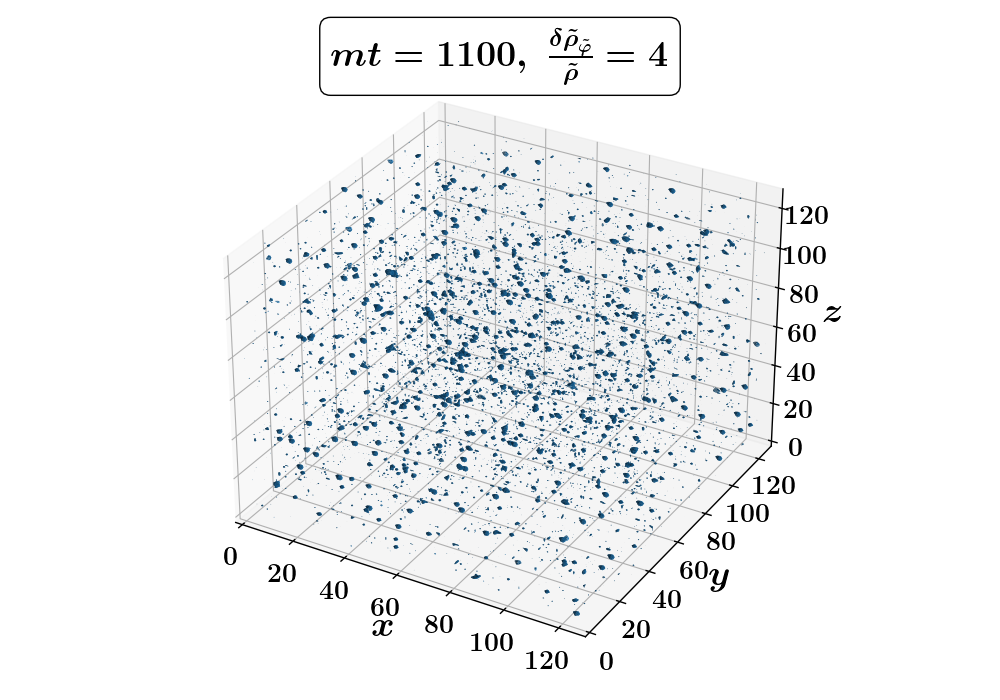}}
    \subfloat{\includegraphics[width = 0.33\textwidth]{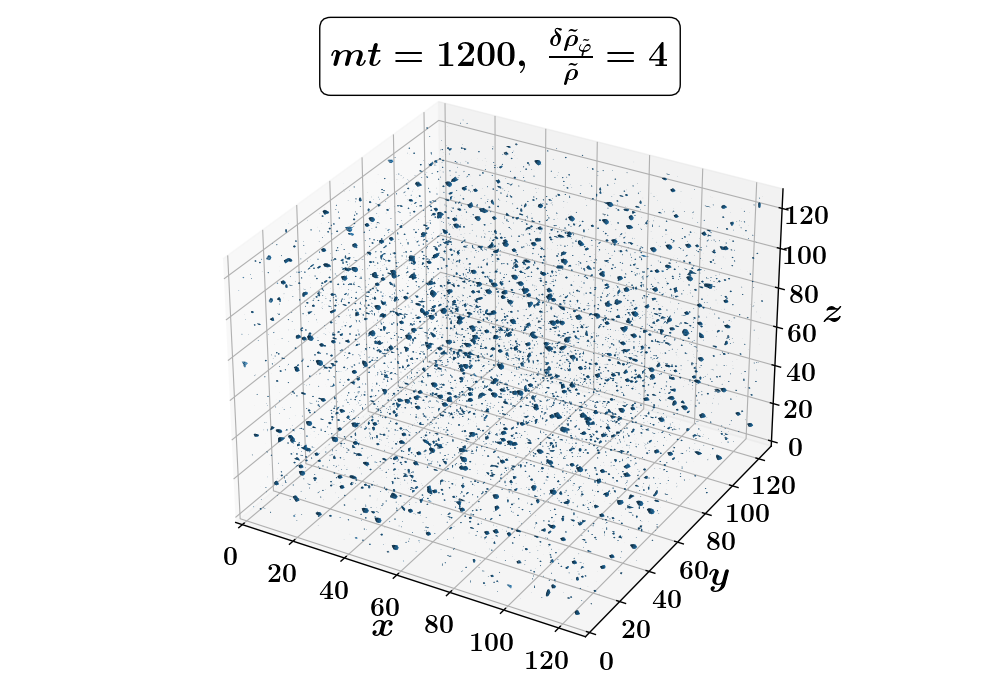}}
    \subfloat{\includegraphics[width = 0.33\textwidth]{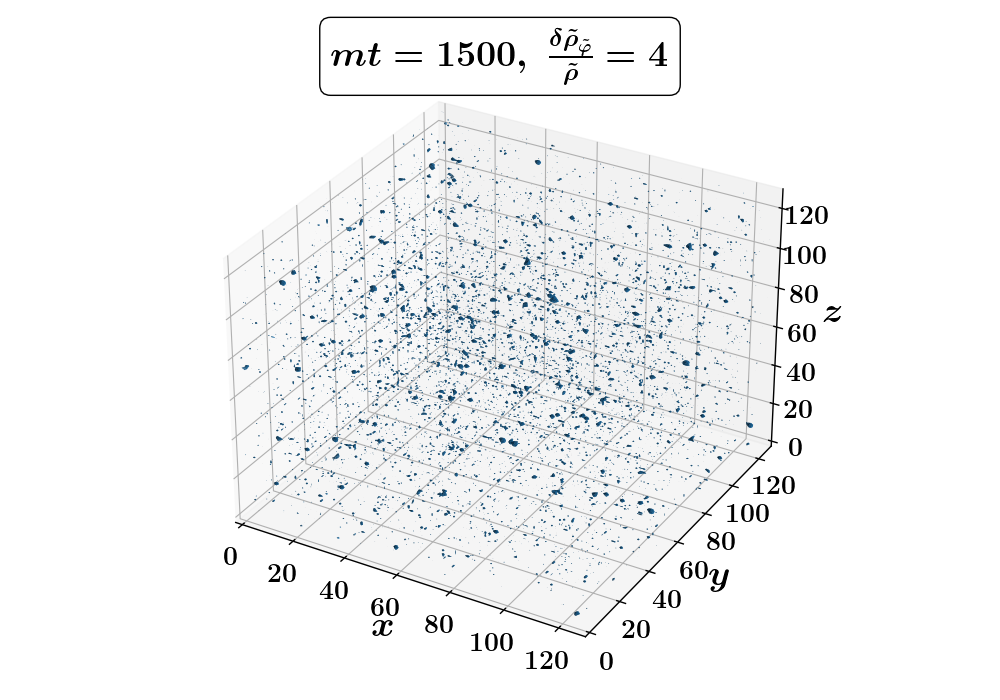}}
    \caption{ Same as Fig.~\ref{fig:oscillon_decay_progression_Emodel_coupling}, but in the presence of a slightly higher external coupling $g^2 = 8.0 \times 10^{-6}$ for the T-model $\lambda_{\text{\tiny T}}= 100\sqrt{2/3}$.}
    \label{fig:oscillon_decay_progression_Tmodel_coupling}
\end{figure}

\begin{figure}[hbt]
    \centering
    \subfloat{\includegraphics[width = 0.48\textwidth]{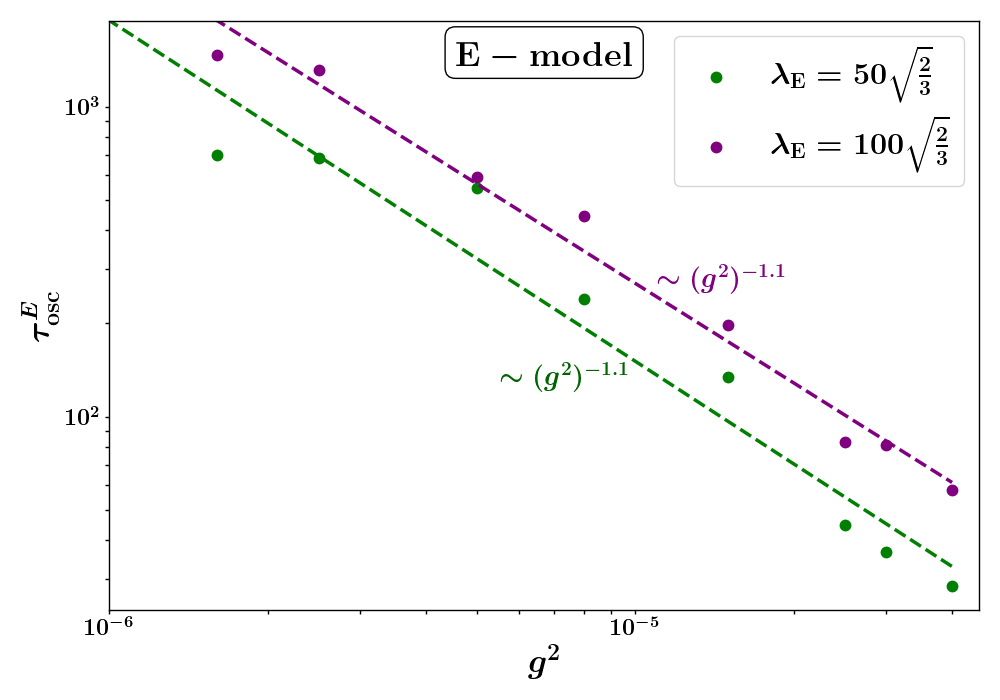}}
    \subfloat{\includegraphics[width = 0.48\textwidth]{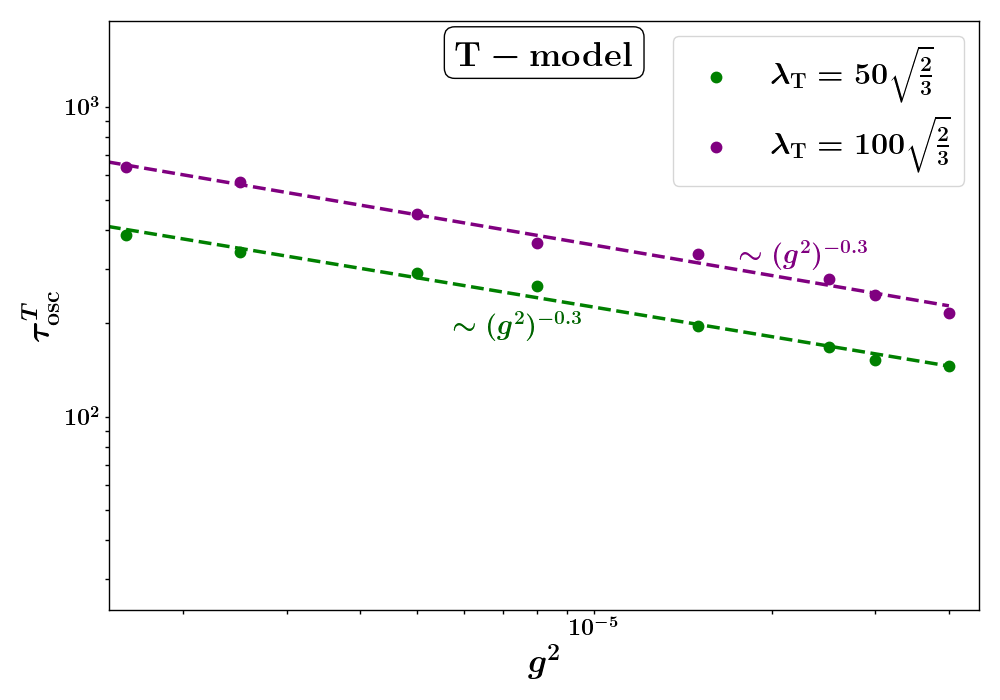}}
    \caption{ The dependence of the  lifetime  of (a population of) oscillons, determined here as the duration during which $\widetilde{G}_{\tilde{\varphi}} \,\propto \, \widetilde{K}_{\tilde{\varphi}} \, (\simeq \widetilde{V}_{\tilde{\varphi}}) \propto a^{-3}$, as a function of $g^2$. We note that the oscillon lifetime decreases following an approximate {\em inverse power-law} dependence on $g^2 \in \left(10^{-6}, \, 10^{-4}\right)$,  which is indicated by the dashed lines (obtained as the best fit curves in the logarithm scale). Furthermore, the (negative) power-law index in the E-model is higher than that in the T-model.}
    \label{fig:osc_life_fit}
\end{figure}

We find that the presence of a non-vanishing external coupling, $g^2\neq 0$, can greatly  reduce the lifetime of oscillons, as shown in Figs.~\ref{fig:oscillon_decay_progression_Emodel_coupling} and \ref{fig:oscillon_decay_progression_Tmodel_coupling}.  In particular, an increase in the value of $g^2$  leads to a decrease in oscillon lifetime for a fixed value of the self-coupling $\lambda_{\text{\tiny E}}, \, \lambda_{\text{\tiny T}}$. However, the exact functional form of this dependence is  difficult to  determine directly from the simulations, and this will be the primary goal of our upcoming paper~\cite{Mishra:2024Part2}. While very small values of $g^2$ do not  have much impact on their lifetime, for intermediate values of $g^2$, say $g^2 \in \left(10^{-6}, 10^{-4}\right)$ the reduction in their lifetimes is quite considerable, as can be seen  in Table.~\ref{tab:oscillon_lifetime}.  In particular we find that the oscillon lifetime decreases following an approximate {\em inverse power-law} dependence on $g^2$, as shown in Fig.~\ref{fig:osc_life_fit}. We also find that the (negative) power-law index is higher for E-model oscillons, with $\tau_{\rm osc}^{\text{\tiny{E}}} \propto \l( g^2 \r)^{-1.1}$, compared to T-model oscillons with $\tau_{\rm osc}^{\text{\tiny{T}}} \propto \l( g^2 \r)^{-0.3}$. 

The reason why the external coupling affects the oscillon lifetime can be understood in terms of decay channels. In the absence of an interaction term $(g^2=0)$, oscillons are only expected to decay via the emission of scalar radiation  which takes place at an exceptionally slow rate due to the  stabilising non-linear structure of the oscillons~\cite{Zhang:2020bec,Zhang:2020ntm,Cyncynates:2021rtf}. However, with $g^2\neq 0$, there is an additional channel~\cite{Hertzberg:2010yz} through which oscillons can decay, namely oscillons decaying into $\chi$-particles, via $\varphi\:\varphi\to\chi\:\chi$ processes, along with scalar radiation. In particular,  Figs.~\ref{fig:oscillon_decay_progression_Emodel_nocoupling} - \ref{fig:oscillon_decay_progression_Tmodel_coupling} show the progression of oscillon decay for different external couplings  by plotting the evolution of the $3d$ constant overdensity isocontours defined as
    \begin{equation}
        \frac{\delta\widetilde{\rho}_{\tilde{\varphi}}}{\widetilde{\rho}_{\tilde{\varphi}}}\left(\tilde{t},\widetilde{\bm{x}}\right) = \frac{\widetilde{\rho}_{\tilde{\varphi}}\left(\tilde{t},\widetilde{\bm{x}}\right) - \langle \widetilde{\rho}_{\tilde{\varphi}} \rangle_{_ \text{\tiny V}}\left(\tilde{t}\right)}{\langle \widetilde{\rho}_{\tilde{\varphi}} \rangle_{_\text{\tiny V}}\left(\tilde{t}\right)} \, ,
    \end{equation}
where the $\langle ... \rangle_{_\text{\tiny V}}$ refers to volume-averaging. The axes in these plots are measured in comoving length dimensions. Although the choice of the value of the overdensity isocontours is rather arbitrary, we follow the standard choice in the literature, namely $\delta\rho/\rho\sim\mathcal{O}(1)$. In Figs.~\ref{fig:oscillon_decay_progression_Emodel_nocoupling} and \ref{fig:oscillon_decay_progression_Tmodel_nocoupling}, we see  the formation and evolution of these inhomogeneities for the E and T-models with $g^2=0$ that persist over a long time-scale. Since resonance and backreaction occur slightly earlier in the T-model, early signs of inflaton fragmentation can be observed in Fig.~\ref{fig:oscillon_decay_progression_Tmodel_nocoupling}. Furthermore, signs of oscillon longevity in the absence of external coupling  can be  noticed more clearly in the top panel of Fig.~\ref{fig:energy_den_comp_smaller_lambda}. 

\begin{figure}[hbt]
    \centering
    \subfloat{\includegraphics[width = 0.75\textwidth]{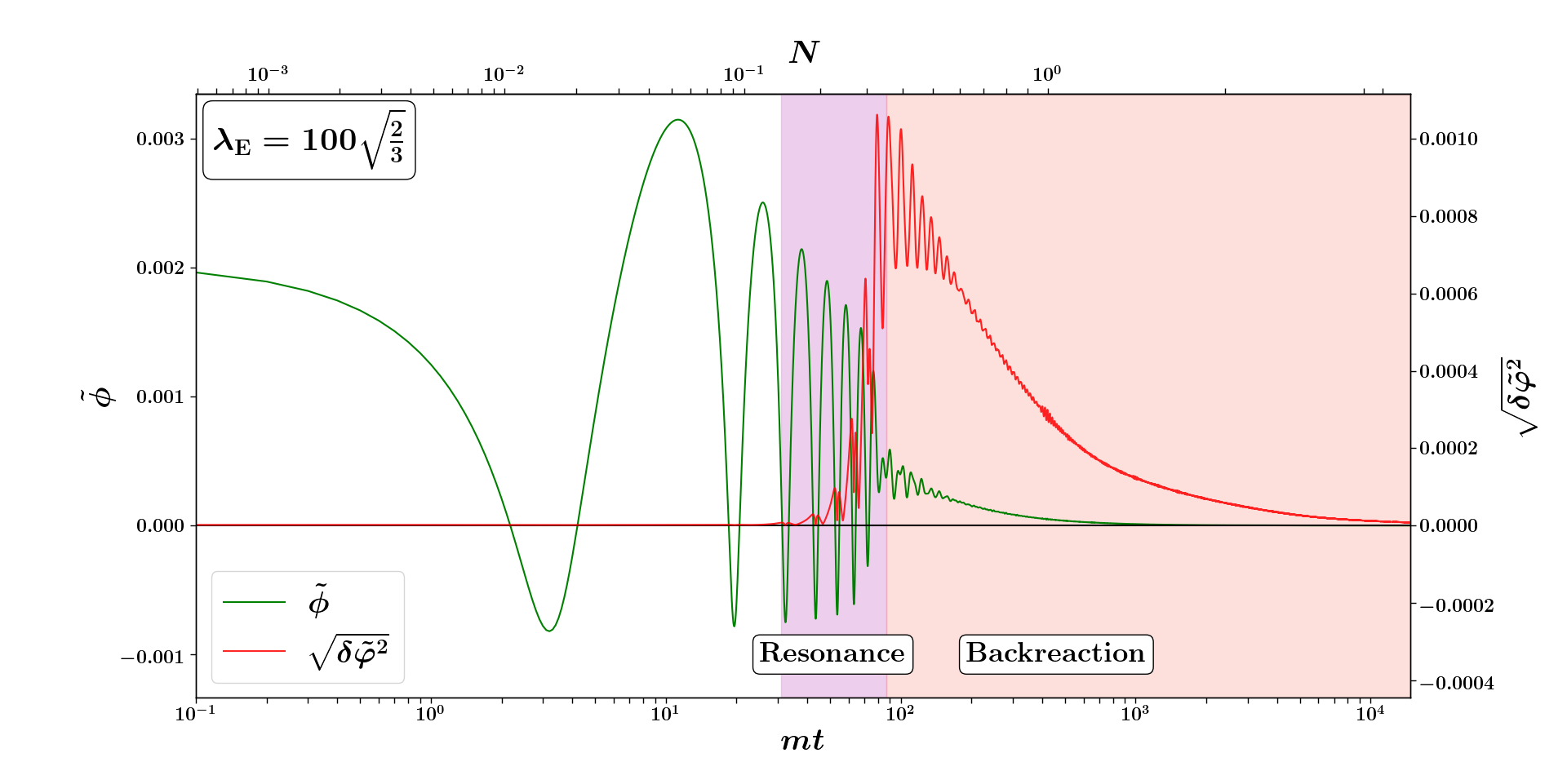}} \\
    \subfloat{\includegraphics[width = 0.75\textwidth]{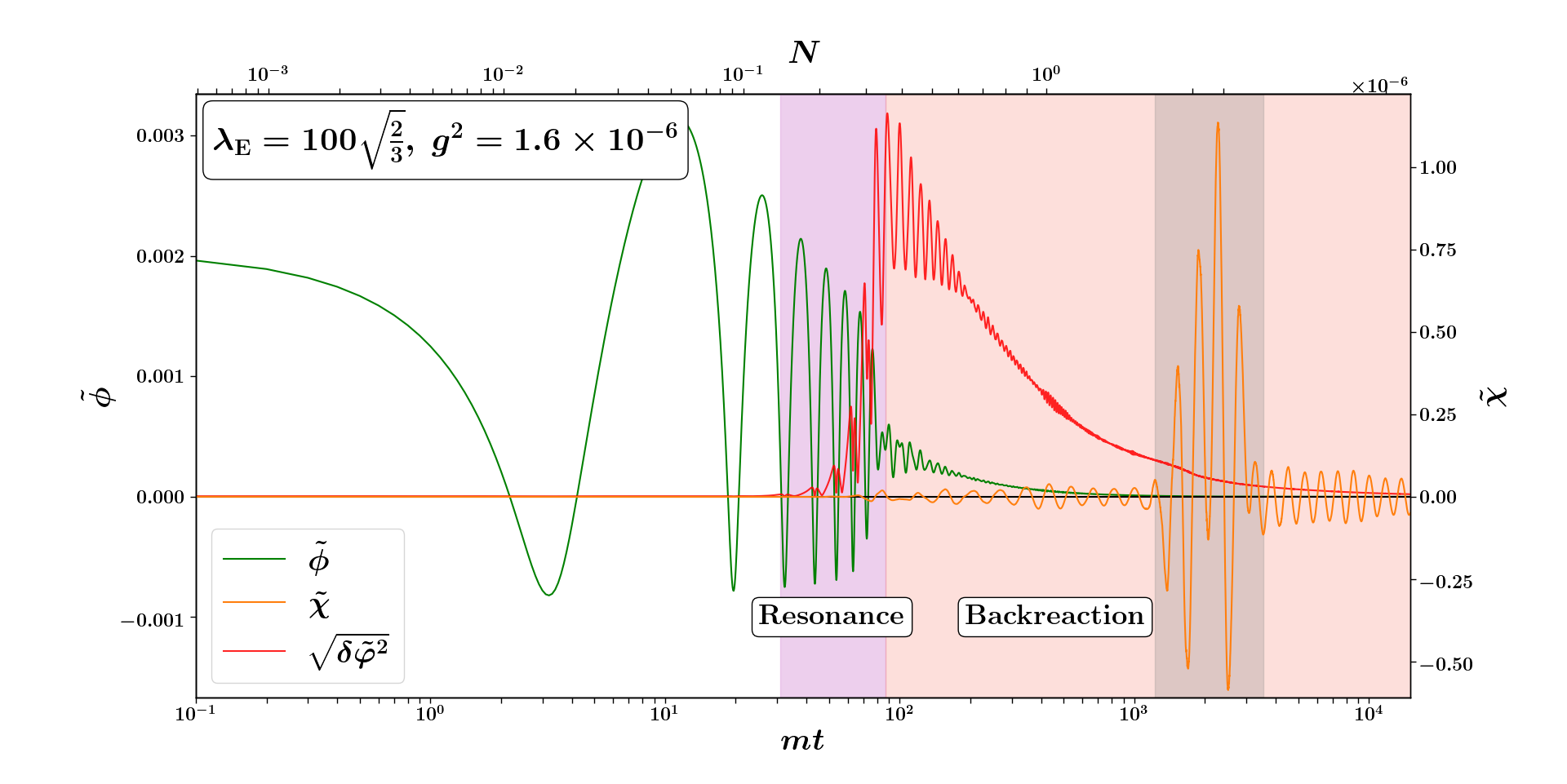}}
    \caption{ Evolution of the volume-averaged inflaton configurations for the E-model potential with  $\lambda_{\text{\tiny E}} = 100 \sqrt{2/3}$ with external couplings $g^2 = 0$ (\textbf{top}) and $g^2 = 1.6 \times 10^{-6}$ (\textbf{bottom}) are shown.  We notice that strong self-resonance leads to the formation of  large inflaton inhomogeneities (oscillons)  by $t \sim 100 \, m^{-1}$, which eventually decay into the $\chi$ fluctuations  starting from $t \sim  10^3 \, m^{-1}$, initially rapidly  (shown in light grey shade), and later in a slow and steady rate. Therefore, the presence of such an external coupling  significantly impacts the lifetime and fractional energy  of oscillons.}
    \label{fig:reh_osc_decay_phi_chi_delphi}
\end{figure}

By comparing the time evolution  of $\widetilde{G}_{\tilde{\varphi}}$ with those of the test matter and radiation fields, we see that the gradient energy scales as $\sim a^{-3}$ for time-scales of $\mathcal{O}(10^3)\:m^{-1}$. The production of $\chi$-particles  from oscillons  when $g^2 \neq 0$ can be explicitly seen to occur at late times in Fig.~\ref{fig:reh_osc_decay_phi_chi_delphi} when the broad band  external resonance of the inflaton condensate is absent, (\textit{i.e.}, long after the condensate fragments into oscillons). The effects of nonvanishing $g^2$ are  also shown in Figs.~\ref{fig:oscillon_decay_progression_Emodel_coupling} and \ref{fig:oscillon_decay_progression_Tmodel_coupling}, where the nonlinear oscillons are observed to dissipate away on much shorter time-scales than in the case of $g^2=0$. This is also evident from the evolution  of $\widetilde{G}_{\tilde{\varphi}}$ in Fig.~\ref{fig:energy_den_comp_smaller_lambda}.  Oscillon decay is also illustrated  in the plot for the  $2d$ slices of inflaton fluctuations in Fig.~\ref{fig:oscillon_formation_progression_failure}. The aforementioned  observations illustrate the strong dependence of the strength of the external coupling which is  summarised in Table~\ref{tab:oscillon_lifetime} and  Fig.~\ref{fig:osc_life_fit}, as discussed  above.

The energy density fraction of oscillons, $f_\text{osc}$, is defined to be the ratio of the total energy contained within nonlinear configurations, above a suitably chosen threshold, over the total energy. Mathematically, this is given by
    \begin{equation}\label{eq:f_osc}
        f_\text{osc}\equiv\frac{\widetilde{E}_\text{osc}}{\widetilde{E}_\text{tot}}=\frac{\int_{\delta\tilde{\rho}>\tilde{\rho}_\text{th}}\d^3 \widetilde{x}\:\widetilde{\rho}\left( \tilde{t},\widetilde{\bm{x}}\right)}{\int_\text{vol}\d^3 \widetilde{x}\:\widetilde{\rho}\left( \tilde{t},\widetilde{\bm{x}} \right)}
    \end{equation}
We perform the integral in Eq.~\eqref{eq:f_osc} for a threshold $\widetilde{\rho}_\text{th}$ four times the volume-averaged energy density. We emphasize that there is no widely accepted choice for $\widetilde{\rho}_\text{th}$,  although it is usually taken to be in between $\widetilde{\rho}_\text{th} \in \l[ 2 \, \widetilde{\rho}, \, 5 \, \widetilde{\rho}\r]$. For example, Refs.~\cite{Amin:2011hj,Mahbub:2023faw} used $\widetilde{\rho}_\text{th}=2 \, \widetilde{\rho}$ in their calculation of the oscillon fraction. The choice of this threshold will, of course, affect the corresponding value of $f_\text{osc}$. In this work, we fix the threshold to be $\widetilde{\rho}_\text{th}=4 \, \widetilde{\rho}$. 
\begin{table}[htb]
    \centering
    \caption{Lifetime of oscillons in the presence of $\f{1}{2}g^2 \varphi^2 \chi^2$ interaction.}
    \label{tab:oscillon_lifetime}
    \begin{tabular}{c|c|c c}
    \hline
    \multirow{2}{*}{$\bm{\lambda_\text{\tiny E}}, \bm{\lambda_\text{\tiny T}}$}  & \multirow{2}{*}{$\bm{g^2}$} & \multicolumn{2}{c}{Oscillon Lifetime ($\bm{m^{-1}}$)} \\
                                         &                        & ${\bm{\tau_{\rm osc}^\text{\tiny E}}}$                 & ${\bm{\tau_{\rm osc}^\text{\tiny T}}}$                \\
    \hline
    \multirow{9}{*}{$50 \sqrt{\frac{2}{3}}$} & 0                      & 833.8                        & 479.3                       \\
                                         & $1.6 \times 10^{-6}$   & 697.3                        & 386.5                       \\
                                         & $2.5 \times 10^{-6}$   & 681.1                        & 340.0                       \\
                                         & $5.0 \times 10^{-6}$   & 544.7                        & 290.2                       \\
                                         & $8.0 \times 10^{-6}$   & 238.7                        & 264.2                       \\
                                         & $1.5 \times 10^{-5}$   & 134.4                        & 196.5                       \\
                                         & $2.5 \times 10^{-5}$   & 44.5                        & 167.9                       \\
                                         & $3.0 \times 10^{-5}$   & 36.5                        & 152.4                       \\
                                         & $4.0 \times 10^{-5}$   & 28.3                        & 145.2                       \\
    \hline
    \multirow{9}{*}{$100 \sqrt{\frac{2}{3}}$} & 0                      & 1797.7                        & 962.7                       \\
                                         & $1.6 \times 10^{-6}$   & 1464.0                        & 637.6                       \\
                                         & $2.5 \times 10^{-6}$   & 1308.2                        & 569.0                       \\
                                         & $5.0 \times 10^{-6}$   & 591.9                        & 450.4                       \\
                                         & $8.0 \times 10^{-6}$   & 444.0                        & 363.0                       \\
                                         & $1.5 \times 10^{-5}$   & 196.9                        & 335.0                       \\
                                         & $2.5 \times 10^{-5}$   & 84.5                        & 277.5                       \\
                                         & $3.0 \times 10^{-5}$   & 81.1                        & 246.5                       \\
                                         & $4.0 \times 10^{-5}$   & 58.1                        & 216.5                       \\
    \hline
    \end{tabular}
\end{table}

\begin{figure}[hbt]
    \centering
    \subfloat{\includegraphics[width = 0.48\textwidth]{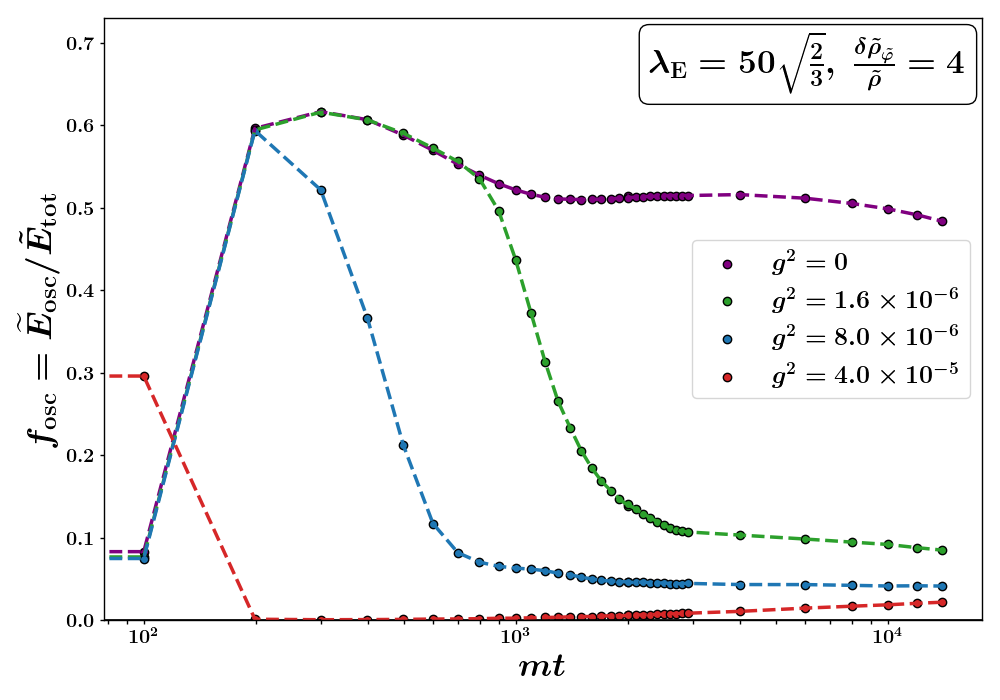}} 
    \subfloat{\includegraphics[width = 0.48\textwidth]{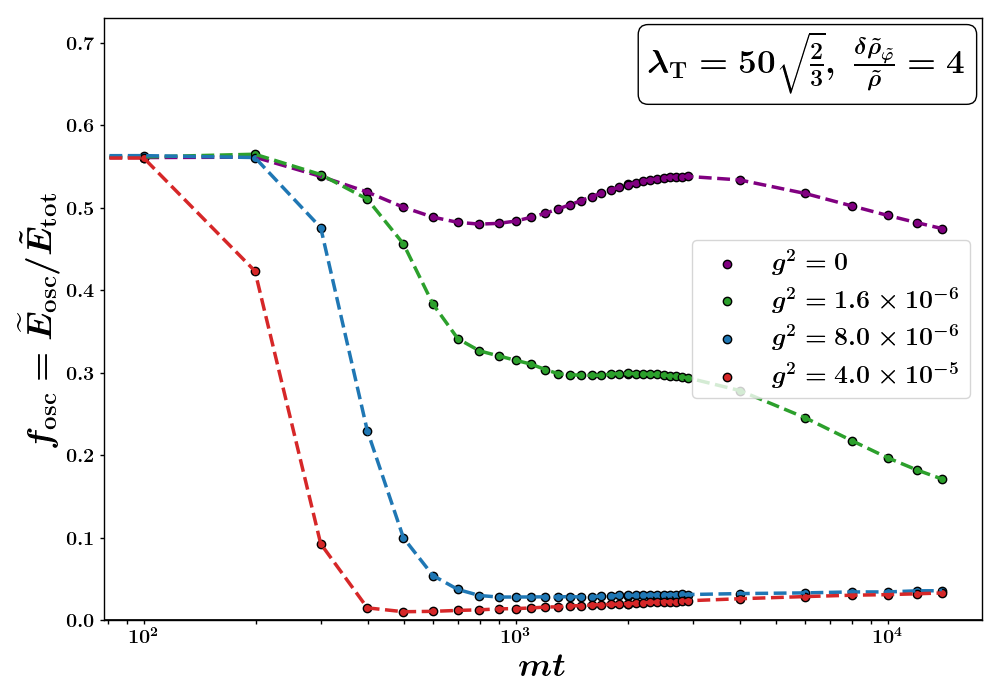}} \\
    \subfloat{\includegraphics[width = 0.48\textwidth]{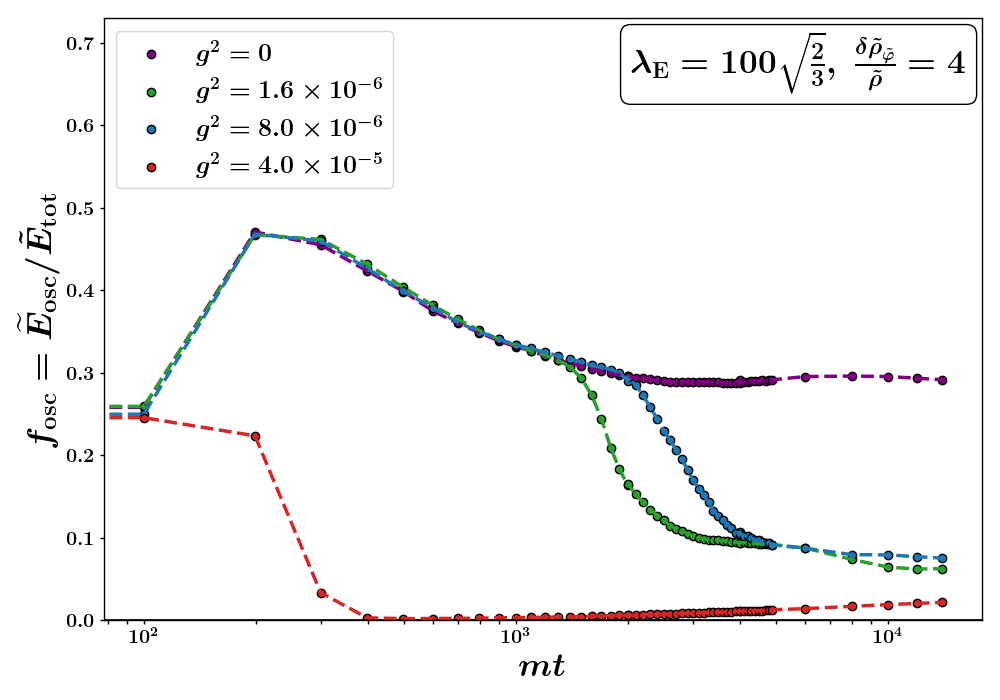}}
    \subfloat{\includegraphics[width = 0.48\textwidth]{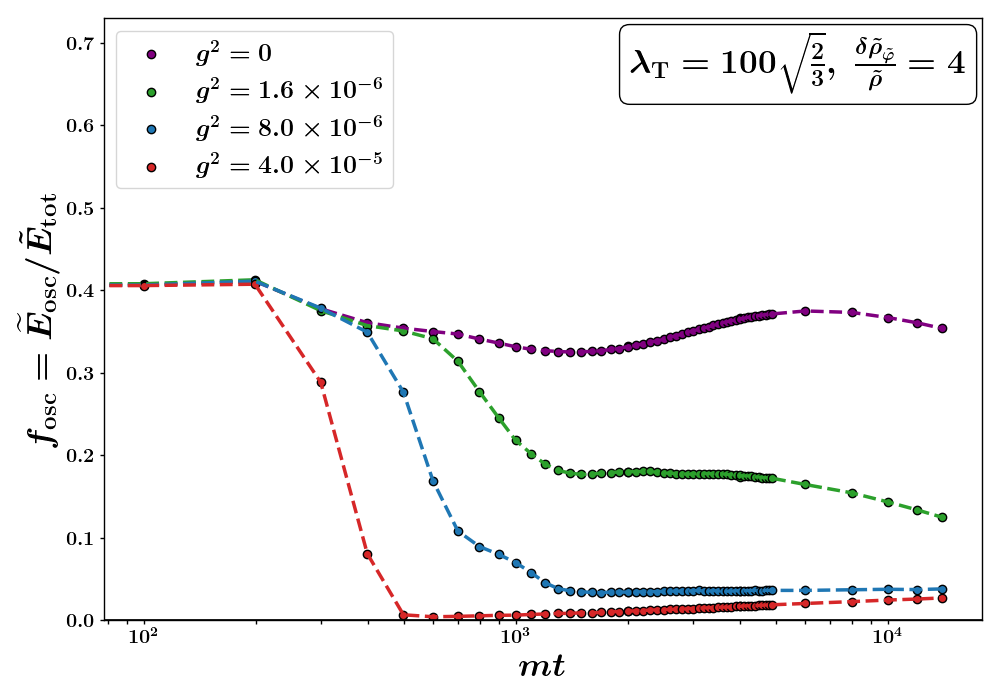}}
    \caption{The fractional energy contained in oscillons $f_\text{osc}$ for different external couplings  associated with the  E-model (\textbf{left}) and T-model (\textbf{right})  is shown for $\lambda = 50\sqrt{2/3}$ ({\bf top}) and $\lambda = 100\sqrt{2/3}$ ({\bf bottom}).  For $g^2 = 0$ (purple curves), after reaching its peak value following the onset of backreaction, the fraction $f_\text{osc}$ for each plot then decreases to a near-constant value at late times. The presence of the external coupling leads to a rapid decay  where the late-time (near-constant) asymptote of $f_\text{osc}$ tends to be  much lower than that in the case of $g^2=0$.} 
    \label{fig:osc_frac}
\end{figure} 

The oscillon energy fraction is plotted in Fig.~\ref{fig:osc_frac} which demonstrates that $f_\text{osc}$ rapidly increases at early times  during the period of strong self-resonance, reaching a maximum around the onset of the backreaction phase marked by $t \sim 100 \, m^{-1}$. This is expected since the gradient term $\widetilde{G}_{\tilde{\varphi}}$ also experiences its maximal amplification during this period, as can be seen from Fig.~\ref{fig:energy_den_comp_smaller_lambda}. In the absence of $g^2$,   the purple curves show that $f_\text{osc}$ can become  as high as $60 \%$ at the onset of the backreaction phase,  which later asymptotes towards a slightly smaller value of $\approx 50\%$. Furthermore,   for $g^2 =0$ we find that the oscillon energy fraction $f_\text{osc}$ saturates to comparatively lower values for larger self-coupling 
 $\lambda_{\text{\tiny E}}, \, \lambda_{\text{\tiny T}}$. Such an observation was first made for the E-model potential in Ref.~\cite{Mahbub:2023faw}.\footnote{In Ref.~\cite{Mahbub:2023faw}, the strength of the self-interaction is controlled by $\alpha$, where $ \lambda_\text{\tiny E}=\sqrt{\frac{2}{3\alpha}} \, . $} 

Noticeably, we observe that  for the E-model, an external coupling strength of $g^2=4.0\times 10^{-5}$ is sufficient enough to prevent abundant growth of inflaton  inhomogeneities in the first place. This can be seen from the red curves in  the left column of Fig.~\ref{fig:osc_frac}, where for $g^2=4.0\times 10^{-5}$,  the peak values of $f_\text{osc} \lesssim 0.3$ ($\lesssim 30 \%$ of the total energy budget), even at the onset of backreaction at $t \sim 100 \, m^{-1}$. This is in stark contrast to the other three curves, corresponding to smaller values of $g^2$ in the top panels of Fig.~\ref{fig:osc_frac}.

Nevertheless, of greater importance is the asymptotic value of $f_\text{osc}$ at late times. In the presence of $g^2 \in (10^{-6}, \, 10^{-4})$, even though $f_\text{osc}$  initially reaches  peak values similar to that of $g^2 = 0$, the presence of the additional decay channel results in reduced asymptotic values of the oscillon fraction, \textit{i.e.} $f_\text{osc} \lesssim 0.3$ at late times. In particular, for $g^2 \gtrsim {\cal O}(10^{-5})$, the final fraction $f_\text{osc}$ can be less than $10\%$ of the total energy budget.  The reduction in the value of $f_\text{osc}$ in between $t \sim ({\rm few}) \,  100 \, m^{-1}$ and  $t \sim ({\rm few}) \, 1000 \, m^{-1}$ corresponds to the rapid decay of oscillons into $\chi$-particles, which is consistent with (the grey shaded region) in Fig.~\ref{fig:reh_osc_decay_phi_chi_delphi}. The remaining oscillons (inflaton lumps) then continue to decay into both $\chi$-particles and scalar radiation at a much slower rate, hence the asymptotic values of $f_\text{osc}$ exhibit a slow-decay trend  towards $t \sim 10^4 \, m^{-1}$, as can be seen in Fig.~\ref{fig:osc_frac}.

Note that for large enough external coupling,  corresponding to $g^2 \gtrsim 4 \times 10^{-5}$, the red curves show that $f_{\rm osc}$ exhibits a slowly increasing (instead of decreasing) trend. This is due to the contribution from the interaction term ${\widetilde {\cal I}}(\tilde{\varphi},\tilde{\chi})$ in Eq.~(\ref{eq:rescaled_en_pressure_phi}), which is usually negligible as compared to that of the gradient term ${\widetilde G}_{\tilde{\varphi}}$ for $g^2 < {\cal O} \l(10^{-5}\r)$. However, the interaction term becomes significant (if not dominant) for $g^2 > {\cal O} \l(10^{-5}\r)$, as can be seen from the brown curves in the bottom panel of Fig.~\ref{fig:energy_den_comp_smaller_lambda}. Hence, for large values of $g^2$, the late time asymptote of $f_{\rm osc}$ is not completely dictated by oscillons.

\section{Discussion and conclusions}
\label{sec:discussion}
Oscillons are  nonrelativistic quasi-solitonic field  configurations, where attractive non-linear self-interactions  enable them  to exhibit lifetimes that are far longer than  any natural time scale appearing in the  Lagrangian. Since  they are often  investigated during the initial stages  of reheating following the end of  inflation with a  shallow potential, a quantitative understanding of the dynamics of their formation and evolution in a realistic post-inflationary scenario is crucial. The majority of papers in the literature have carried out the study of the formation, evolution, and decay of oscillons in the post-inflationary epoch for an isolated self-interacting scalar field, in the absence of any external interactions\footnote{See Ref.~\cite{Hertzberg:2010yz} for a study of the quantum decay of an oscillon \textit{via} an external coupling.} (with the exception of minimal gravitational coupling). However, if such objects were to form in the early universe  \textit{via} fragmentation of the inflaton condensate, they can potentially store  a significant fraction of the inflaton energy density that was supposed to  decay into other particles in order to reheat the universe. This raises a number of important questions,  such as, \textit{(i)} do oscillons form in the presence of (non-negligible) external couplings\,?, \textit{(ii)} if yes, then how does the strength of the external coupling affect the lifetime of oscillons\,?, and \textit{(iii)} how does the  efficiency of preheating get modified in such a decaying  inflaton-oscillon scenario\,? It is therefore of great interest, both  from the phenomenological standpoint of understanding the role of oscillons in reheating, and  in establishing in general how couplings to other fields affect their lifetime, to carry out a dedicated quantitative study in this direction. 
 
In this paper, we have taken the first steps towards such an analysis by studying oscillon formation and decay during preheating in two classes of  asymptotically flat inflaton potentials where the inflaton ($\varphi$) is coupled to a single massless scalar  ($\chi$)  via an interaction ${\cal I} = \f{1}{2}\,g^2\varphi^2\chi^2$. We started out with the  (semi-)analytical study  of preheating in the linear regime \textit{via} parametric resonance in order to establish the parameter space for which an exponential growth of inflaton fluctuations is observed at the earliest stages of preheating which is a necessary condition for oscillon formation. In order to firmly establish the formation of oscillons, we then moved on to solve the fully non-linear coupled field  equations numerically, using the \cosmolattice~framework, which enabled us to carry out a comprehensive study of the formation and  decay of oscillons, along with estimating their  lifetime and fractional energy density.

Our results are, perhaps, best illustrated by Fig.~\ref{fig:par_space_lambda_g^2} where we show the interplay  between the self-coupling $(\lambda)$ and the external coupling $(g^2)$ in determining the dynamics of preheating. In particular, we have identified and demarcated five  distinct regions of interest  in the parameter space $\lbrace \lambda, \, g^2 \rbrace$ of the couplings. For both E- and T-models, consistency with inflationary predictions for primordial GWs rule out  a region of the parameter space corresponding to low values of the self-interaction coupling ($\lambda$). This includes cases where $\lambda<\mathcal{O}(10^{-1})$ for both potentials. On the other hand,  extremely large self-couplings of the order $\lambda \gtrsim 4 \times \mathcal{O}(10^2)$ correspond to situations where the mass term of the inflaton becomes negligible compared to the higher-order self-interaction terms.   In such a situation, preheating leads to the formation of fragmented transients which do not clump together as oscillons, since oscillons (by nature) are non-relativistic configurations with vanishing EoS parameter. In fact, if $\lambda$ is large enough, then the oscillating inflaton field spends most of its time in the plateau region of the potential, leading to an EoS $\langle w_\phi \rangle < -1/3$, in which case the universe continues to accelerate (on the average) for a few post-inflationary oscillations~\cite{Iacconi:2023mnw}.  Such a scenario is termed as \textit{oscillating inflation}~\cite{Damour:1997cb,Liddle:1998pz}. 
 
More importantly, we identify the regions in $\{ \lambda, \, g^2 \}$ parameter space which give rise to inflaton fragmentation,  with and without the formation of oscillons, since  fragmentation does not always guarantee the formation of oscillons. In particular, we find that oscillon formation occurs in  both  E- and T-models  above a certain threshold value of $\lambda\sim\mathcal{O}(10)$,  which agrees well with the previous studies in Refs.~\cite{Kim:2021ipz,Lozanov:2017hjm,Mahbub:2023faw} for $g^2 =0$. In the presence of external coupling, we observe the formation of robust oscillons for $g^2 \lesssim 10^{-4}$. Of equal significance is the fact that the lifetime of these oscillons is  strongly dependent on  the magnitude of $g^2$, as summarised in Table~\ref{tab:oscillon_lifetime}. Oscillon lifetimes for both E- and T-model potentials can then be suitably fitted in the interval of interest $g^2\in\left( 10^{-6},10^{-4} \right)$ and are shown in Fig.~\ref{fig:osc_life_fit}, where it can be seen that $\ln g^2$ and $\ln\tau_\text{osc}$ are approximately linear. Stated differently, $g^2$ and $\tau_\text{osc}$ satisfy an approximate power-law  relation with a negative index,  \textit{i.e.} an  {\em inverse power-law} relation. We found that for both values of self couplings $\lambda_\text{\tiny E} = \lambda_\text{\tiny T}=50\sqrt{2/3}$ and $100\sqrt{2/3}$, the oscillon lifetimes fall as 
 \begin{equation}
     \tau_\text{osc}^\text{\tiny E} \propto \left( g^2 \right)^{-1.1} \, ; \quad\quad \tau_\text{osc}^\text{\tiny T}\propto \left( g^2 \right)^{-0.3} \, .
 \end{equation}
 
The aforementioned  observations raise the obvious question of why the lifetimes are different between the two models with the E-model producing longer living oscillons while also displaying more efficient decay with the increase in $g^2$. In the absence of the external coupling, we find that $\tau^\text{\tiny E}_\text{osc}\approx 2 \tau^\text{\tiny T}_\text{osc}$ in units of $m^{-1}$, which is  a noticeable  difference. Similarly, one may wonder whether or not an approximate functional form exists relating $\tau_\text{osc}$ and $g^2$ (and whether or not it stays a power-law of the form $\tau_\text{osc}\sim (g^2)^{-\gamma}$) for other inflationary models. Admittedly,  it is difficult to explain these nuances in this work  (based on lattice simulations),   which rather warrants a  dedicated  separate study. We wish to address them in our upcoming paper~\cite{Mishra:2024Part2}, which will be devoted to an analytical study of oscillon decay in the presence of such an external coupling.
 
Regardless of these subtleties, we  have demonstrated that the post-inflationary universe can be populated with these oscillons  and that, in the correct circumstances, they can constitute a significant fraction of the energy density, which will be of cosmological importance if they are  sufficiently long-lived. Even if they are short-lived, the prospects of studying reheating through oscillon decay appears to be fascinating.  Since the primary goal of this work is to study the formation and decay of oscillons,  we  did not carry out a study of the end stages of reheating marked by $\langle w \rangle\rightarrow1/3$, which necessarily requires the inclusion of additional couplings, and involves  numerical codes simulating the  Boltzmann equations~\cite{Emond:2018ybc,Garcia:2023dyf}. (Although, we did carry out a thorough analysis of the evolution of the EoS during preheating via inflaton and oscillon decay, which can be found in App.~\ref{subsec:EOS_oscillons}.)  
 
Nevertheless, it remains to be seen how different external couplings (Yukawa and trilinear couplings) and (or) the presence of several massless scalar fields modify the preheating dynamics.  Various stages of (p)reheating, along with the associated complex high energy dynamics, are of great phenomenological interest to cosmologists, since they carry crucial prospects for the detection of a stochastic gravitational wave (GW) background. In fact, both oscillon formation and decay are expected to seed second-order GWs because of the  large source terms  arising from non-vanishing anisotropic stresses. It is well-known that oscillon formation (or even inflaton fragmentation in general~\cite{Garcia:2023eol,Garcia:2024zir}) is accompanied by high-frequency (GHz-scale) stochastic GWs~\cite{Lozanov:2019ylm,Hiramatsu:2020obh}.  Even though such GWs are beyond the sensitivities of the current and proposed detectors, potentially significant progress in the development of resonant cavity detectors~\cite{Kanno:2023whr,Aggarwal:2020olq,Berlin:2021txa,Ito:2022rxn} is expected in the upcoming decade, which will enable them  to  achieve enough sensitivity to detect the high-frequency cosmological GWs~\cite{Herman:2022fau}. Of greater interest is the fact that rapid decay of oscillons carries its own GW signatures that can be well within the sensitivities of current and proposed detectors~\cite{Lozanov:2022yoy}.

Before concluding, it is worth mentioning that, as with most cosmological lattice studies,  our analysis has been carried out without incorporating the effects of  metric perturbations that are important for the evolution of long wavelength fluctuations at relatively late times. As a result, we do not quantify the influence that gravitational interactions have on the preheating dynamics and the formation of oscillons (and importantly, on oscillon lifetimes~\cite{Zhang:2020ntm}). Metric perturbations, if accounted for,  usually present themselves  in the form of the gauge-invariant Bardeen potential $\Phi_k$ in the equation of motion of the inflaton fluctuations, and  can induce additional  Floquet instability in the system, including in potentials that otherwise would not exhibit resonant growth of fluctuations. For example, the inflaton fluctuations at the bottom of a purely quadratic potential can be written in a way that resembles the Mathieu equation in the presence of metric perturbations, giving rise to the \textit{metric preheating}~\cite{Jedamzik:2010dq,Easther:2010mr,Martin:2020fgl} scenario. In fact, working in the framework of numerical relativity\footnote{\textsf{GRChombo}~\cite{Andrade:2021rbd} is an example of an open-source numerical relativity code that can be used to study scalar field dynamics in the early universe.} affords one greater control of computing   the amplification of density contrast,  along with the usual metric perturbations, which is ideal in refining the selection criteria for  nonlinear compact structures such as the oscillons and oscillatons~\cite{Kou:2019bbc,Aurrekoetxea:2023jwd}; even allowing us to study the formation of primordial black holes (PBHs)~\cite{Nazari:2020fmk,Cotner:2018vug,del-Corral:2023apl}  through the detection of apparent horizon formation~\cite{Nazari:2020fmk}.

\section{Acknowledgments}
MS was supported by the  INSPIRE scholarship of the Department of Science and Technology (DST), Govt.~of India during his Master's thesis work during which a significant portion of this work was carried out. SSM and EJC are supported by  STFC Consolidated Grant [ST/T000732/1]. EJC is also supported by a Leverhulme Research Fellowship [RF- 2021 312]. SSM thanks IUCAA, Pune for their hospitality. We thank Paul Saffin for useful discussions. Numerical simulations were carried out on the  Padmanabha HPC cluster at IISER TVM.  For the purpose of open access, the authors have applied a CC BY public copyright license to any Author Accepted Manuscript version arising. \\

{\bf Data Availability Statement:} This work is entirely theoretical and has no associated data. The data files for the lattice simulations (with the exception of the $3d$ configuration files) and other codes can be found in the following GitHub repository: \href{https://github.com/RM503/Oscillons}{\faGithubSquare} .

\appendix 
\section{Inflationary predictions of the E- and T-model potentials}
\label{app:CMB_T_E_models}
The parameters derived in Tables~\ref{tab:Sim_Parameters_emodel}~and~\ref{tab:Sim_Parameters_tmodel} are fixed by taking into account inflationary constraints arising from the CMB.  In this appendix, we explain how the CMB constraints can be used to set the parameters for the lattice simulations analytically. Derivation of the expressions for  various inflationary quantities used below can be found in Refs.~\cite{Baumann_TASI,Mishra:2024axb}.  The E- and T-model potentials   are given by
\begin{align*}
    V(\phi)\big\lvert_{\text{\tiny E-model}}&=V_0\left( 1-e^{-\lambda_{\text{\tiny E}}\frac{\phi}{m_p}} \right)^2 \, , \nonumber \\
    V(\phi)\big\lvert_{\text{\tiny T-model}}&=V_0\tanh^2\left( \lambda_{\text{\tiny T}}\frac{\phi}{m_p} \right) \, .
\end{align*}
The initial field value $\phi_\text{in}$ is conventionally chosen to be the value of the inflaton field at which the inflationary phase terminates $\phi_\text{end}$. We can analytically determine this using the first potential slow-roll parameter\footnote{The first potential slow-roll parameter is approximately equal to $\epsilon_H$ defined in Eq.~\eqref{eq:epsilon_H} during slow-roll.}
\begin{equation}\label{eq:epsilon_V}
    \epsilon_V=\frac{m_p^2}{2}\left( \frac{V_{,\phi}}{V} \right)^2 \, ,
\end{equation}
where the end of  inflation corresponds to $\epsilon_V(\phi_\text{end})=1$. This condition can then be used to solve for $\phi_\text{end}$. Using Eq.~\eqref{eq:epsilon_V} for the E- and T-model potentials, we arrive at the following expressions
    \begin{align}
        \phi_\text{end}=\begin{dcases}
            \frac{m_p}{\lambda_{\text{\tiny E}}}\ln\left( \sqrt{2}\lambda_{\text{\tiny E}} + 1 \right) \, ; & \text{E-model} \, , \\
            \frac{m_p}{2\lambda_{\text{\tiny T}}}\text{arccosech}\left( 2\sqrt{2}\lambda_{\text{\tiny T}} \right) \, ; & \text{T-model} \, .
        \end{dcases}
    \end{align}
The inflaton  mass  $m$ defined in Sec.~\ref{sec:simulations} can be determined using the CMB normalisation of the curvature power spectrum at the pivot scale, which we denote as $\mathcal{P}_{\zeta\star}$. In the slow-roll regime, this can be calculated using~\cite{Baumann_TASI,Mishra:2024axb}
\begin{equation}
    \mathcal{P}_{\zeta\star}\simeq \f{1}{12\pi^2 } \, \frac{V^3(\phi_\star)}{V_{,\phi}^2(\phi_\star) m_p^6} \, ,
\end{equation}
where the `$\star$' subscripts represent evaluations at the pivot scale $k_\star=0.05\:\text{Mpc}^{-1}$. In order to determine the field value $\phi_\star$ at the Hubble exit of the pivot scale,  we use  the relation between the elapsed number of $e$-folds and the inflaton field  excursion (following directly from Eq.~\eqref{eq:efolds})
    \begin{equation}
        N\left( \phi_\star \right)\equiv\Delta N=\frac{1}{m_p^2}\int_{\phi_\text{end}}^{\phi_\star}\d\phi \, \frac{V}{V_{,\phi}} \, ,
    \end{equation}
the number of $e$-folds can be related to $\phi_\star$. After the integral is carried out, the resulting expression can be inverted and $\phi_\star$ can be calculated for the required elapsed $e$-folds $\Delta N$. For the E-model potential, we obtain
    \begin{equation}\label{eq:phi_star_E}
        \lambda_{\text{\tiny E}}\frac{\phi_\star}{m_p}=-2\lambda_{\text{\tiny E}}^2\Delta N+\lambda_{\text{\tiny E}}\frac{\phi_\text{end}}{m_p}-e^{\lambda_{\text{\tiny E}}\frac{\phi_\text{end}}{m_p}}-W_{-1}\left[ -\exp\left( -2\lambda_{\text{\tiny E}}^2\Delta N+\lambda_{\text{\tiny E}}\frac{\phi_\text{end}}{m_p}-e^{\lambda_{\text{\tiny E}}\frac{\phi_\text{end}}{m_p}} \right) \right] \, ,
    \end{equation}
where $W_{-1}$ is the `$-1$ branch' of the Lambert $W$ function~\cite{NIST:DLMF}. For the T-model potential, we have
    \begin{equation}\label{eq:phi_star_T}
        \lambda_{\text{\tiny T}}\frac{\phi_\star}{m_p}=\frac{1}{2}\text{arccosh}\left[ 8\lambda_{\text{\tiny T}}^2\Delta N+\cosh\left( 2\lambda_{\text{\tiny T}}\frac{\phi_\text{end}}{m_p} \right) \right] \, .
    \end{equation}
Eqs.~\eqref{eq:phi_star_E} and \eqref{eq:phi_star_T}, along with the CMB normalisation $\mathcal{P}_{\zeta\star}=2.1\times 10^{-9}$, can then be used to constrain $V_0$ for different choices of $\lambda$ thereby constraining $m$. 
    
Although  the aforementioned analytical approach was presented based on the slow-roll approximations, in practice, given that slow-roll approximations break down towards the end of inflation,  we determined the parameters in Tables~\ref{tab:Sim_Parameters_emodel}~and~\ref{tab:Sim_Parameters_tmodel} numerically, in order to be more accurate.

\section{Relation between number of $e$-folds and number of oscillations}
\label{app:N_osc_vs_DeltaN}
For  homogeneous scalar field oscillations around a quadratic potential, the time-averaged EoS vanishes, \textit{i.e.}  $\langle w_{\phi} \rangle = 0$, leading to a matter dominated expansion. The same is also true when the scalar field fragments to form oscillons. So in both the cases, we have 
$$ \langle \rho_\phi \rangle = \rho_i \, \l( \f{a_i}{a}\r)^3 \, .$$
Substituting this into the Friedmann equation 
$$ \f{\dot{a}}{a} = \sqrt{ \f{\rho_i }{3m_p^2} \, \l( \f{a_i}{a}\r)^3 } \, ,$$
and integrating from initial time $t_i$ to some time $t$, we obtain
$$ a(t) = a_i \, \l[ 1 \, + \, \f{\sqrt{3}}{2} \, \l( \f{\rho_i}{m_p^2} \r)^{1/2}  \l( t-t_i \r)\r]^{2/3} \, $$
which leads to
$$ t-t_i = \sqrt{\f{4}{3}} \, \l( \f{m_p^2}{\rho_i} \r)^{1/2} \l[ \l(\f{a}{a_i}\r)^{3/2} - 1\r] \, . $$
Since the period of oscillations is given by $T = 2\pi/m$,  where $m$ is the mass of the field, the number of oscillations is given by
\beq
\f{t-t_i}{T}  \equiv  n_{\rm osc} = \f{1}{\sqrt{3\pi^2}} \, \f{m \, m_p}{\sqrt{\rho_i}} \, \l[ e^{\f{3}{2} \, \Delta N} - 1 \r] \, .   
\label{eq:n_osc_1}
\eeq
Noting that  close to the minimum of the potential, $\rho_i \simeq \rho_e = 3/2 \, V(\phi) = 3/4 \, m^2 \phi_i^2$, we get the final expression for the number of oscillations to be
\beq
 n_{\rm osc} = \f{2}{3\pi} \, \l(\f{m_p}{\phi_i}\r) \, \l[ e^{\f{3}{2} \, \Delta N} - 1 \r] \, ,  
\label{eq:n_osc_2}
\eeq
or equivalently 
\beq
 \Delta N  = \f{2}{3}  \ln{\l[  1 + \f{3\pi}{2} \l( \f{\phi_i}{m_p} \r) n_{\rm osc}\r] }\, .  
\label{eq:n_osc_3}
\eeq
For example, assuming $\phi_i = 2 m_p/(3\pi)$, the number of $e$-folds of expansion after $10^3$ oscillations is given by $\Delta N \simeq 4.6$, indicating that oscillons which decay after $1000$ oscillations live up to $4.6$ $e$-folds of expansion.  {Note that Eq.~(\ref{eq:n_osc_3}) is not applicable when the EoS deviates substantially from $w=0$.}

\section{Post-inflationary tachyonic oscillations}
\label{app:tachyonicity}
 The  Fourier modes of the inflaton fluctuations $\delta \varphi_k$ satisfy the equation of a damped oscillator with a time-dependent frequency of the form (see Eq.~\eqref{eq:delphi_k_linear})
\begin{equation}
    \label{eq:app_fluc_EOM}
    \ddot{\delta \varphi}_k + 3H\dot{\delta \varphi}_k + \l[\f{k^2}{a^2}+ V_{,\phi\phi} (\phi)\r] \delta \varphi_k= 0 \, .
\end{equation}
 After the end of inflation, as the inflaton rolls down towards the minimum of the potential, there may be a region  in the field space where its effective mass-squared term becomes negative, \textit{i.e.},
    \begin{equation}
        m^2_\text{eff} (\phi) \equiv V_{,\phi\phi} (\phi) <0 \, .
    \end{equation}
In such a region,  the infrared modes of $\delta \varphi_k$ for which $k^2/a^2+V_{,\phi\phi}<0$ grow exponentially, triggering \textit{tachyonic instability} which offers a preheating channel distinct from the standard parametric resonance (where particle production occurs due to non-adiabatic time variation of the effective mass-squared term). During the course of oscillations, the inflaton condensate exhibits repeated field excursions to either sides of the inflection point, marked by $V_{,\phi\phi}\l( \phi_\text{inflection} \r) = 0$, giving rise to \textit{tachyonic oscillations}.  The number of tachyonic oscillations  depend on $\phi_{\rm in}$ (and hence on $\lambda$) as well as on the evolution of the Hubble friction term. Tachyonic instability is a common feature of many hilltop-type models~\cite{Antusch:2017vga} and asymptotically flat potentials~\cite{Tomberg:2021bll,Koivunen:2022mem}.

\begin{figure}[htb]
    \centering
    \subfloat{\includegraphics[width = 0.48\textwidth]{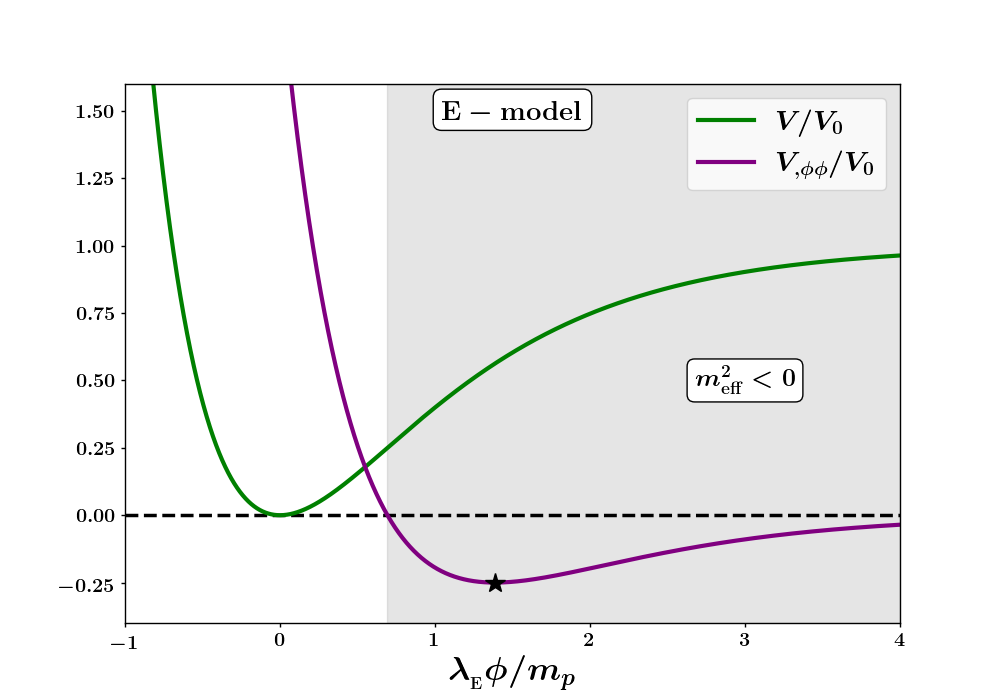}} 
    \subfloat{\includegraphics[width = 0.48\textwidth]{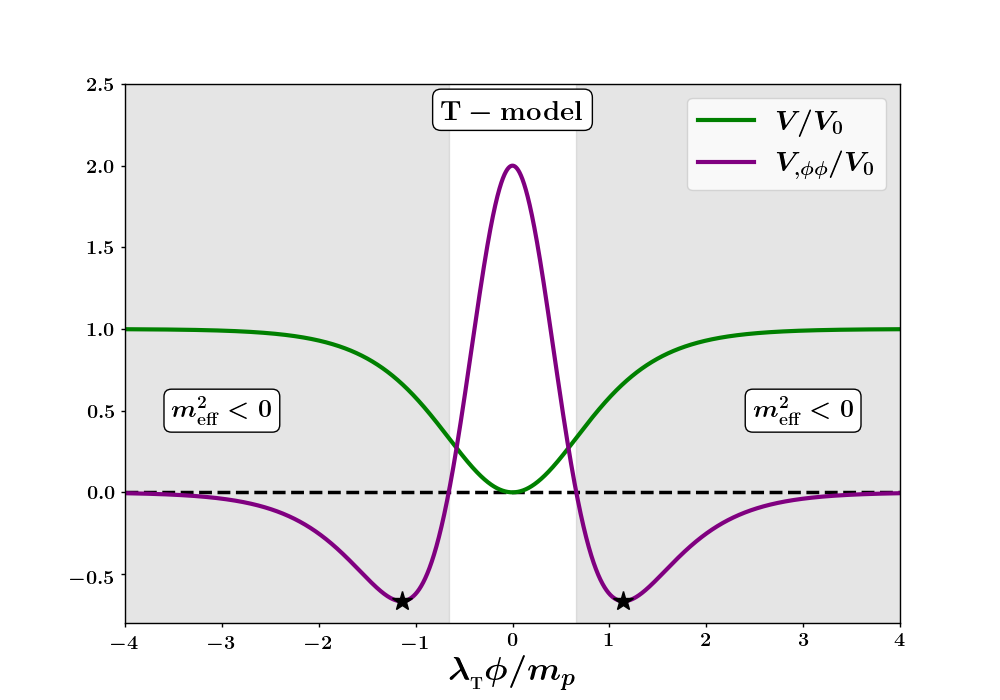}}
    \caption{An illustration of tachyonic regions in the E-model (\textbf{left}) and T-model (\textbf{right}) potentials for which  $m^2_\text{eff} < 0$. The  `$\star$' marks indicate the points for which the corresponding $m^2_\text{eff}$ values are at their minima (most-negative).}
    \label{fig:tachyonic_potentials}
\end{figure}

The  tachyonic instability region  for the E-model potential~(\ref{eq:inf_pot_E-model}) is shown in the left panel of Fig.~\ref{fig:tachyonic_potentials}. The grey shaded region indicates the presence of a negative $m_\text{eff}^2(\phi)$, with the  boundary given by 
 \begin{equation}
    \label{eq:app_tachyonic_bound_Emodel}
    \frac{\lambda_{\text{\tiny E}}\phi}{m_p} = \ln 2 \, .
\end{equation}
Similarly, for the T-model potential~(\ref{eq:inf_pot_T-model}), the grey-shaded regions in the right panel of Fig.~\ref{fig:tachyonic_potentials} indicate a  negative $m_\text{eff}^2$, with boundaries
\begin{equation}
    \label{eq:app_tachyonic_bound_Tmodel}
    \frac{\lambda_{\text{\tiny T}}\phi}{m_p}=\frac{1}{2}\ln\left( 2\pm\sqrt{3} \right) \, .
\end{equation}
The  `$\star$' marks indicate  points in the field space where the corresponding $m^2_\text{eff}$ values are at their minima (maximally negative).

\begin{figure}[htb]
    \centering
    \subfloat{\includegraphics[width = 0.48\textwidth]{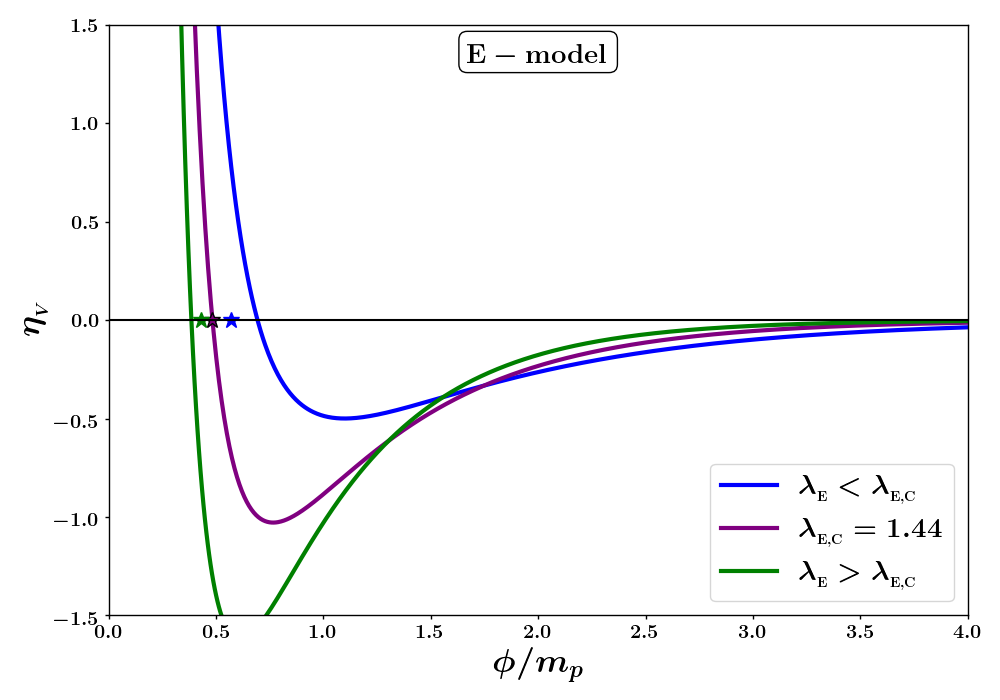}} 
    \subfloat{\includegraphics[width = 0.48\textwidth]{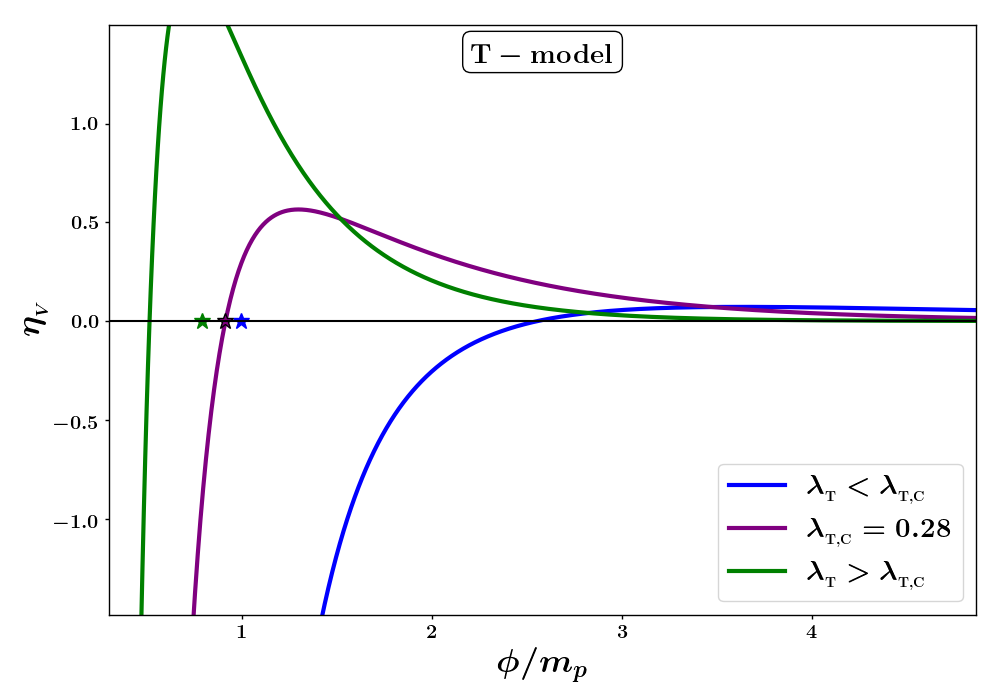}}
    \caption{ The figures illustrate how the possibility of exhibiting tachyonic oscillations in the E-model (\textbf{left}) and T-model (\textbf{right}) potentials depends upon the value of $\lambda$. Post-inflationary oscillations are tachyonic when the inflaton makes excursions to either side of $\eta_{_V} \l( \phi_\text{inflection} \r) = 0$. The `$\star$' marks correspond to field values ($\phi_\text{end}$) at the end of inflation, \textit{i.e.} $\epsilon_H(\phi_\text{end}) = 1$. The corresponding condition for post-inflationary tachyonic oscillations becomes $\phi_\text{inflection} < \phi_\text{end}$, which is achieved for  $\lambda > \lambda_\text{\tiny E,C} = 1.44$ (for E-model) or $\lambda > \lambda_\text{\tiny T,C} = 0.28$ (for T-model).} 
    \label{fig:tachyonic_eta_lambda}
\end{figure}

\begin{figure}[htb]
    \centering
    \subfloat{\includegraphics[width = 0.85\textwidth]{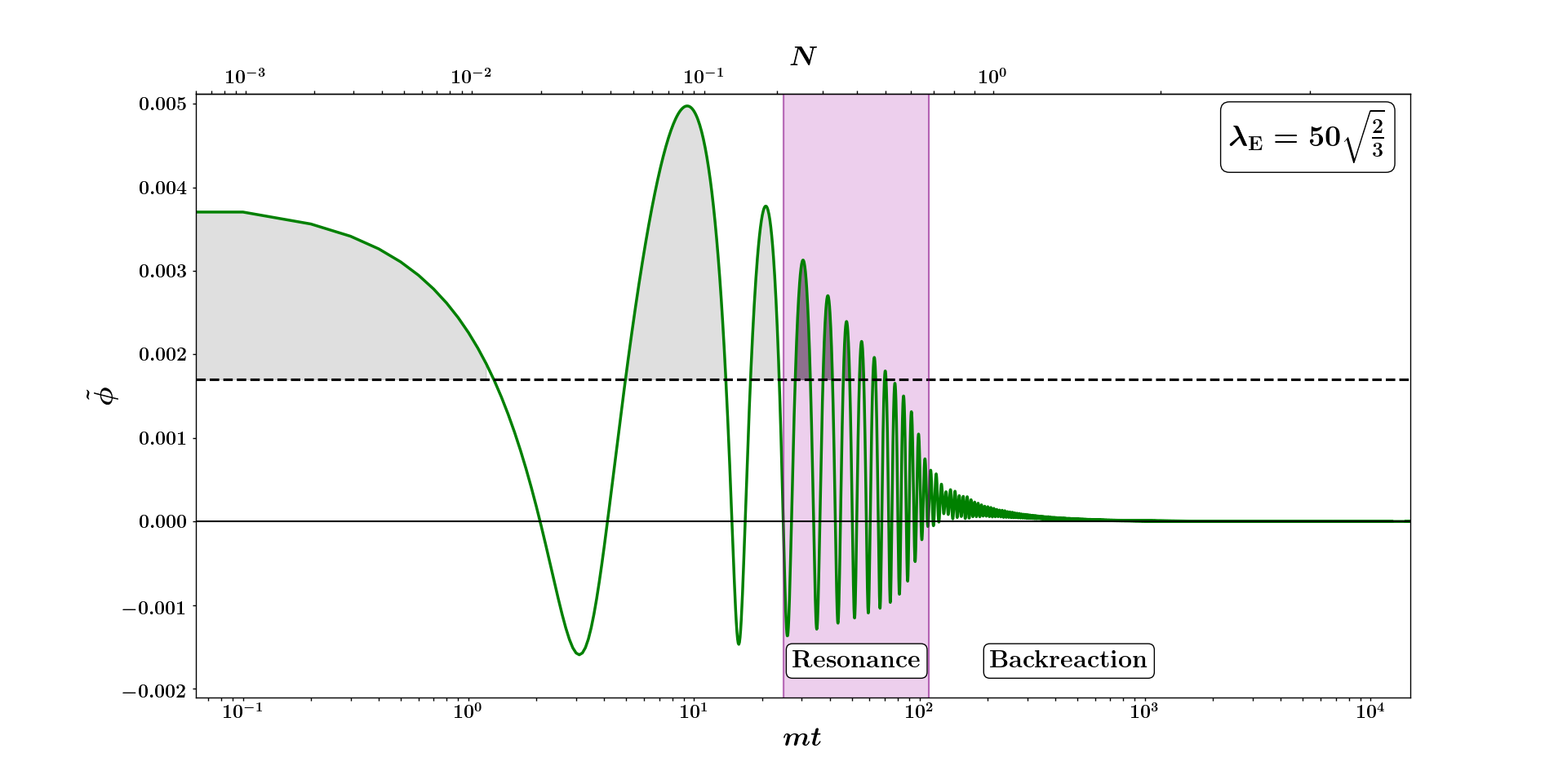}} 
    \caption{ Time  evolution of the homogeneous inflation condensate (with the upper horizontal axis being the number of $e$-folds $N$ elapsed since the end of inflation) is shown here for the E-model with $\lambda_\text{\tiny E} = 50 \sqrt{2/3}$. The grey shaded regions above the dashed line correspond to the regimes of tachyonic oscillations, which  are absent at late times, since the field amplitude decreases due to Hubble friction (as well as due to self-resonance).}
    \label{fig:phi_field_tachy}
\end{figure}

 A simple way to infer the existence of tachyonic oscillations is the following. The inflaton oscillations are tachyonic  if it performs field excursions to either sides of the inflection point, defined  by
\beq
\eta_{_V}\l( \phi_\text{inflection} \r) \equiv m^2_p \left( \frac{V_{, \phi\phi}}{V} \right)\bigg\vert_{\phi=\phi_\text{inflection}} =  0 \, ,
\eeq
after the end of inflation. This is possible  only if the inflection point satisfies $|\phi_\text{inflection}| < |\phi_\text{end}|$. Hence, the existence of tachyonic oscillations in the $\alpha$-attractor models ultimately depends upon  the value of $\lambda$ in Eqs.~\eqref{eq:inf_pot_E-model} and \eqref{eq:inf_pot_T-model}. Figure~\ref{fig:tachyonic_eta_lambda} demonstrates  that for   $\lambda_\text{\tiny E,C} > 1.44$ for the E-model, and for $\lambda_\text{\tiny T,C} > 0.28$ for the T-model, the first few   post-inflationary oscillations are tachyonic. Tachyonic oscillations  for the E-model potential with $\lambda_{\text{\tiny E} } = 50\sqrt{2/3}$ is shown by the grey shaded regions in Fig.~\ref{fig:phi_field_tachy}. We note that the oscillations cease to be tachyonic  at late times, since the field amplitude decreases due to Hubble friction (as well as due to self-resonance and particle production in general).
\section{The impact of the formation and decay of oscillons on the equation of state}
\label{subsec:EOS_oscillons}
The  asymptotic  value towards which the equation of state (EoS) of the system eventually settles depends on the nature of the self-interaction potential near the minimum (quadratic or not, up to the leading order) as well as on the presence of inhomogeneities. Much like the way a system of self-gravitating masses virialises, it can be shown that the post-inflationary scalar field dynamics,  under the influence of an attractive self-interaction, also eventually virialises,\footnote{In practice,  the system virialises only approximately, in the sense that there will always be residual deviations away from the virial theorem  for the averaged energy density components. However,  the virial theorem can still be used to explain some of the details of the behaviour of the EoS. The energy density components satisfy the virial theorem to a high degree once the system has evolved for a sufficiently long time.} obeying~\cite{Lozanov:2017hjm,Antusch:2022mqv}
\begin{equation}\label{eq:virial_theorem}
    \Bigg\langle \left( \frac{\partial\tilde{f}}{\partial\tilde{t}} \right)^2 \Bigg\rangle_\text{\tiny V,T} = \Bigg\langle \left(\frac{ \widetilde{\grad}\tilde{f}}{a} \right)^2 \Bigg\rangle_\text{\tiny V,T} + \Bigg\langle \tilde{f} \, \frac{\partial\widetilde{V}}{\partial\tilde{f}} \Bigg\rangle_\text{\tiny V,T}; \quad \tilde{f}=\{\widetilde{\varphi},\widetilde{\chi}\} \, ,
\end{equation}
where $\langle \cdots \rangle_\text{\tiny V,T}$ refers to volume and time-averaging and $\widetilde{V}$ takes into account the interaction term as well (here, it is a shorthand for $\widetilde{V}+\widetilde{\mathcal{I}}$). Since both the E- and T-model potentials  behave as $V(\varphi)\propto \varphi^{2n}$  (with $n=1$) near their minima (up to the leading order), we find that the virialisation condition in Eq.~\eqref{eq:virial_theorem} can be expressed as
    \begin{align}
        \frac{1}{2}\Bigg\langle \left( \frac{\partial\widetilde{\varphi}}{\partial\tilde{t}} \right)^2 \Bigg\rangle_\text{\tiny V,T} &= \frac{1}{2}\Bigg\langle \left( \frac{\widetilde{\grad}\widetilde{\varphi}}{a} \right)^2 \Bigg\rangle_\text{\tiny V,T} + n\Big\langle \widetilde{V}\left( \widetilde{\varphi} \right) \Big\rangle_\text{\tiny V,T} + 2\Big\langle \widetilde{\mathcal{I}}\left( \widetilde{\varphi},\widetilde{\chi} \right) \Big\rangle_\text{\tiny V,T} \, , \nonumber \\
        \Big\langle \widetilde{K}_{\tilde{\varphi}} \Big\rangle_\text{\tiny V,T} &= \Big\langle \widetilde{G}_{\tilde{\varphi}} \Big\rangle_\text{\tiny V,T}+n\Big\langle \widetilde{V}\left( \widetilde{\varphi} \right) \Big\rangle_\text{\tiny V,T} + 2\Big\langle \widetilde{\mathcal{I}}\left( \widetilde{\varphi},\widetilde{\chi} \right) \Big\rangle_\text{\tiny V,T} \, ,
        \label{eq:virial_phi}
    \end{align}
    and
    \begin{align}
        \frac{1}{2}\Bigg\langle \left( \frac{\partial\widetilde{\chi}}{\partial\tilde{t}} \right)^2 \Bigg\rangle_\text{\tiny V,T} &= \frac{1}{2}\Bigg\langle \left( \frac{\widetilde{\grad}\widetilde{\chi}}{a} \right)^2 \Bigg\rangle_\text{\tiny V,T} \, , \nonumber \\
        \Big\langle \widetilde{K}_{\tilde{\chi}} \Big\rangle_\text{\tiny V,T} &= \Big\langle \widetilde{G}_{\tilde{\chi}} \Big\rangle_\text{\tiny V,T} \, .
        \label{eq:virial_chi}
    \end{align}
    In the limit where the gradient and the interaction energy densities are subdominant, we notice, using the definition in Eq.~\eqref{eq:rescaled_EoS_total}, that Eqs.~\eqref{eq:virial_phi} and \eqref{eq:virial_chi} yield the familiar relation for the EoS parameter
    \begin{equation}
        \langle \widetilde{w} \rangle_\text{\tiny V,T} = \frac{n-1}{n+1} \, ,
    \end{equation}
    as  expected from a homogeneous condensate. 
    \begin{figure}[htb]
        \centering
        \subfloat{\includegraphics[width = 0.48\textwidth]{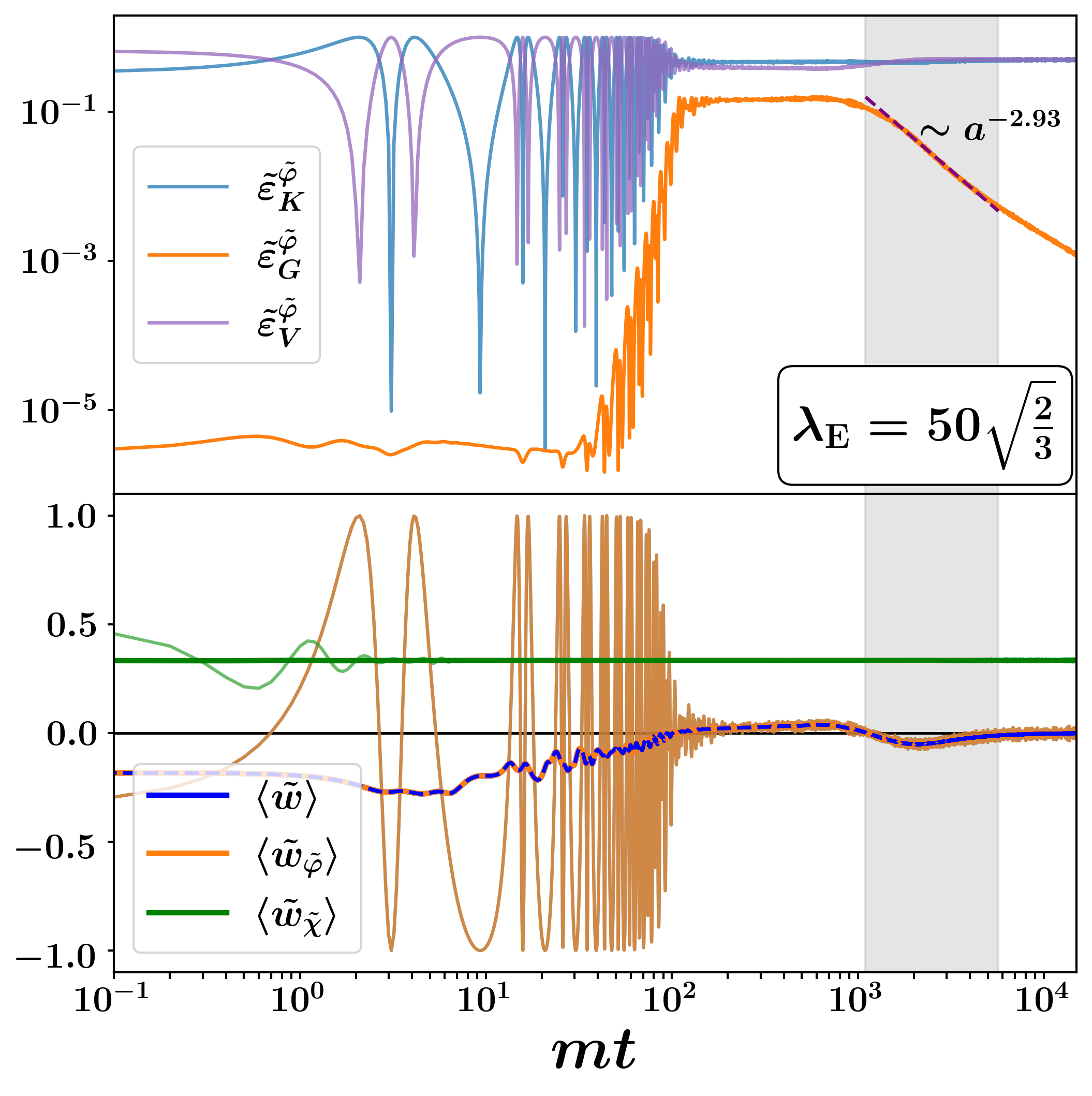}} 
        \subfloat{\includegraphics[width = 0.48\textwidth]{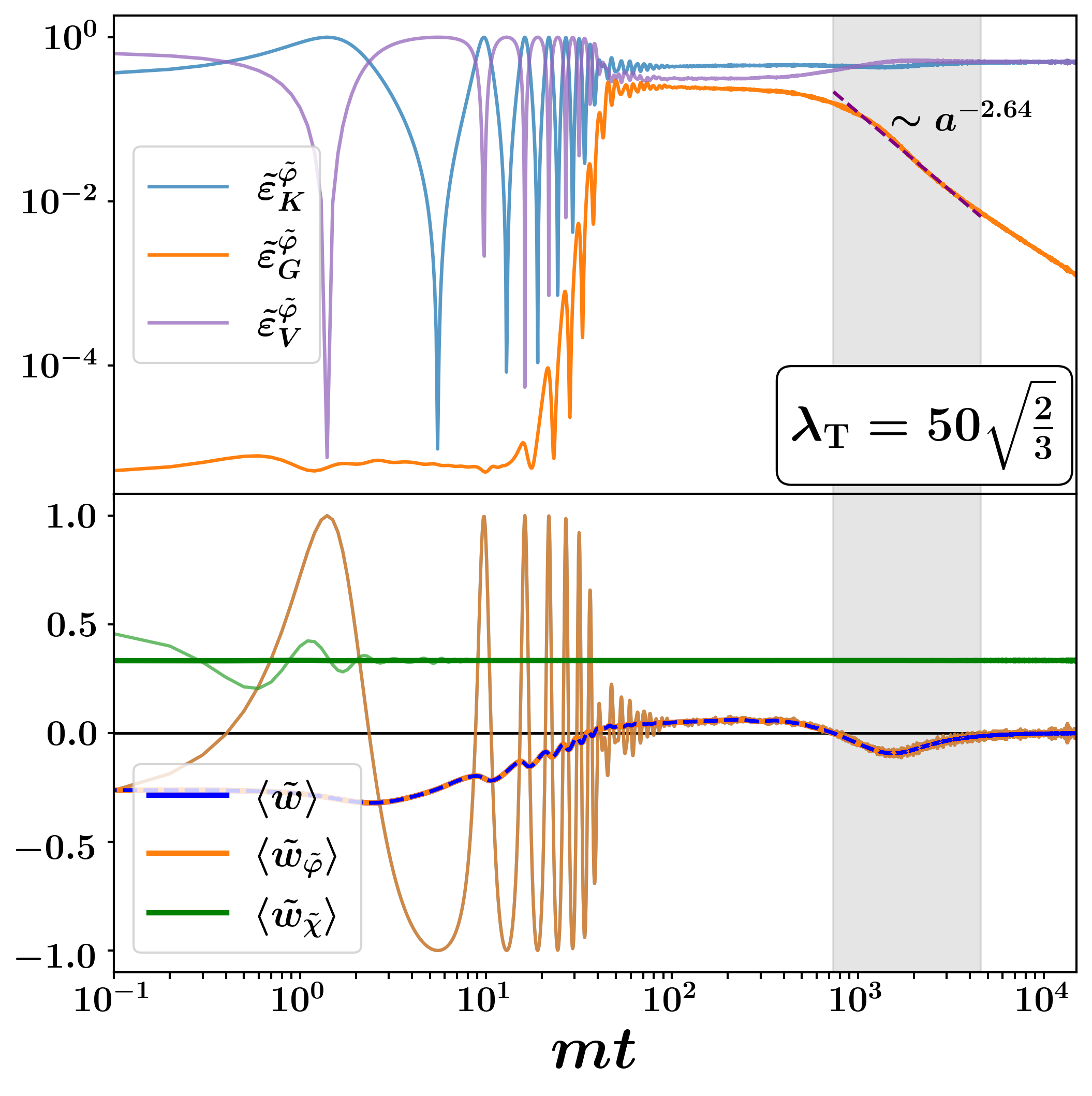}} \\ 
        \caption{ Evolution of the time-averaged equation of state $\left< {\widetilde w} \right>$ in the absence of an external coupling for the E-model (\textbf{left}) and the T-model (\textbf{right}) are shown, superposed with the corresponding fractional energy density components $\tilde{\varepsilon}_i$. The positive deviations from $\left< {\widetilde w} \right> \simeq 0$ are due to  the potential term (purple curve) falling below the kinetic term (blue curve) transiently. We also highlight regions where the EoS display negative deviations (in the shaded gray regions) due to the fact that the potential term becomes higher than the kinetic term for a short period (although the sub-dominant gradient term falls faster), prior to achieving virialisation. Nonetheless, as the system evolves, $\left< {\widetilde w} \right>$ asymptotically approaches zero.}
        \label{fig:EoS_no_ext_coupling}
    \end{figure} 
    
\begin{figure}[hbt]
        \centering
            \subfloat{\includegraphics[width = 0.48\textwidth]{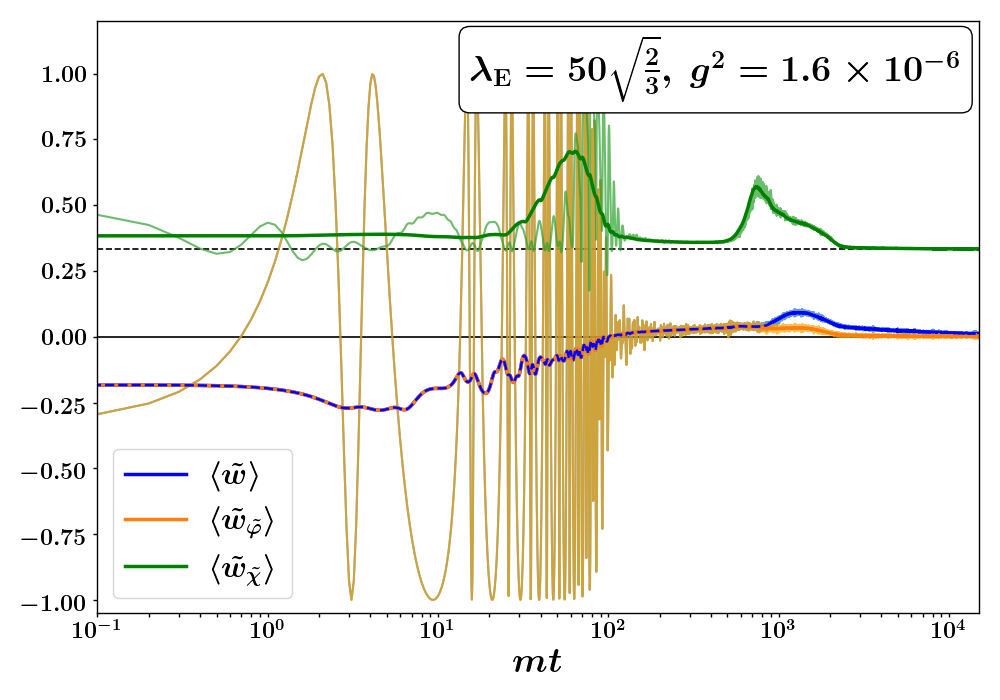}} 
            \subfloat{\includegraphics[width = 0.48\textwidth]{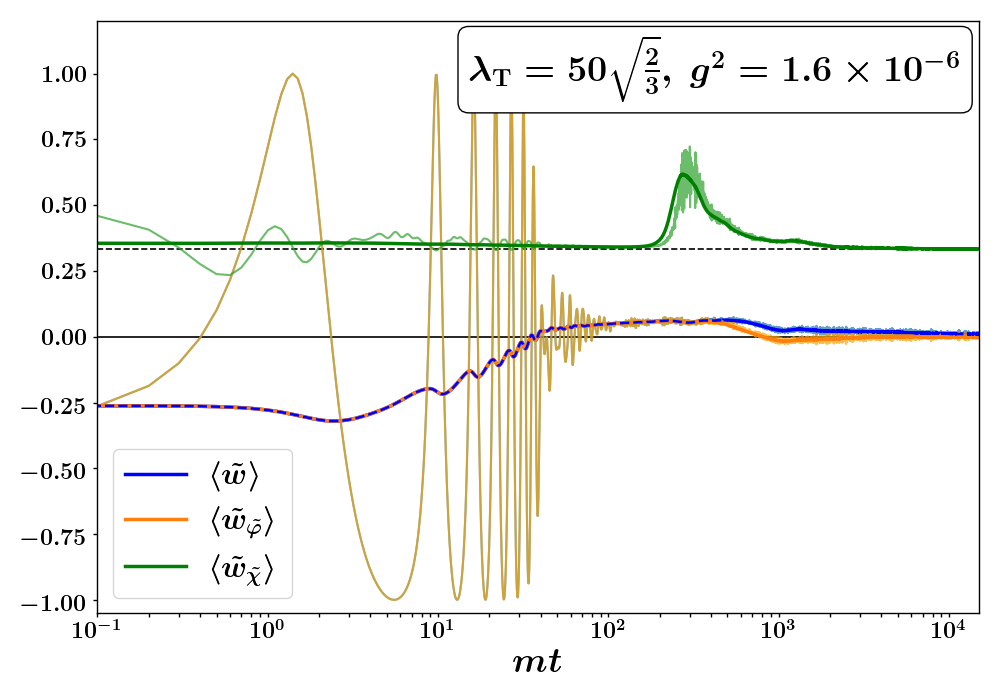}} \\
            \subfloat{\includegraphics[width = 0.48\textwidth]{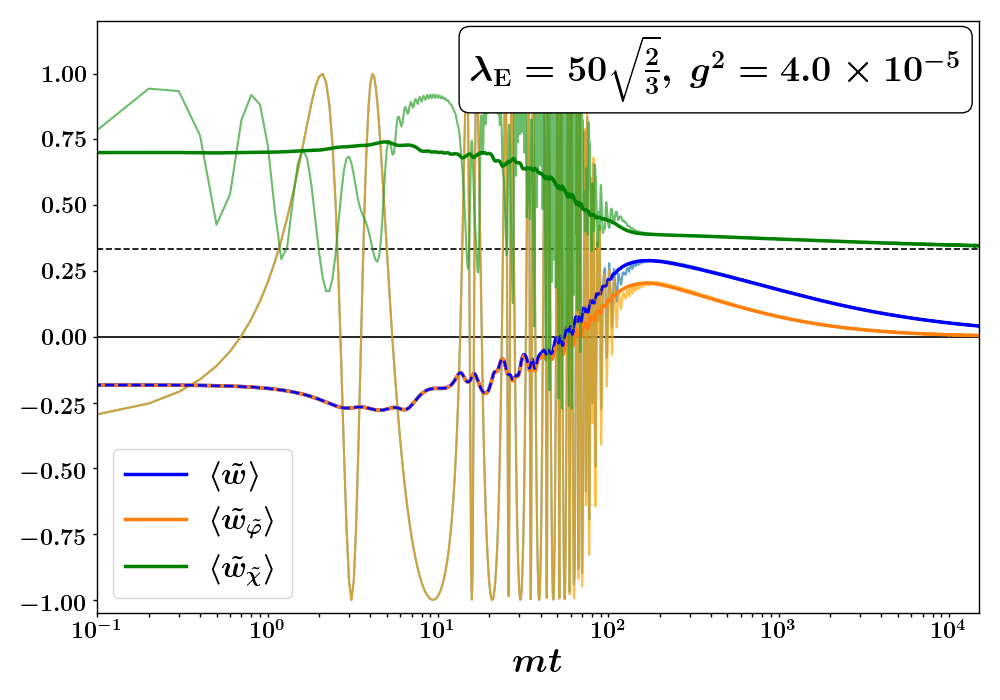}} 
            \subfloat{\includegraphics[width = 0.48\textwidth]{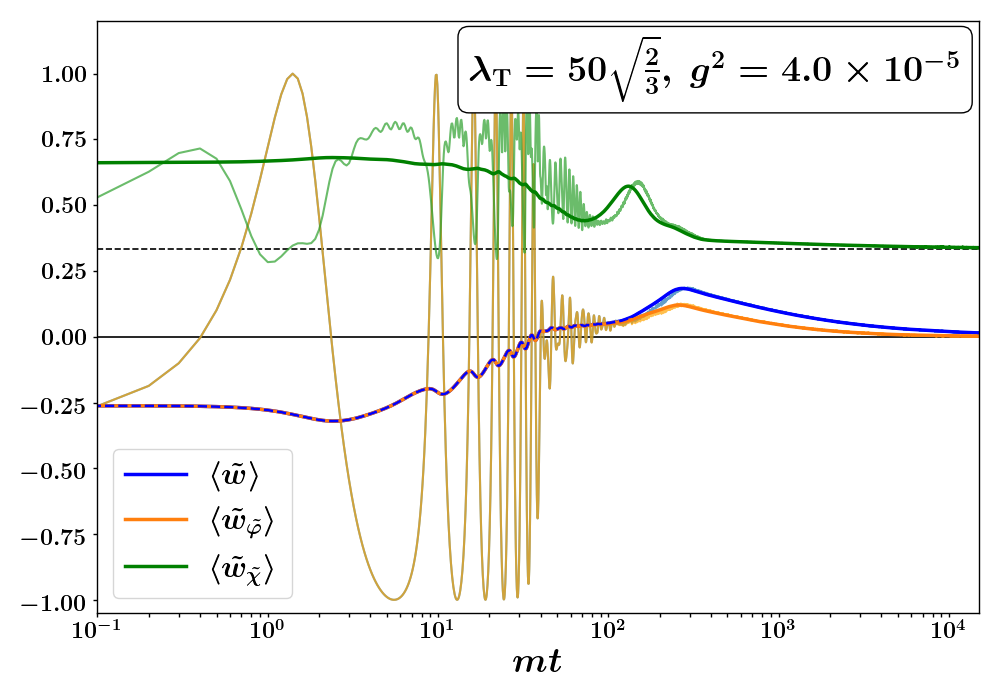}} 
        \caption{Evolution of the time-averaged equation of state $\left< {\widetilde w} \right>$ in the presence of an external coupling $g^2$, for the E-model (\textbf{left}) and the T-model (\textbf{right})  potentials are shown. The presence of such an external coupling can disrupt the formation (or accelerate the decay) of oscillons. Again, as the system evolves with time, $\left< {\widetilde w} \right>$ asymptotically approaches zero.}
        \label{fig:EoS_ext_coupling}
    \end{figure}

Evolution of the average EoS obtained for different $\{ \lambda, \, g^2 \}$ can then be explained by using the virial condition along with examining the relative variations in the different energy components.  Since the self-interaction potentials used in this work possess quadratic minima, \textit{i.e.} they correspond to $n=1$, the observed late-time asymptote $\langle \widetilde{w} \rangle_{\text{\tiny V,T}}\rightarrow 0$ (pressureless dust behaviour)  in Figs.~\ref{fig:EoS_no_ext_coupling} and \ref{fig:EoS_ext_coupling}  can be inferred from the relative equal proportions of kinetic and potential energies. This is particularly true at late times when all gradient and interaction terms become subdominant,  while the kinetic and potential terms of $\widetilde{\varphi}$ balance each other out. The EoS of the $\chi$-field does not contribute to the overall EoS, even though $\langle \widetilde{w}_{\tilde{\chi}} \rangle_\text{\tiny V,T}=1/3$, since   the production of $\chi$-particles is inefficient at late times  in the absence of parametric resonance, and the number density of  existing $\chi$-particles gets redshifted away. Tiny fluctuations in their number densities may still be  observed  due to the slow decay of the inhomogeneities in the $\varphi$-field into the $\chi$ field fluctuations.

Of course, in Fig.~\ref{fig:EoS_no_ext_coupling} there are deviations away from the $\langle \widetilde{w} \rangle_\text{\tiny V,T}=0$ behaviour, which is particularly evident during the onset of resonance and backreaction phases and for some period of time afterwards. When $g^2=0$, during the resonance and backreaction phases, we know that the gradient energy of the $\widetilde{\varphi}$-field becomes comparable to the kinetic and potential energies. Hence, we find that
    \begin{equation}
        \langle \widetilde{w} \rangle_\text{\tiny V,T} \approx \frac{1}{3} \left[ 1 + \frac{\langle \widetilde{V} \rangle_\text{\tiny V,T}}{\langle \widetilde{G}_{\tilde{\varphi}} \rangle_\text{\tiny V,T}} \right]^{-1}, \quad (g^2=0) \, .
    \end{equation}
    Considering that $\langle \widetilde{V} \rangle_\text{\tiny V,T} > \langle \widetilde{G}_{\tilde{\varphi}} \rangle_\text{\tiny V,T}$ during resonance and backreaction, as can be seen in the top panels of Fig.~\ref{fig:energy_den_comp_smaller_lambda}, we can conclude that during these phases
    \begin{equation}
        0 < \langle \widetilde{w} \rangle_\text{\tiny V,T} < \frac{1}{3}  \, .
    \end{equation}
    
To explain the negative deviations in the EoS shortly after backreaction, we refer to the top panel of Fig.~\ref{fig:EoS_no_ext_coupling} which displays the evolution of the different energy components $\langle \widetilde{E}_i \rangle$, where $i=\{ K,G,V \}$, as a fraction of the total energy. It can be observed that $\widetilde{G}_{\tilde{\varphi}}\propto a^{-3}$ for the duration of the backreaction phase (indicative of the inhomogeneities behaving like pressureless matter). Afterwards, however, $\widetilde{G}_{\tilde{\varphi}}$ experiences a rather sharp decline. During this period $\langle \widetilde{V}_{\tilde{\varphi}} \rangle_\text{\tiny V,T} > \langle \widetilde{K}_{\tilde{\varphi}} \rangle_\text{\tiny V,T}$ (only slightly) which we can use to account for the small negative values of the EoS. Approximate scaling relations of the gradient energies are shown in Fig.~\ref{fig:EoS_no_ext_coupling}. Since $\widetilde{\rho}_{\tilde{\varphi}}\propto a^{-3}$, we find that  
    \begin{equation}
        \widetilde{G}_{\tilde{\varphi}}\propto 
            \begin{dcases}
                a^{-5.93},\quad\quad\text{(E-model)}  \, ;\\
                a^{-5.64},\quad\quad\text{(T-model)} \, .
            \end{dcases}
    \end{equation}
\indent For $g^2\neq 0$, even though the system tends to a matter-dominated phase in the long run, there are more interesting transient behaviour owing to the presence of coupling between the two fields. The $\widetilde{\chi}$-field  becomes virialised after the first few oscillations such that $\langle \widetilde{w}_{\tilde{\chi}} \rangle_\text{\tiny V,T}\sim1/3$. This can also be seen in the middle and bottom panels of Fig.~\ref{fig:energy_den_comp_smaller_lambda} where $\langle \widetilde{K}_{\tilde{\chi}} \rangle_\text{\tiny V,T}$ and $\langle \widetilde{G}_{\tilde{\chi}} \rangle_\text{\tiny V,T}$ track each other throughout their evolution. There are periods when $\langle \widetilde{w}_{\tilde{\chi}} \rangle_\text{\tiny V,T}$ become stiff, coinciding with periods when $\widetilde{\mathcal{I}}(\widetilde{\varphi},\widetilde{\chi})$ peaks. However, since the $\widetilde{\chi}$-field only provides a subdominant contribution to the energy of the system, the overall EoS does not become stiff and any period in its evolution, and any positive deviations that arise relax to the EoS of a matter-dominated system.

\printbibliography

\end{document}